\documentclass[12pt,preprint]{aastex}

\shorttitle{X-ray Substructure Studies of Four Galaxy Clusters}
\shortauthors{Zhang et al.}

\begin{document}


\title{X-ray Substructure Studies of Four Galaxy Clusters using
\emph{XMM-Newton} Data\altaffilmark{5}}


\author{Yu-Ying Zhang\altaffilmark{1}} 

\author{Thomas H. Reiprich\altaffilmark{1}}

\author{Alexis Finoguenov\altaffilmark{2,3}}

\author{Daniel S. Hudson\altaffilmark{1}}

\and 

\author{Craig L. Sarazin\altaffilmark{4}}


\altaffiltext{1}{Argelander-Institut f\"ur Astronomie, Universit\"at Bonn, Auf dem H\"ugel 71, 53121 Bonn, Germany}
\altaffiltext{2}{Max-Planck-Institut f\"ur extraterrestrische Physik, Giessenbachstra\ss e, 85748 Garching, Germany}
\altaffiltext{3}{University of Maryland, Baltimore County, 1000 Hilltop Circle, Baltimore, MD 21250, USA}
\altaffiltext{4}{Department of Astronomy, University of Virginia, P.O.~Box 400325, Charlottesville, VA 22904, USA }
\altaffiltext{5}{This work is based on
  observations made with the \emph{XMM-Newton}, an ESA science mission
  with instruments and contributions directly funded by ESA member
  states and the USA (NASA).}


\begin{abstract} Mahdavi et al. find that the degree of agreement between weak
  lensing and X-ray mass measurements is a function of cluster radius.
  Numerical simulations also point out that X-ray mass proxies do not work
  equally well at all radii. The origin of the effect is thought to be
  associated with cluster mergers. Recent work presenting the cluster maps
  showed an ability of X-ray maps to reveal and study cluster mergers in
  detail.  Here, we present a first attempt to use the study of substructure in
  assessing the systematics of the hydrostatic mass measurements using
  two-dimensional (2D) X-ray diagnostics. The temperature map is uniquely
  able to identify the substructure in an almost relaxed cluster which would
  be unnoticed in the ICM electron number density, and pressure maps. We
  describe the radial fluctuations in the 2D maps by a
  cumulative/differential scatter profile relative to the mean profile
  within/at a given radius. The amplitude indicates $\sim 10$\% fluctuations
  in the temperature, electron number density, and entropy maps, and $\sim
  15$\% fluctuations in the pressure map. The amplitude of and the
  discontinuity in the scatter complement 2D substructure diagnostics, e.g.,
  indicating the most disturbed radial range. There is a tantalizing link
  between the substructure identified using the scatter of the entropy and
  pressure fluctuations and the hydrostatic mass bias relative to the expected
  mass based on the $M$-$Y_{\rm X}$ and $M$-$M_{\rm gas}$ relations
  particularly at $r_{500}$.  \emph{XMM-Newton} observations with $\sim
  120,000$ source photons from the cluster are sufficient to apply our
  substructure diagnostics via the spectrally measured 2D temperature,
  electron number density, entropy, and pressure maps.
\end{abstract}





\section{Introduction}

The robustness of cluster mass estimates has become more and more
important as galaxy clusters have been widely used as important
cosmology tools (e.g., Ebeling et al. 1998, 2000; Allen et al. 2002;
B\"ohringer et al. 2004; Vikhlinin et al. 2009a, 2009b). Precision
cluster cosmology experiments using the mass function are based on
accurately calibrated mass-observable scaling relations in terms of
their shape, scatter, and evolution (e.g., Vikhlinin et al. 2006;
Arnaud et al.  2007).  To calibrate the mass-observable scaling
relations, the first necessary task is to obtain well-understood
measurements of the cluster mass and observables (e.g., B\"ohringer et
al.  2007; Zhang et al. 2006, 2008). X-ray measurements provide an
important estimate of the cluster mass.  With deep X-ray observations
from \emph{XMM-Newton} and \emph{Chandra}, one can precisely trace
both temperature and electron number density distributions of the
intracluster medium and thus measure the mass distributions with
statistical uncertainties below 15\% up to $r_{500}$ (e.g., Vikhlinin
et al.  2006).  However, the accuracy of X-ray cluster mass estimates
is limited by additional physical processes in the ICM and projection
effects.  Although the current total cluster mass calibration between
two independent approaches, weak lensing and X-ray, shows an agreement
(e.g., Mahdavi et al.  2008, Zhang et al.2008), a radial dependence is
found in the ratio of weak lensing and X-ray mass measurements (e.g.,
Mahdavi et al.  2008). Such a radial dependence is thought to be due
to a bias in the hydrostatic mass estimates (e.g., Nagai et al. 2007).

Cluster merging is one of many effects causing biases in the X-ray
mass estimates. Previous results on X-ray cluster maps (e.g.,
Vikhlinin et al. 2001, Markevitch et al. 2003, Schuecker et al.  2004,
Finoguenov et al. 2005) show that clusters are not simple hydrostatic
equilibrium systems. Both relaxed and unrelaxed clusters may be
affected by additional non-thermal pressure processes. Particularly,
merging clusters of galaxies are often not in a hydrostatic
equilibrium state. Cluster mergers change the X-ray luminosities and
temperatures of clusters, both in a transient sense and in the long
term (e.g.,  Ricker \& Sarazin 2001; Poole et al. 2006, 2007), and also
dramatically affect the properties of their galaxies (e.g., Sun et al.
2007). The temperature distribution, as an important input in the
X-ray mass estimate, can cause biases in the X-ray measured mass
distribution. Mergers seriously affect both mass estimates and
observables, and thus the scaling relations of galaxy clusters
(e.g., Evrard et al.  2008).  Elimination of systematic uncertainties
from the scaling relation calibration demands that major cluster
mergers are identified and effects of major mergers on cluster mass
estimates are quantified. Substructure can be used to identify and
trace the merging process and the substructure fraction can be used to
link the cluster mass systematics with the mass assembly history (e.g.,
Smith \& Taylor 2008).

Substructure studies are enormously important to understand cluster
mass estimates and the drivers of the scaling relations. Cluster
mergers create disturbances associated with both shocks and mixing of
the stripped gas. As parameters controlling their relative importance,
e.g., the viscosity, are not so well constrained (Sijacki \& Springel
2006), it is unclear how much each contributes and at which scales
each effect dominates. Observationally, we are able to provide 
better constraints using spatial fluctuations of the temperature,
electron number density, entropy and pressure maps. Observational
results can be compared with numerical simulations with different
prescriptions to reveal more details of the merging physics.

We aim to perform quantitative substructure studies using X-ray
spectrally measured two-dimensional (2D) maps to access the systematic
errors in cluster mass measurements due to departures from hydrostatic
equilibrium.  Substructures in galaxy clusters have been intensely
investigated since the \emph{ROSAT} era using the X-ray surface
brightness distribution from observations and simulations (e.g., Jones
\& Forman 1984; B\"ohringer et al. 2007; Ventimiglia et al. 2008;
Piffaretti \& Valdarnini 2008). However, substructures are less
obvious in the X-ray surface brightness distribution than in the
temperature distribution (Riemer-Sorensen et al. 2009; Jee \& Tyson
2009; Andersson et al.  2009). The \emph{Chandra} and
\emph{XMM-Newton} telescopes, with their high spatial resolution,
conveniently provide us the opportunity to perform substructure
studies using also the temperature map.  Most such studies derive
approximate X-ray temperature maps via X-ray hardness ratio maps
(e.g., Fabian et al.  2001, 2002; Churazov et al.  2003; Markevitch et
al.  2001, 2005; Finoguenov et al.  2005; Zhang et al.  2005; Forman
et al.  2007).  An alternative and more reliable way to derive X-ray
temperature maps with high precision is to perform a spectral analysis
in each spatial bin.  This method avoids, for instance, spurious
temperature variations due to underlying metallicity variations
because the metallicity is determined simultaneously (e.g., Henry et
al. 2004; Reiprich et al.  2004, 2009; Belsole et al.  2004; Pratt et
al.  2005; Sanderson et al. 2005; Sakelliou \& Ponman 2006; Sanders \&
Fabian 2007; Simionescu et al.  2007; Kapferer et al. 2008).

In this paper, we use the spatial fluctuations in the ICM temperature,
electron number density, entropy, and pressure in the 2D maps as substructure
indicators and the deviation of the mass-observable data pair from the
mass-observable relations of relaxed clusters as an estimate of the mass bias.
In Sect.~\ref{s:method}, we describe the key steps in the data reduction,
particularly emphasizing the background subtraction. Our technique to measure
the spectral temperature is shown in Sect.~\ref{s:spe} and how to derive the
2D maps using the spectral analysis is shown in Sect.~\ref{s:map},
respectively. We briefly describe the mass modeling in Sect.~\ref{s:mass}, and
show our results based on spectrally measured X-ray 2D maps in
Sect.~\ref{s:scatter}. We summarize our conclusions in
Sect.~\ref{s:conclusion}.  Unless explicitly stated otherwise, we adopt a flat
$\Lambda$CDM cosmology with the density parameter $\Omega_{\rm m}=0.3$ and the
Hubble constant $H_{\rm 0}=70$~km~s$^{-1}$~Mpc$^{-1}$. We adopt the solar
abundance table of Anders \& Grevesse (1989). Confidence intervals correspond
to the 68\% confidence level. The Orthogonal Distance Regression package
(ODRPACK~2.01\footnote{http://www.netlib.org/odrpack and references therein},
e.g., Boggs et al. 1987) taking into account measurement errors on both
variables is used for example, to derive
correlations between observationally derived parameters. We use Monte
Carlo simulations to evaluate the propagation of the errors in the
X-ray mass modeling on all quantities of interest.

\section{Data Reduction}
\label{s:method}

Spectrally measured 2D X-ray maps using most existing techniques
require high photon statistics and are only applied to a few very
nearby clusters/galaxies (e.g., Henry et al.  2004; Reiprich et al.
2004; Belsole et al.  2004; Pratt et al.  2005; Sakelliou \& Ponman
2006; Sanders \& Fabian 2007; Simionescu et al.  2007). In most
previous studies, a relatively simple blank sky background subtraction
was often applied. Therefore, high signal-to-noise ratio (S/N) data are
required to avoid large uncertainties caused by the background
modeling. These technical limitations make this approach applicable
only to targets with extremely good photon statistics, requiring 1-2
orders of magnitude higher exposures than the typical archival data
for nearby clusters. It is a challenge to carry out such studies on
medium quality data.

A precise background subtraction method could make spectrally measured
map analysis possible also for medium quality \emph{XMM-Newton} data
to increase the size of the cluster sampling. Snowden et al.  (2008)
developed a precise background modeling method but only for the MOS
data and only to measure the radial temperature profile. We adopted
their method and developed an advanced background modeling pipeline,
which is applicable to both pn and MOS data and which can be used to
measure the spectral temperature for both the radial analysis and the
map analysis. It allows us to perform reliable spectral analysis in
each spatial bin to derive the X-ray maps, but for clusters with
\emph{XMM-Newton} data with $\ge$~120,000 source counts in total.

The XMMSAS version 7.1.0 software combined with our in-house-developed pipeline is
used for data reduction.

\subsection{Data Selection}

To demonstrate the robustness of the method and to determine the S/N
threshold for such substructure studies, we composed a sample of four
clusters showing different morphologies as well as different photon
statistics in their \emph{XMM-Newton} data. Four clusters of galaxies
are selected from the HIghest X-ray FLUx Galaxy Cluster Sample
(HIFLUGCS\footnote{The HIFLUGCS sample consists of 64 X-ray brightest
  galaxy clusters in the extragalactic sky. They were selected from
  the \emph{ROSAT} All-Sky Survey (RASS), irrespective of their
  morphology, simply applying an X-ray flux limit.}; Reiprich \&
B\"ohringer 2002) according to the following criteria. (1) The
$r_{2500}$\footnote{$r_{\Delta}$ is the radius within which the
  density contrast to the critical density is $\Delta$.  $M_{\Delta}$
  is the total mass within $r_{\Delta}$. For example, for
  $\Delta=200$, $r_{200}$ is the radius within which the density
  contrast is 200 and $M_{200}$ is the total mass within $r_{200}$.
  The $r_{200}$ used here is derived from the cluster global
  temperature in Reiprich \& B\"ohringer (2002) and the $M_{200}$-$T$
  relation from simulations in Evrard er al.  (1996).} fits the
\emph{XMM-Newton} field of view (FOV). (2) The photon statistics are
sufficient for the map analysis using spectral measurements but varies
in a small range which gives 10-60 bins, of which the uncertainty of
the spectrally measured temperature in each spatial bin is $\simeq
10$\%. (3) The background is mildly contaminated by flares. (4) The
map analysis has not already been published.  The first criterion is
important to measure the temperature distribution, and thus to
guarantee reliable X-ray mass modeling to derive the mass bias. The
second and third criteria are required to guarantee robust X-ray
background modeling, particularly in the spectral analysis for the map
analysis.  In addition, the reason we chose those four clusters with
slightly different photon statistics is to test how far from the
cluster center and how reliably one can perform substructure studies
with a range of data quality. Such an investigation is important to
justify the required photon statistics for substructure studies of
galaxy clusters using X-ray maps on different levels.  Our empirical
results will be useful for the community to sample clusters and to
perform X-ray observations for such substructure studies. We set the
fourth criterion in order to get new scientific results out of our
tests.

\subsection{Data Preparation} 
\label{s:screen}

To prepare the data, we apply iterative screening using a 2$\sigma$ clipping
as described in Zhang et al. (2006, 2007) using both the soft band
(0.3-10~keV) and the hard hand (10-12~keV for MOS and 12-14~keV for pn) to
filter flares. Hereafter, we call those light-curve screened events for the
clusters the target observations (TOs). The properties of the four clusters
are presented in Table~\ref{t:basic}.

\subsection{Point-Like Source Identification and Subtraction}
\label{s:psrc}

The ``edetect\_chain'' command is used to detect point-like sources. Point
sources in the outskirts of the cluster are subtracted. In the cluster center,
only these detected point-like sources carefully checked by eye and identified
with detected point-like sources in Chandra (Hudson et al. 2009) are
subtracted. 

There is good agreement between \emph{XMM-Newton} and \emph{Chandra} detected central
point-like sources. For IIIZw54, a point source is detected by both
\emph{XMM-Newton} and \emph{Chandra} at the center (03:41:17.54, +15:23:47.61), where
a cD galaxy sits. A3391 also has a point source at the center (06:26:20.45,
-53:41:35.80) coincident with the cD, detected by both \emph{XMM-Newton} and
\emph{Chandra}.  For EXO0422, there are no evident point-like sources detected by
\emph{Chandra} at the center. The \emph{XMM-Newton} image shows extremely peaked
X-ray emission similar to a point-like source (04:25:51.25, -08:33:36.97) at
the position of the cluster galaxy C1G~0422-09 (also see Belsole et al. 2005).
Conservatively, we identify it as a point-like source and subtract it.  For
A0119, there are no evident point-like sources in the cluster center in either
\emph{XMM-Newton} or \emph{Chandra} data, co-spatial with a cD.

\subsection{Background Treatment}
\label{s:allbkg}

As also described in Snowden et al. (2008), the following four background
components have been taken into account in our background treatment. The first
is the quiescent particle background (QPB). The second is the fluorescent
X-ray background (FXB). The third is the soft proton - caused background
(SPB).  The fourth is the cosmic X-ray background (CXB).

\subsubsection{QPB and FXB} 
\label{s:qpb}

The treatment of the QPB and FXB has been documented using the filter wheel
closed (FWC) observations for MOS in Snowden et al.  (2008) and for pn in
Freyberg et al. (2006).

As a first step to model the QPB and FXB, we extract the spectra using
events out of the FOV\footnote{An expression of ``\#XMMEA\_16'' in the
  SAS command ``evselect'' means to select the events out of the FOV.}
from the FWC
observations\footnote{http://xmm.vilspa.esa.es/external/xmm\_sw\_cal/background/\#EPIC}
to investigate the properties of the QPB and FXB using the 2-12 keV
band as done in Snowden et al.  (2008) and Freyberg et al. (2006). The
FWC MOS1/MOS2 (pn) spectrum can be well fitted by a ``powerlaw/b''
model together with six (eight) ``Gaussian/b'' models to account for
the FXB lines. The photon index of the ``powerlaw/b'' model, $\Gamma$,
is $0.154 \pm 0.006$ for MOS1 (reduced $\chi^2=1.14$ for 645 degrees
of freedom (dof)), $0.138\pm 0.008$ for MOS2 (reduced $\chi^2=1.09$
for 645 dof), and $0.345\pm 0.012$ for pn (reduced $\chi^2=0.89$
for 630 dof), respectively.  The best fit provides reliable
measurements of the ``LineE'' parameter with a few percent precision
and of the ``Sigma'' parameter within a factor of 1.4 for the
``Gaussian/b'' model.  Across the detectors we found the slope of the
``powerlaw/b'' is $\sim 0.15$ for MOS and $\sim 0.35$ for pn, both
varying by at most 15\%. The properties of the FWC observations are
consistent with the results found in de Plaa et al.  (2006) and
Freyberg et al. (2006)

De Luca \& Molendi (2004) pointed out that a simple renormalization of
the QPB using the high energy band (e.g., 8-12~keV) count rate may
lead to systematic errors in both the continuum and the lines. We thus
checked the out of FOV - extracted spectra of $\sim 60$ TOs. The slope
of the individual TOs is indeed inconsistent (up to $\sim$50\%) with
the stack FWC observations when the full 2-12~keV band is used. A
consistency of the slope appears when the 3-10~keV band is fitted. We
therefore re-normalize the FWC observations for the QPB and FXB
subtraction using a broad band of 3-10~keV.  The best fit of the
photon index ($\Gamma$) of the stack FWC observations using the
3-10~keV band is $0.144 \pm 0.016$ for MOS1 (reduced $\chi^2=1.07$ for
640 dof), $0.140\pm 0.017$ for MOS2 (reduced $\chi^2=1.03$ for 637
dof), and $0.341\pm 0.039$ for pn (reduced $\chi^2=0.89$ for 498 dof),
respectively.

As a second step, for both MOS and pn, we model and subtract the QPB for
individual observations using the stacked FWC observations with the same mode
as for the TOs. To determine the normalization, we extract the
spectra using events out of the FOV (\#XMMEA\_16) and outside of a
$15.4^{\prime}$ radius from the detector center for both FWC observations and
TOs. As De Luca \& Molendi (2004) found, both X-ray photons and low energy
particles can reach CCD~2 and CCD~7 of the MOS cameras. We exclude both CCDs
for MOS. For later observations with the MOS1 camera, we exclude CCD~6 in the
FWC data to match the loss of MOS1 CCD~6 in the TOs. As also found in Snowden
et al. (2008), some observations show occasional deviations of CCD~4 and CCD~5
for MOS1 and CCD~5 for MOS2. These CCDs are then excluded as well for those
observations. Freezing the photon index of the ``powerlaw/b'' model and the
``LineE'' and ``Sigma'' parameters of the ``Gaussian/b'' model to the best fit
derived above using all events out of the FOV from the FWC observations in the
3-10~keV, we obtained the normalization from the best fit.  The
renormalization factor ($n_{\rm QPB}$) of the continuum component is derived
as the ``powerlaw/b'' normalization ratio of the TO to FWC observations. The
FWC spectrum ($S_{\rm FWC}$) is normalized by this renormalization factor,
$n_{\rm QPB}$, and subtracted from the TO spectrum ($S_{\rm TO}$) for each
instrument.

\subsubsection{SPB}

The screening procedure described in Sect.~\ref{s:screen} using both
the hard band and the soft band to prepare the data has filtered all
of the significant SPB component for most observations. The
observations with significant residual SPB found in the spectral
analysis shall be excluded. Luckily, none of the observations for the
four clusters show significant residual SPB.
  
\subsubsection{CXB} 
\label{s:cxb}

Both RASS data and PSPC pointed data can be used to model the CXB. The
latter, of higher statistical quality, are preferred. The \emph{ROSAT}
PSPC calibration shows an accuracy of better than 5\% even for
energies lower than 0.28~keV (Beuermann 2008). Therefore we use the
\emph{ROSAT} PSPC pointed data in the 0.1-2.4~keV band to model the
CXB. The spectrum was extracted from the region just beyond $r_{200}$
for each cluster.  The best fit of the spectrum shows that the CXB can
be well described by a combined model,
``mekal$+$wabs$*$(mekal$+$powerlaw)''.  The temperature of the
unabsorbed thermal component is often $\sim 0.1$~keV, and of the
absorbed thermal component is often between 0.1 and 0.2~keV,
respectively. To avoid large background fluctuations, we have excluded
regions showing bright sources identified by eye. The absorbed
``powerlaw'' model, with its slope set to 1.41, accounts for
unresolved point sources (De Luca \& Molendi 2004). The ``wabs'' model
is set to the hydrogen column density from the LAB survey\footnote{The
  Leiden/Argentine/Bonn Galactic HI Survey (LAB survey) contains the
  final data release of observations of $\lambda$21-cm emission from
  Galactic neutral hydrogen over the entire sky, merging the
  Leiden/Dwingeloo Survey (LDS; Hartmann \& Burton 1997) of the sky
  north of $\delta = -30^{\circ}$ with the Instituto Argentino de
  Radioastronomía Survey (IARS; Arnal et al.  2000; Bajaja et al.
  2005) of the sky south of $\delta = -25^{\circ}$.  The angular
  resolution of the combined survey is half-power beamwidth (HPBW)
  $\sim 0.6^{\circ}$.
  http://www.astro.uni-bonn.de/$\sim$webrai/english/tools\_labsurvey.php}
(Hartmann \& Burton 1997; Arnal et al. 2000; Bajaja et al. 2005;
Kalberla et al. 2005) at the \emph{XMM-Newton} determined cluster
center. We repeated the CXB modeling with a
``mekal$+$wabs$*$(mekal$+$mekal$+$powerlaw)'' model, which includes a
second absorbed thermal emission component, and which shows no
significant improvement of the fit.  Therefore, we use the
``mekal$+$wabs$*$(mekal$+$powerlaw)'' model for the CXB.

With increasing radial distance from the cluster center, the CXB
becomes dominant over the cluster emission. We therefore use the
outskirts to model the CXB. We extract the \emph{XMM-Newton} spectra
from the outermost region ($9^{\prime}.17<R<10^{\prime}$ from the
cluster center) in the \emph{XMM-Newton} FOV, $S_{\rm TO}$.  The FWC
spectrum is extracted from the same detector coordinates as for the TO
spectrum, and normalized by $n_{\rm QPB}$ (derived in
Sect.~\ref{s:qpb}). We call these spectra, $S_{\rm TO} - n_{\rm
  QPB}S_{\rm FWC}$, the secondary observational (SO) spectra. To
derive the normalization of the CXB and to measure the cluster
emission, we made a joint fit of the above \emph{ROSAT} PSPC spectrum
by ``mekal$+$wabs$*$(mekal$+$powerlaw)'', and the three
\emph{XMM-Newton} EPIC spectra by
``wabs$*$mekal$+$mekal$+$wabs$*$(mekal$+$powerlaw)$+$powerlaw/b''.
Note in this co-fit analysis, we link the temperatures and
normalizations of the two ``mekal'' model and the normalization of the
``powerlaw'' model for the CXB between \emph{ROSAT} PSPC and
\emph{XMM-Newton} EPIC. The first ``wabs$*$mekal'' component in the
model for the \emph{XMM-Newton} EPIC spectra takes into account the
hydrogen column density absorption (frozen to the value from the LAB
survey) and cluster emission with its metallicity fixed to 0.3 solar
metallicity. The ``powerlaw/b'' component takes into account the
residual SPB in the \emph{XMM-Newton} spectra, which normalization
should be consistent with zero. For some TOs, the ``powerlaw/b''
normalization can be significantly higher, which is inconsistent with
zero. Due to the SPB contamination, the spectra from the corners out
of the FOV for such TOs often show completely inconsistent shape (i.e.
the slope of the ``powerlaw'' component) with that for the FWC
observations. Therefore, the designed QPB background treatment using
the FWC observations will fail for such TOs, and they should not be
used for our analysis. Luckily, the normalization of the
``powerlaw/b'' model for all four clusters is consistent with zero. This
confirms that the light curve screening procedure in
Sect.~\ref{s:screen} has removed all of the significant flares for
these four clusters.

De Luca \& Molendi (2004) derived the normalization of
0.00345~photons~keV$^{-1}$~cm$^{-2}$~s$^{-1}$~deg$^{-2}$ at 1~keV for
the ``powerlaw'' model of the CXB with a photon index of 1.41 using
\emph{XMM-Newton} EPIC data.  The agreement is better than 40\%
between their value and the best-fit value from our co-fit of
\emph{ROSAT} PSPC pointed data and \emph{XMM-Newton} EPIC data for
each annulus for each cluster, i.e. within 28\% for IIIZw54, within
14\% for A3391, within 40\% for EXO0422, and within 30\% for A0119.
And the agreement becomes better with increasing radial distance from
the cluster center as the cluster emission becomes less dominant in
the outskirts.  Setting the normalization to the value in De Luca \&
Molendi (2004), we observe no pronounced change in the $\chi^2$ and
measured parameters of the best fit. Note that the best fits of the
spectra also provide reasonable cluster temperatures in comparison to
previous published results.

\section{Spectral analysis for temperature profile} 
\label{s:spe}

\subsection{Point-Spread Function and Vignetting}

Using the \emph{XMM-Newton} point-spread function (PSF) calibrations
in Ghizzardi (2001) we estimate the redistribution fraction of the
flux. It is 20\% for bins with widths of about $0.5^{\prime}$ and less
than 10\% for bins with widths $\ge 1^{\prime}$ neglecting energy- and
position\footnote{All four observations roughly centered on the
  cluster centers.}- dependent effects.  We thus require annular width
$\ge 0.5^{\prime}$ in the radial spectral analysis. The PSF effect is
important within $0.5^{\prime}$, which corresponds to $\le
0.038r_{500}$ for our four nearby clusters. The PSF effect introduces an
uncertainty only to the radial temperature measurement in the inner
bins. We made an attempt to correct for the PSF effect for
RXCJ2228$+$2037 in Jia et al.  (2008), and found the PSF correction is
important mainly in the inner radii and causes effects well within
10\% level on the temperature measurements. A similar conclusion was
reached in Snowden et al. (2008) for A1795. Therefore we skip the PSF
correction in our radial spectral analysis.

X-ray telescopes often have non-azimuthally symmetric PSFs. In the temperature
map, the structure due to effects of the non-azimuthally symmetric PSF might
be interpreted as actual structure in the cluster. However, those effects
become important only an off-axis radii of larger than $10^{\prime}$. The
regions used for our studies are well within a off-axis radius of
$6^{\prime}$. Note that the regions used for 2D diagnostics are often
$\le 6^{\prime}$ for the four clusters. In addition, these effects are
significant only for regions of $1^{\prime}$ size along the radial-axis in
such outer regions. The radial axis width of the bins are all much larger than
$1^{\prime}$, particularly using the \emph{Mask-V} (defined in
Sect.~\ref{s:mask}), in the outer regions.

For both images and spectra, the vignetting is taken into account in the
extraction using the ``evigweight'' created column in the events.

\subsection{Radial Bin Size}

The blank sky accumulations of the archival \emph{XMM-Newton} observations in
the \emph{Chandra} Deep Field South (CDF-S) can be used as a rough estimate of the
background. Note that in the spectral analysis, the background is properly
treated as described in Sect.~\ref{s:allbkg}. The CDF-S observations are only
used as a rough background estimate to determine the radial bin size for the
spectral analysis.

We screen the CDF-S observation using the same threshold as for the TOs, and
normalize them to the TOs using the hard band (10-12~keV for MOS and
12-14~keV for pn) as described in Zhang et al.  (2004). This former step
guarantees similar QPB levels, and the latter step guarantees similar SPB
levels.  The residual SPB and the difference in the CXB and FXB are ignored in
the determination of the bin size. 

For a cluster with a temperature of $\sim 4$~keV, the uncertainty in
the spectrally measured temperature is $\le $5\% (10\%) using both pn
and MOS spectra, giving net source counts of $\ge 72,000$ (24,000)
after the background subtraction. Therefore the annuli for spectral
analysis are determined by requiring (1) that the width of each
annulus is larger than $0.5^{\prime}$, (2) that the net source counts
is $\ge C$ per MOS2 spectrum in the 0.5-7.8~keV band.  The threshold
$C$ is 18,000 except for the clusters with less than four annuli in total
for which $C$ is 6000. We include an outermost bin which does not
fulfill the threshold of $C$, with an outer radius truncated to give a
maximum net source counts.

\subsection{Spectral Fitting} 

To obtain the projected temperature profile, the three EPIC spectra
for each annulus are normalized to the solid angle of that annulus
taking into account corrections for gaps, bad pixels and point
sources. We performed the background modeling and co-fit as described
for the outermost region in Sect.~\ref{s:cxb}. Note that we firstly
fit the \emph{ROSAT} PSPC spectrum together with one of the
\emph{XMM-Newton} EPIC spectrum and found that the fitting parameters
(temperature, abundance, and normalization) agree to within a few per
cent between different EPIC instruments. We then fit the parameters
simultaneously to the \emph{ROSAT} PSPC spectrum together with all
three \emph{XMM-Newton} EPIC spectra.

To obtain the radial temperature profile for the mass modeling, we
deproject the spectra (e.g., Zhang et al. 2007), in which the spectral
models for the background components are renormalized to the volume of
the radial shell. The deprojected EPIC spectra and \emph{ROSAT} PSPC
spectrum are then fitted simultaneously to derive the radial
temperature measurements.

\section{Mass Modeling}
\label{s:mass}

\begin{figure*}
\begin{center}
\includegraphics[angle=270,width=15cm]{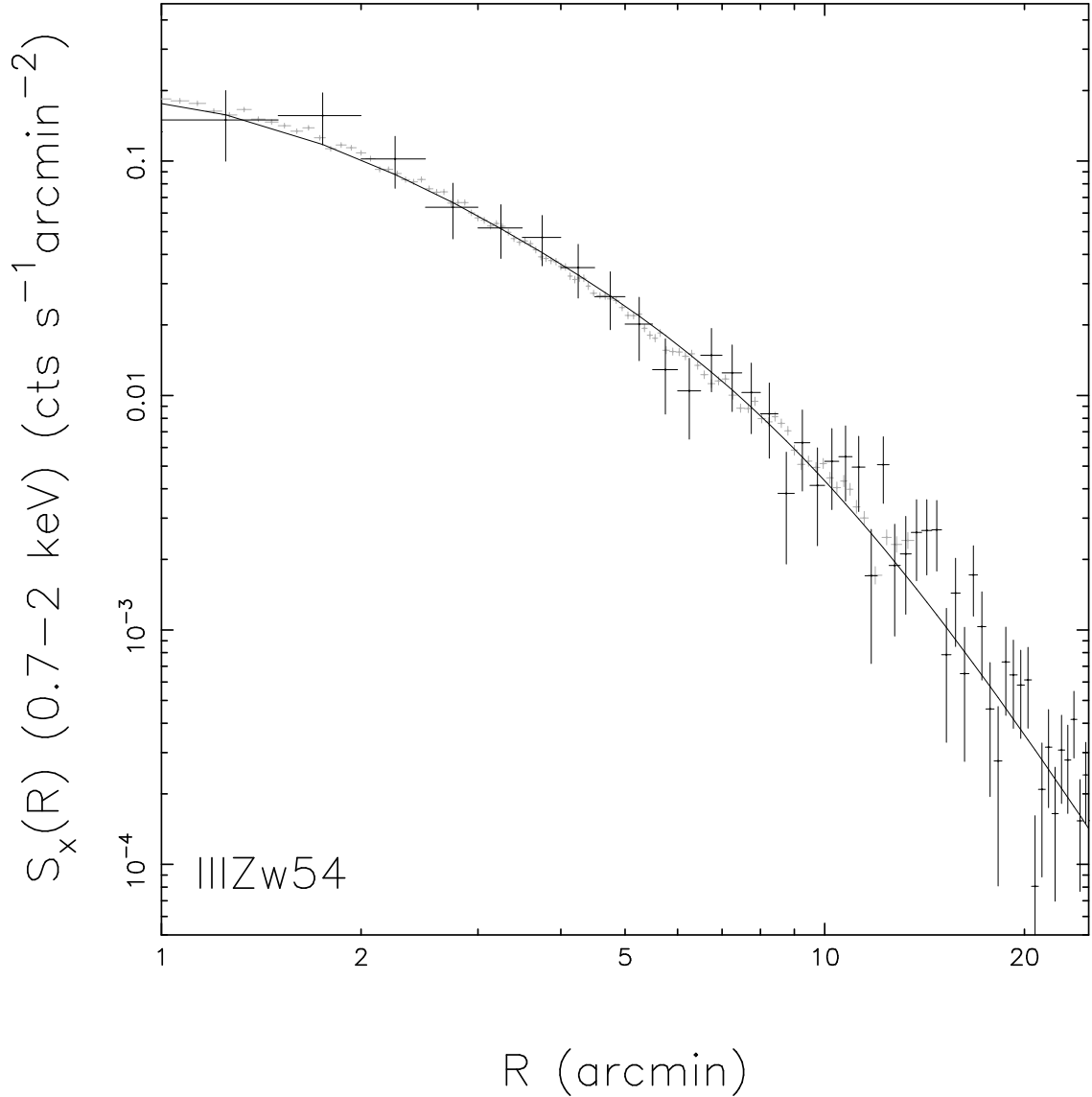}
\end{center}
\caption{Observed surface brightness profile, with the \emph{ROSAT}
observed surface brightness profile (black) converted to match the
\emph{XMM-Newton} observed surface brightness profile (gray). The
continuous curve presents the best fit of the observed surface
brightness profile using a double-$\beta$ model for the electron
number density profile.
\label{f:sx}}
\end{figure*}

The soft band (0.7-2~keV) X-ray surface brightness profile model $S_{\rm
  X}(R)$, in which $R$ is the projected radius, is linked to the ICM electron
number density profile $n_{\rm e}(r)$ and emissivity function as an integral
performed along the line of sight,
\begin{equation}
  S_{\rm X}(R)\propto\int_{-\infty}^{\infty} n_{\rm p} n_{\rm e} d\ell.
\label{e:sx}
\end{equation}
The \emph{XMM-Newton} observed surface brightness profile is derived by
subtracting the renormalized (by $n_{\rm QPB}$) FWC surface brightness profile
and the CXB in the 0.7-2~keV band derived in Section~\ref{s:spe} from the TO
surface brightness profile. The truncation radii (S/N~$\ge 3$, see
Table~\ref{t:basic}) of the \emph{XMM-Newton} observed surface brightness
profiles are rather small ($<r_{500}$). The \emph{ROSAT} observed surface
brightness profiles cover radii well beyond $r_{500}$ with S/N~$\ge 3$,
although with sparse data points in the cluster core.  We thus combine the
\emph{XMM-Newton} observed surface brightness within its truncation radius
($r_{\rm t}$) with the \emph{ROSAT} converted observed surface brightness
profile\footnote{The \emph{ROSAT} converted observed surface brightness
  profile can be derived using the \emph{ROSAT} surface brightness model in
  Reiprich \& B\"ohringer (2002) with the following two steps: (1) calculating
  the electron number density profile from the \emph{ROSAT} surface brightness
  model using the \emph{ROSAT} emissivity function, (2) projecting the
  electron number density profile to obtain the \emph{XMM-Newton}-like surface
  brightness profile using the \emph{XMM-Newton} response and convolving the
  \emph{XMM-Newton} PSF. In this procedure, the scatter and the error of each
  bin of the \emph{ROSAT} observed surface brightness profile are propagated.}
beyond $r_{\rm t}$ as the observed surface brightness profile (e.g., IIIZw54
in Figure~\ref{f:sx}). The observed surface brightness profile is fitted by
Eq.~\ref{e:sx} convolved with the \emph{XMM-Newton} PSF matrices to obtain the
parameters of the double-$\beta$ model of the electron number density profile,
$n_{\rm e}(r)=n_{\rm e01}(1+r^2/r_{\rm c1}^2)^{-3\beta/2}+n_{\rm
  e02}(1+r^2/r_{\rm c2}^2)^{-3\beta/2}$.

We assume that, (1) the ICM is in hydrostatic equilibrium within the
gravitational potential dominated by dark matter (DM), and (2) the DM
distribution is spherically symmetric. The cluster mass is then
calculated from the X-ray measured ICM density and temperature
distributions by,
\begin{equation}
  \frac{1}{\mu m_{\rm p} n_{\rm e}(r)}\frac{d[n_{\rm e}(r) k T(r)]}{dr}=
  -\frac{GM(<r)}{r^2}~,
\label{e:hyd}
\end{equation}
where $\mu=0.62$ is the mean molecular weight per hydrogen atom. $k$ is the
Boltzmann constant, and $T$ is the temperature. Following the method in Zhang
et al. (2007), we use a set of input parameters of the approximation
functions, in which $\beta$, $n_{\rm e0i}$, $r_{\rm ci}$ ($i=1,2$) represent
the double-$\beta$ electron number density profile $n_{\rm e}(r)$ and $P_{\rm
  i}$ ($i=1,...,7$) represent the deprojected temperature profile
$T(r)=P_3\;\exp[-(r-P_1)^2/P_2]+P_6(1+r^2/P_4^2)^{-P_5}+P_7$ (e.g.,  for
IIIZw54 and EXO0422 in Figure~\ref{f:tprof}), respectively, to compute the mean
cluster mass. The mass uncertainties are propagated using the uncertainties of
the electron number density and temperature measurements by Monte Carlo
simulations as described in Zhang et al. 2007, 2008). The cluster masses
$M_{2500}$ and $M_{500}$ are used in studying the scaling relations in
Sect.~\ref{s:scaling}.

\section{Spectrally Measured 2D Maps}
\label{s:map}

In our procedure, the 2D temperature, electron number density, entropy
and pressure maps\footnote{They will be made publicly available
  through the German Astrophysical Virtual Observatory (GAVO) under
  Multivariate Archive of X-Ray Images, http://www.g-vo.org/MAXI/~.}
are created based on the spectral measurements in each spatial bin.
The binning methods described below allow for less biased definition
of the zones for the spectral extraction compared with the zones
determined in the wavelet approach in e.g., Finoguenov et al.  (2005).
The available statistics of our data are sufficient to provide
detailed 2D diagnostics, and the radial study of the fluctuations for
individual clusters - a new complementary tool to measure the
substructure.

\subsection{Mask Determination}
\label{s:mask}

\begin{figure*}
\begin{center}
\includegraphics[angle=0,width=14cm]{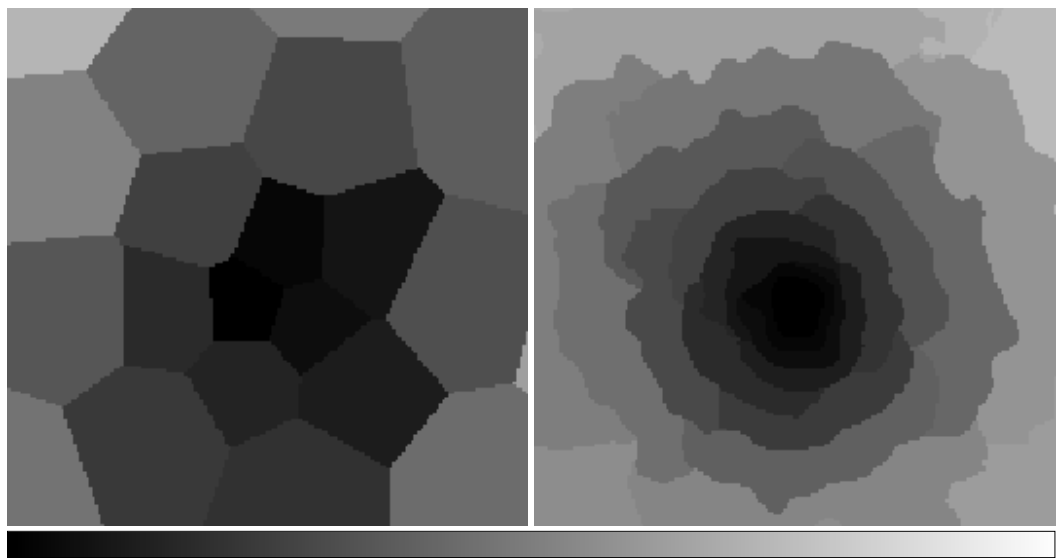}
\end{center}
\caption{Masks using the \emph{Mask-V} (left) 
and \emph{Mask-S} (right) method for IIIZw54. The
  image size is $11^{\prime} \times 11^{\prime}$. 
Each gray scale (from 0 to 34) denotes one bin, but has no physical meaning. 
\label{f:mask}}
\end{figure*}

\begin{figure*}
\begin{center}
\includegraphics[angle=0,width=14cm]{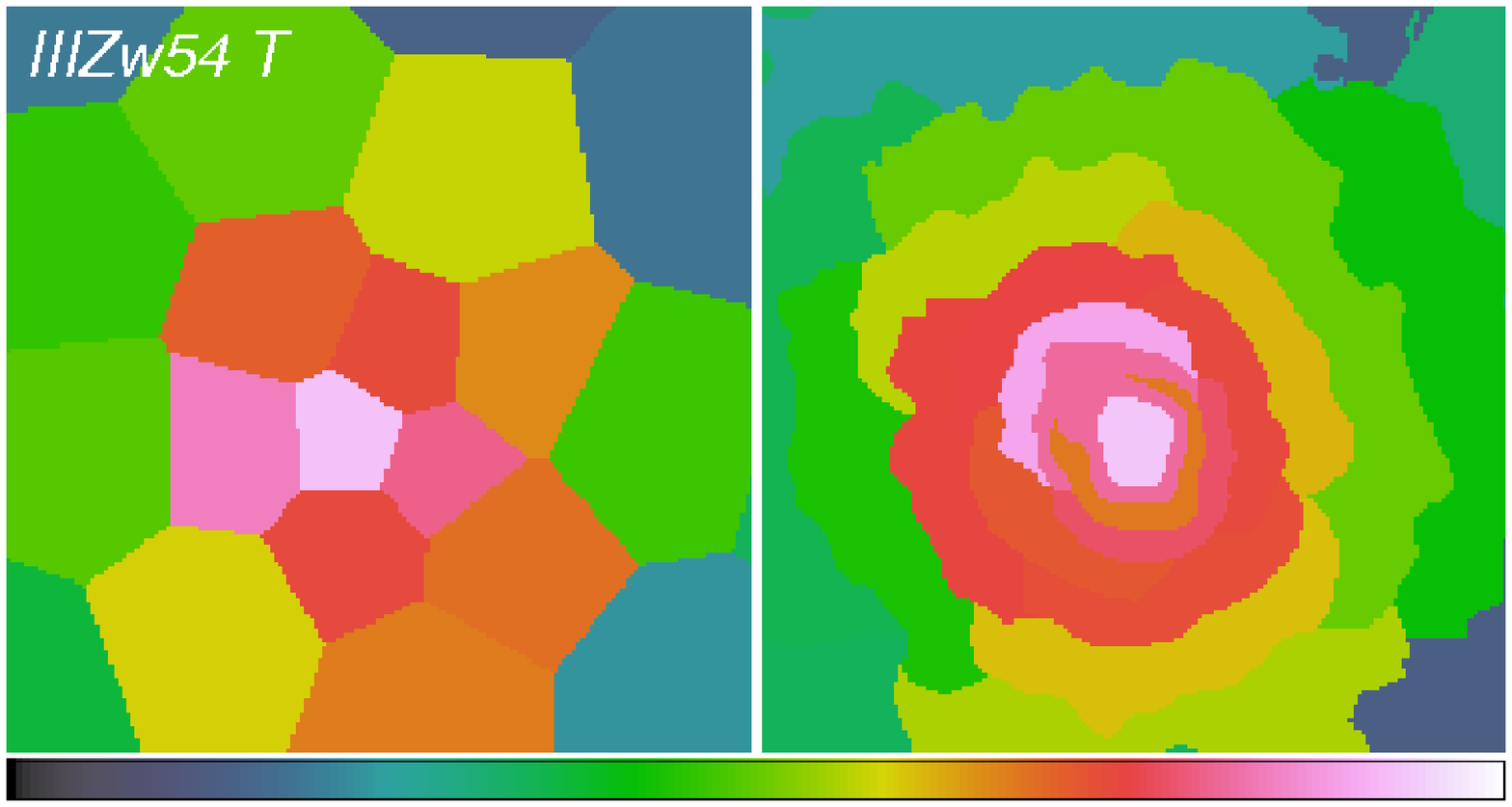}
\includegraphics[angle=0,width=14cm]{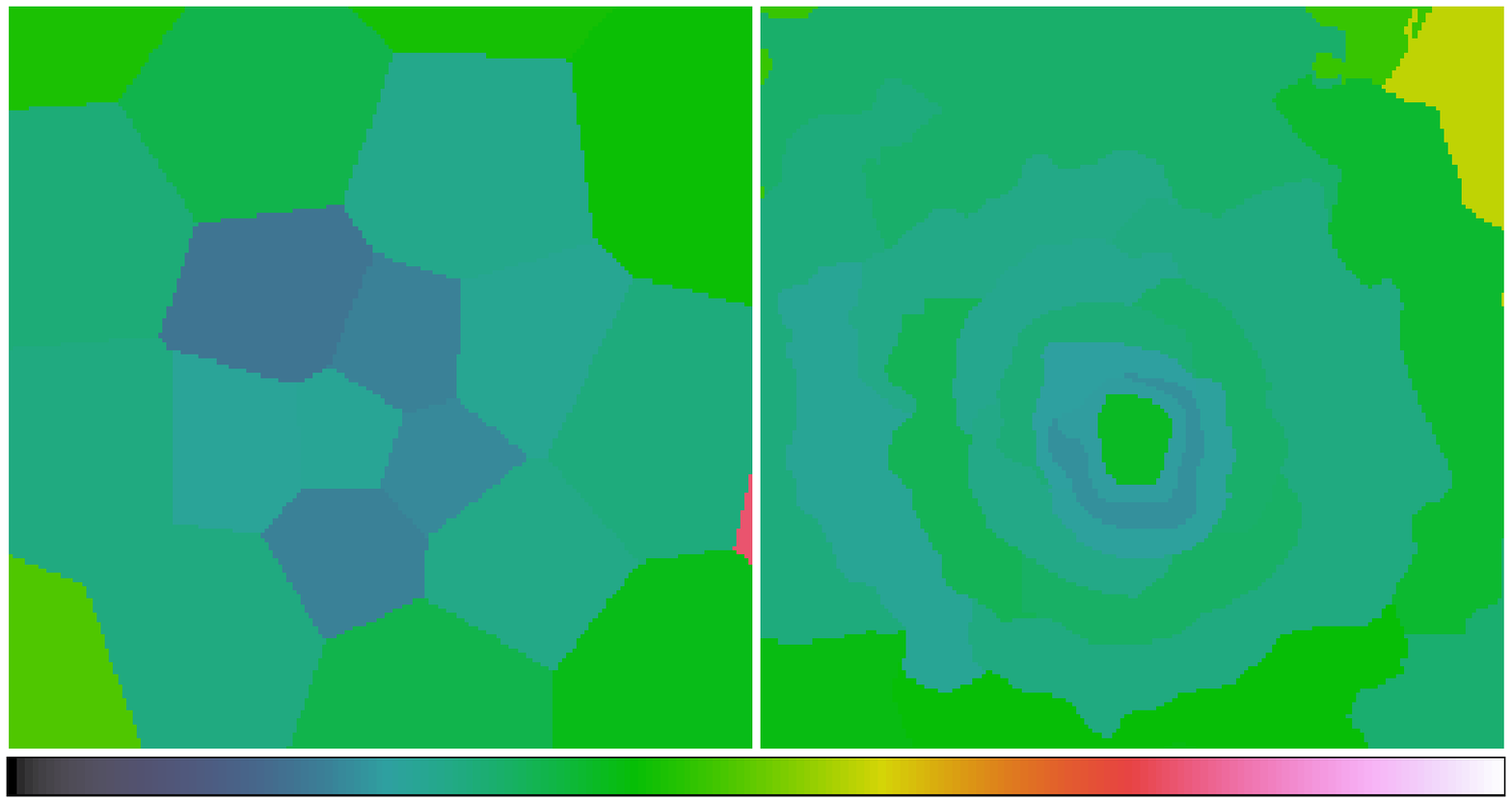}
\end{center}
\caption{Temperature maps (top) and their error maps (bottom) for IIIZw54
  using \emph{Mask-V} (left) and \emph{Mask-S} (right). The color
  bar is in the range of 1.5-3.2~keV in the top panels, and 0-0.2~keV in the
  bottom panels.  The image size is $11^{\prime} \times 11^{\prime}$.
  \label{f:tmap}}
\end{figure*}

We use the MOS2 data to determine the spatial bins in the mask
(e.g., for IIIZw54 in Figure~\ref{f:mask}) for the following two reasons:
(1) the pn data are seriously affected by gaps which can complicate
the analysis of cluster structure, and (2) the CCD6 is missing for
MOS1 for recent observations. We use the 0.5-2~keV band MOS2 image
binned in $4^{\prime \prime}\times 4^{\prime \prime}$ pixels to
determine the mask regions for the spectral analysis. The image is
binned to give an S/N of $\ge33$ for each spatial bin in the mask.
For a cluster with a temperature of $\sim 4$~keV, the uncertainty in
the spectrally measured temperature is $\sim$10\% (e.g., for IIIZw54,
lower panels in Figure~\ref{f:tmap}).

We adopted two methods to determine the bins (e.g., for IIIZw54 in
Figure~\ref{f:mask}), which are both based on the brightness criteria. One is
the weighted Voronoi tessellation method (Cappellari \& Copin 2003; Diehl \&
Statler 2006), whose binning shapes are geometrically unbiased giving a quasi
circle-like shape. This binning scheme is sensitive to local brightness
fluctuations (e.g., Simionescu et al. 2007). The other method was developed by
Sanders (2006), and bins the brightest pixels of the remaining region. The
mask therefore extends along the isophotal annulus centered at the cluster
core.  This binning scheme is sensitive to the detection of shocks and cold
fronts (e.g., Sanders \& Fabian 2007). Hereafter we call the mask defined with
the former method as \emph{Mask-V}, and the mask defined with the latter
method as \emph{Mask-S}.  The advantage of the \emph{Mask-S} binning is that
mixing of different temperature components due to a radial temperature
gradient is minimal, while the \emph{Mask-V} binning is more sensitive to
features like bright spots.

\subsection{Temperature Maps}
\label{s:2dspec}

The spectra are extracted for each bin in the mask, and normalized to the
solid angle of that bin taking into account corrections for gaps, bad pixels
and excluded point sources. The QPB and CXB models derived in
Sect.~\ref{s:spe} in the spectral fit are normalized to the solid angle of the
bin as frozen models.  The MOS and pn spectra are fitted simultaneously by a
``wabs*mekal'' model for hydrogen column density absorption and cluster
emission, with the frozen models to account for the background.  The best-fit
temperature and its error bar for each bin are used to create the temperature
($T$) map and its error map (e.g., for IIIZw54 shown in Figure~\ref{f:tmap}).

\subsection{Electron Number Density, Entropy and Pressure Maps}

The spectral normalization in each spatial bin can be used to derive a
quasi deprojected estimate of the electron number density ($n_{\rm
  e}$) in that spatial bin (e.g., Henry et al. 2004). In XSPEC, the
normalization of the ``mekal'' model is given as $K=10^{-14} / [4\pi
D_{\rm A}^2 (1+z)^2] \int n_{\rm e} n_{\rm H} dV$, where $D_{\rm A}$
is the angular diameter distance, $z$ is the redshift, and the volume
corresponding to that spatial bin is approximated by $V \approx
(4/3)D_{\rm A}^3 \Omega \sqrt{\theta_{\rm out}^2-\theta_{\rm in}^2}$.
Here $\Omega$ is the solid angle of the corresponding spatial bin, and
$\theta_{\rm out}$ and $\theta_{\rm in}$ are the angles of the
outermost and innermost radii of that bin from the cluster center. As
mentioned in Simionescu et al.  (2007), it provides a quasi
deprojection using an approximation of the three-dimensional extent
of each spatial bin and assuming a constant temperature along the line
of sight. As most emission in the bin is from the densest gas near the
innermost radius of that bin, the electron number density derived from
the spectral normalization can be used as the measurement of the
electron number density at the projected radius.

\begin{figure*}
\begin{center}
\includegraphics[angle=0,width=8cm]{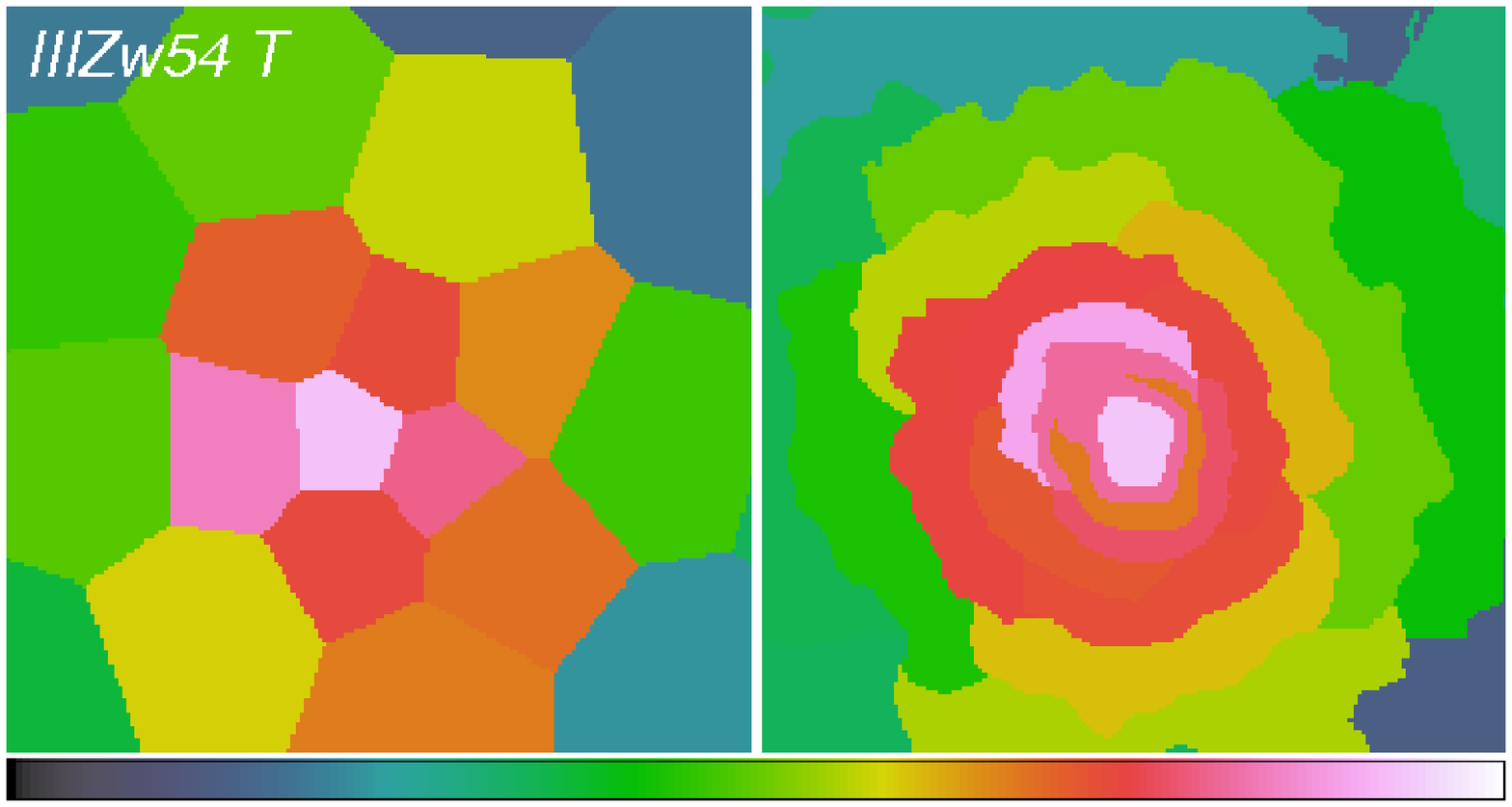}
\includegraphics[angle=0,width=8cm]{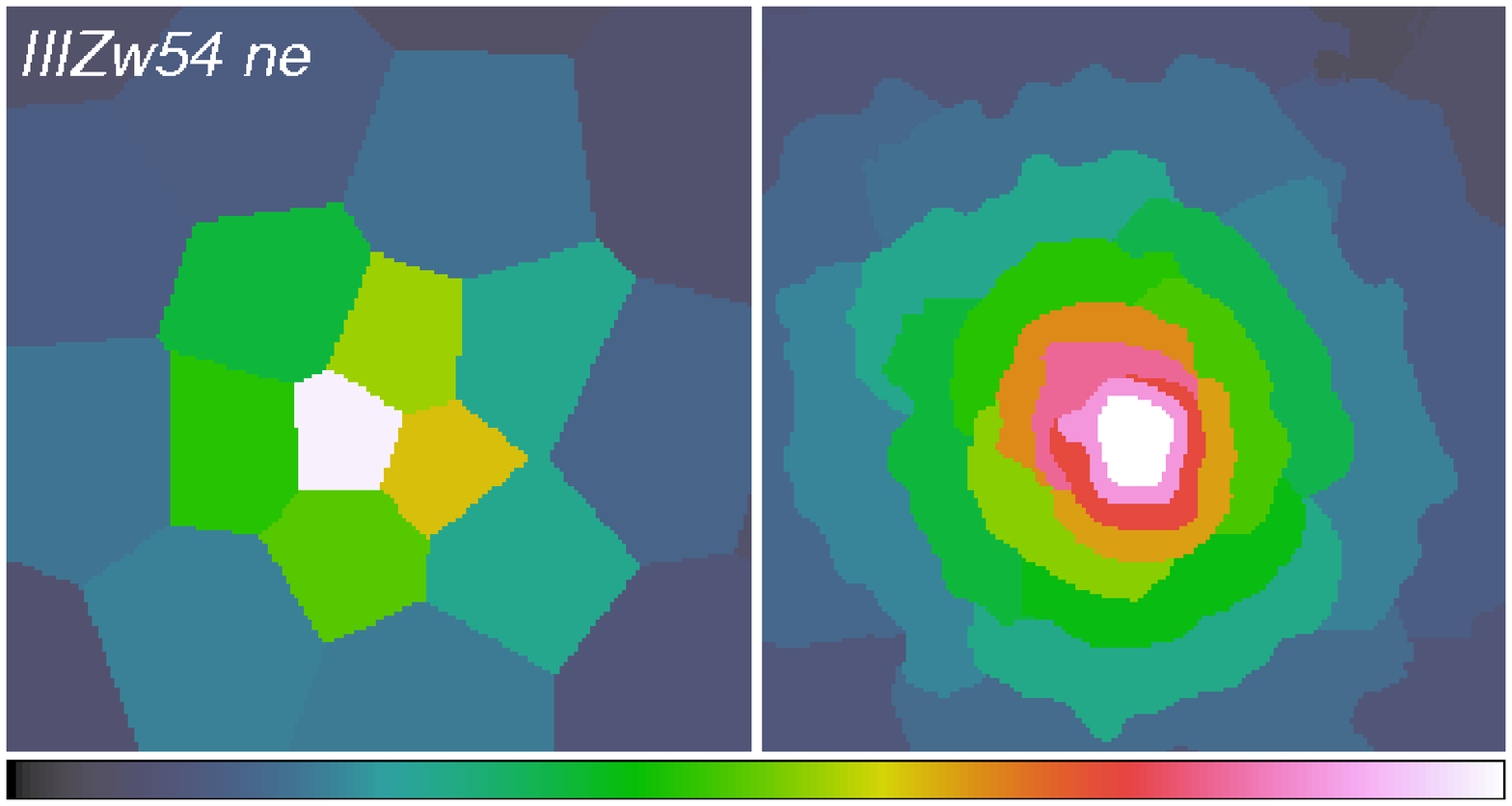}

\includegraphics[angle=0,width=8cm]{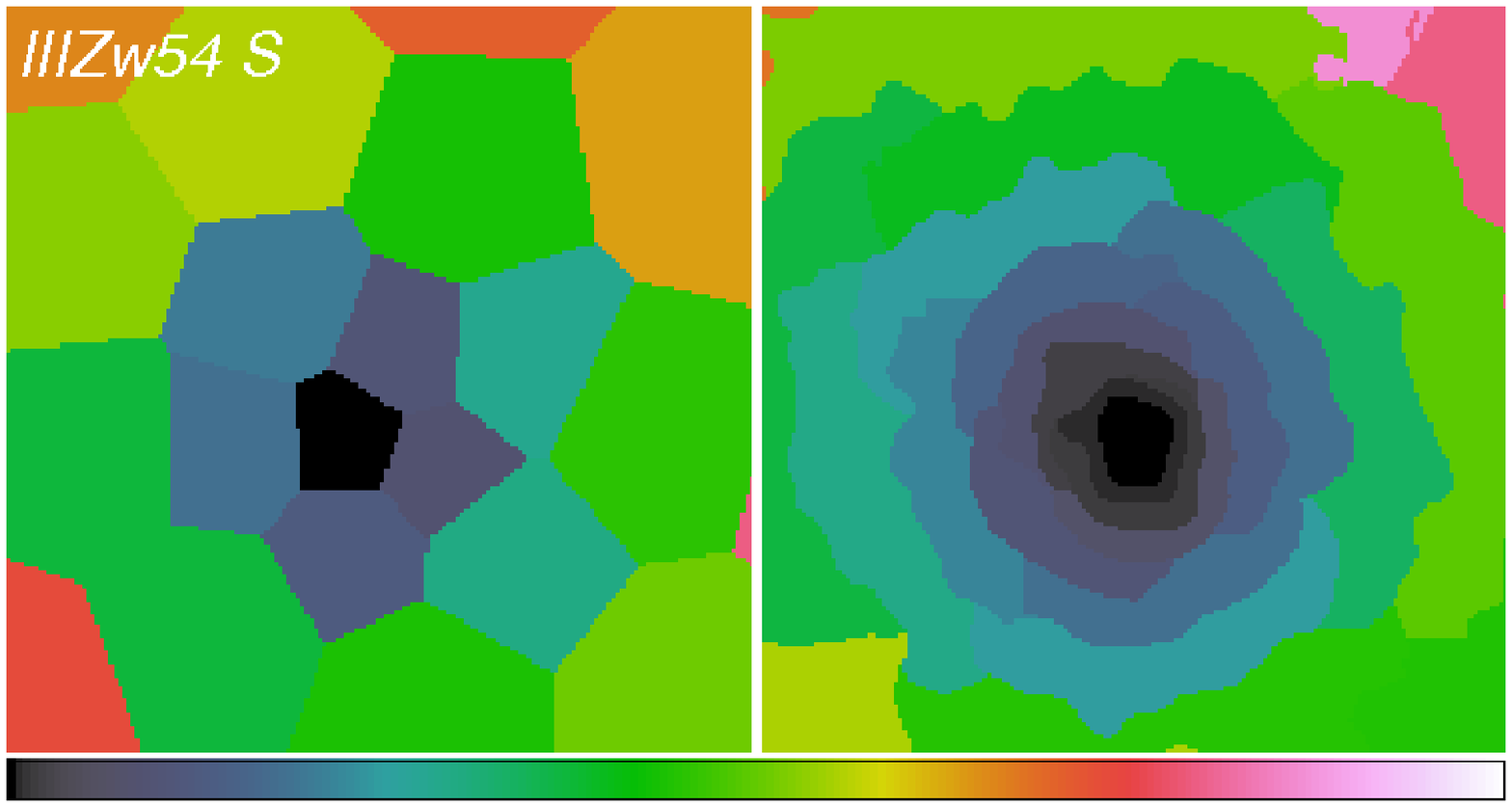}
\includegraphics[angle=0,width=8cm]{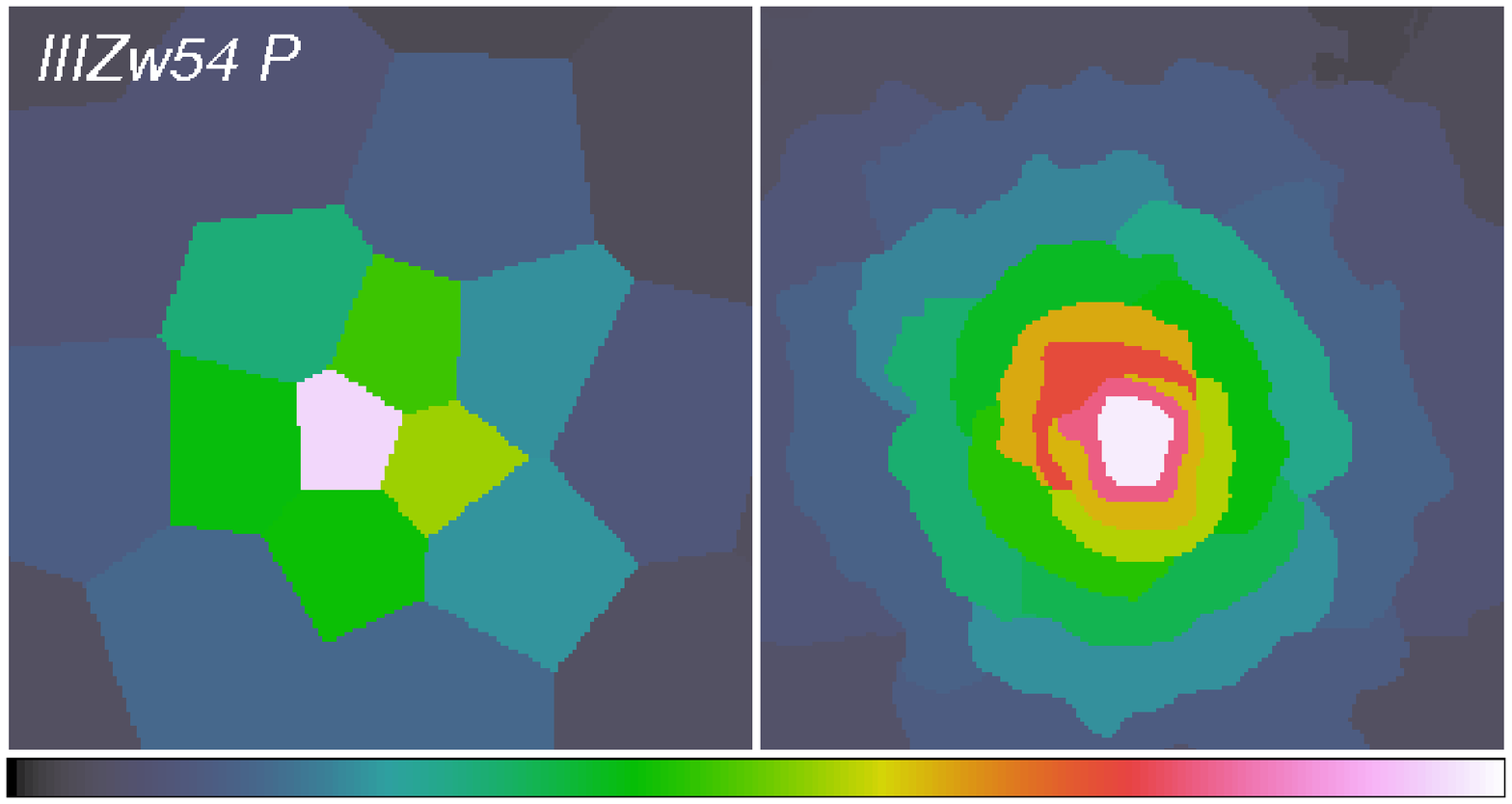}
\end{center}
\caption{Temperature ($T$), electron number density ($n_{\rm e}$), entropy
  ($S$), and pressure ($P$) maps for IIIZw54 with the \emph{Mask-V} on the 
   left and the \emph{Mask-S} on the right in each panel. The color bars 
   are in the range of 1.5-3.2~keV,
  0-0.0063~cm$^{-3}$, 90-400~keV~cm$^2$, and 0-0.02~keV~cm$^{-3}$ in the
  $T$, $n_{\rm e}$, $S$, and $P$ panels, respectively.  The image size is 
  $11^{\prime} \times 11^{\prime}$. 
  \label{f:iiizw54map}}
\end{figure*}

\begin{figure*}
\begin{center}
\includegraphics[angle=0,width=8cm]{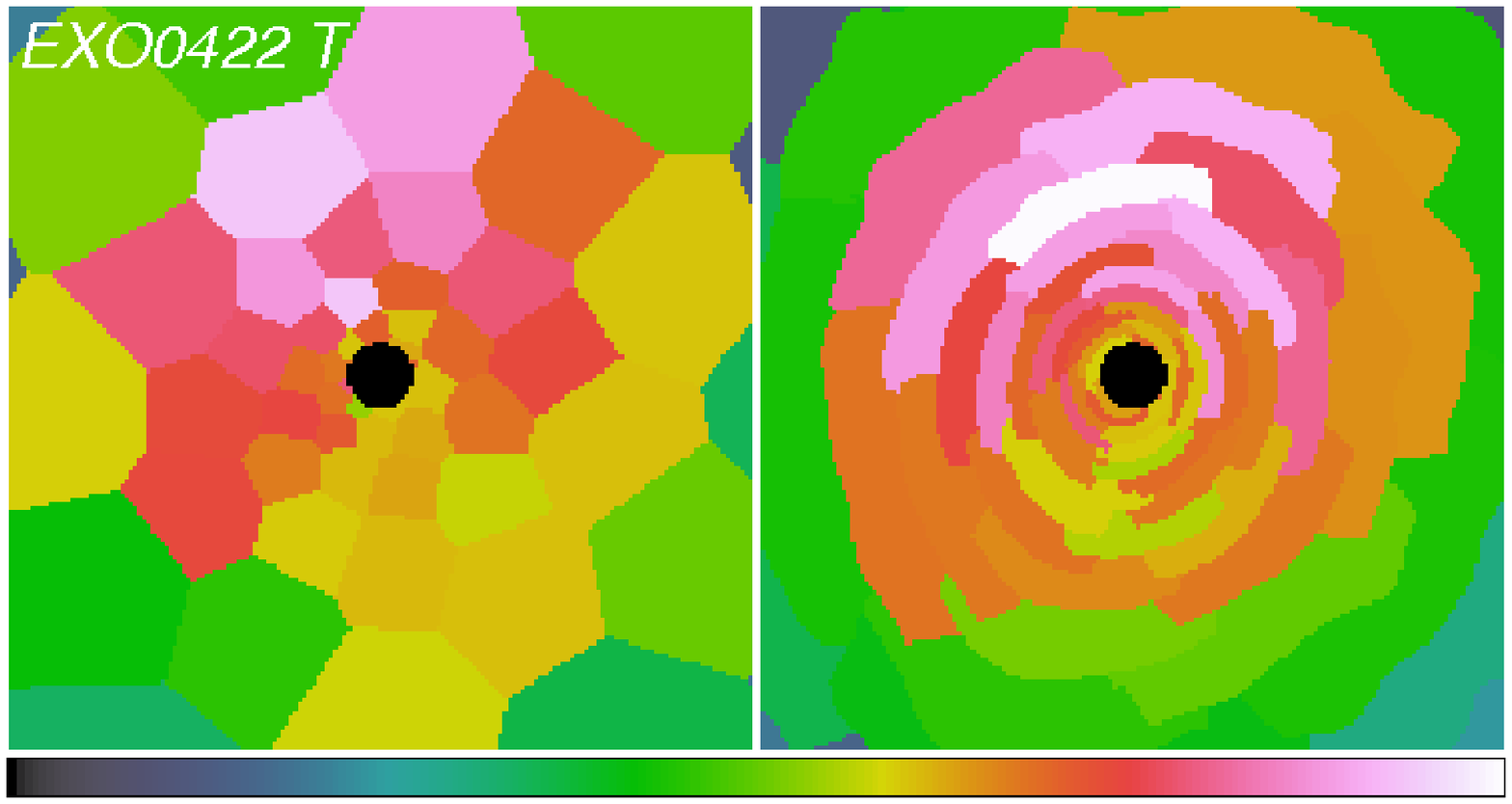}
\includegraphics[angle=0,width=8cm]{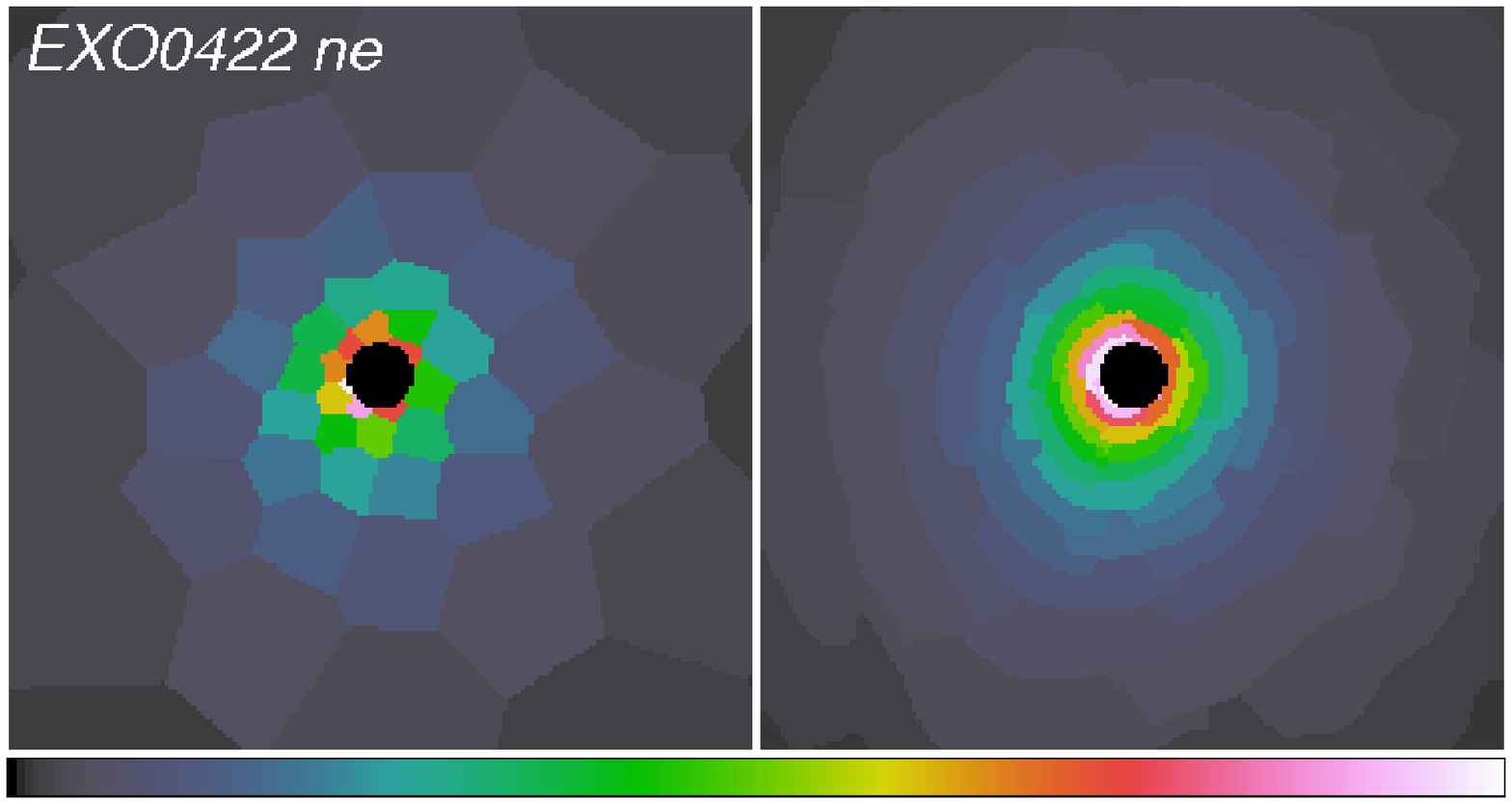}

\includegraphics[angle=0,width=8cm]{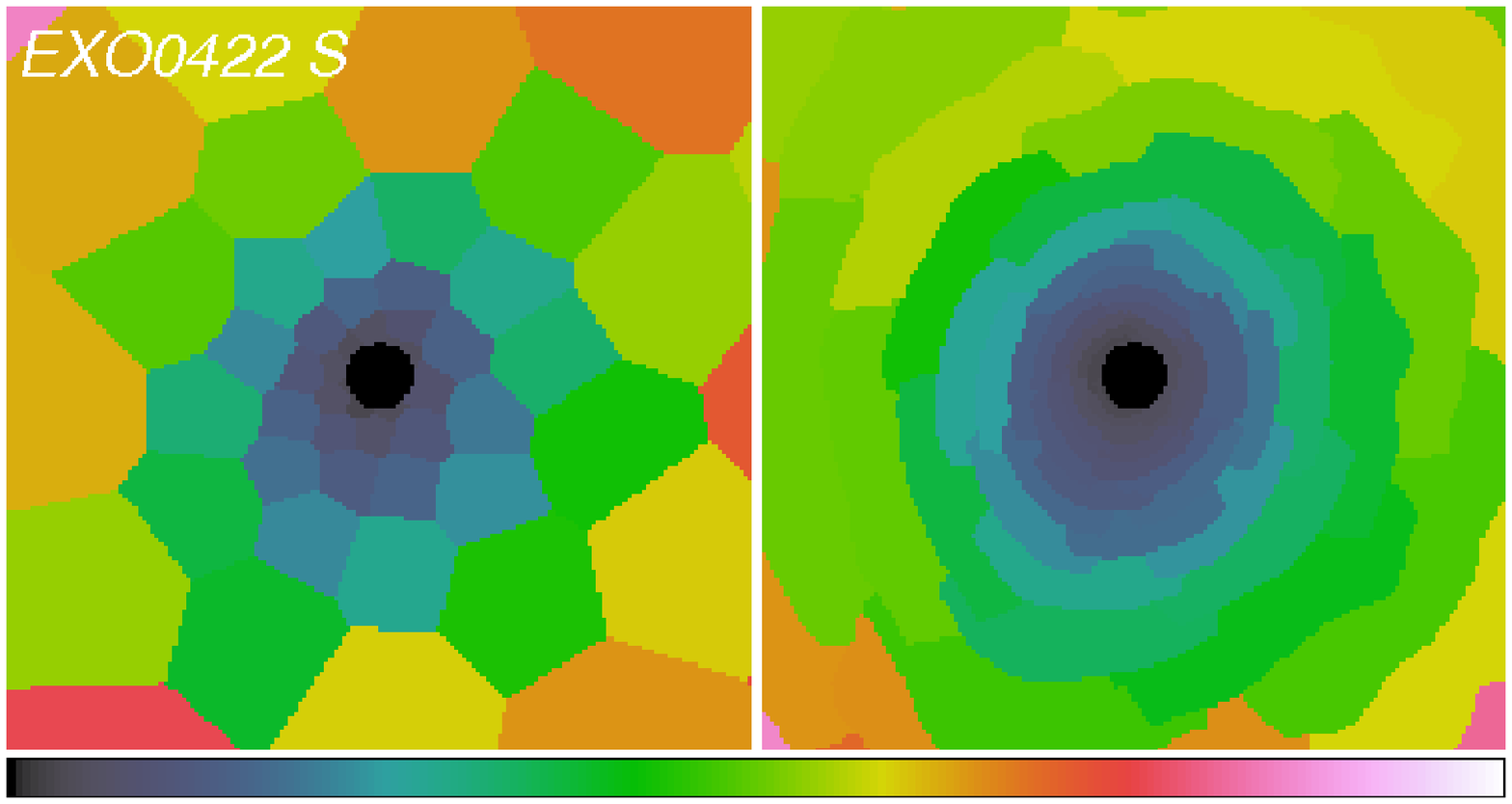}
\includegraphics[angle=0,width=8cm]{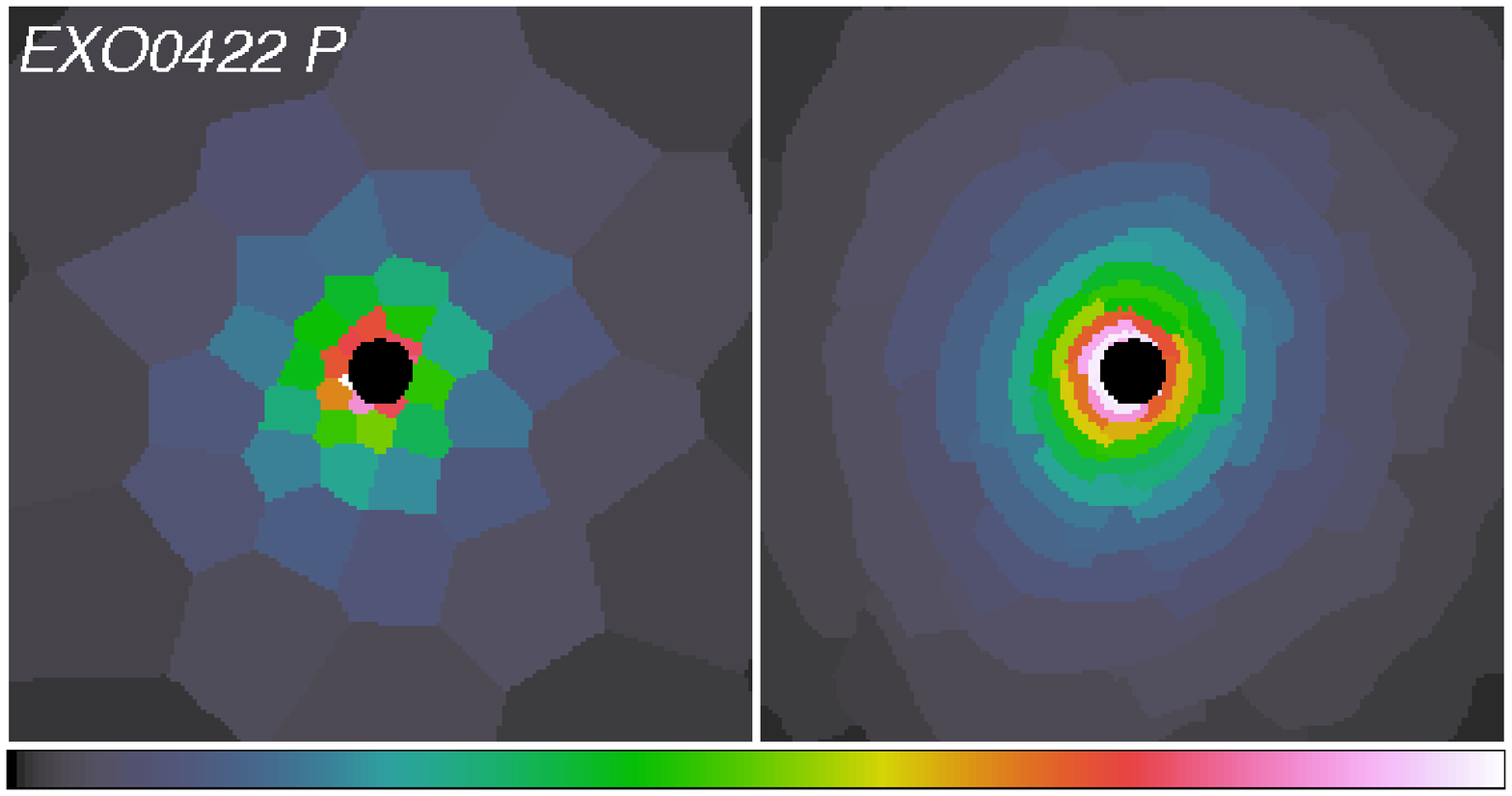}
\end{center}
\caption{Temperature ($T$), electron number density ($n_{\rm e}$), entropy
  ($S$), and pressure ($P$) maps for EXO0422 with the \emph{Mask-V} on the 
  left and the \emph{Mask-S} on the right in each panel. The 
  color bars are in the range of 1.5-4.2~keV,
  0-0.023~cm$^{-3}$, 20-580~keV~cm$^2$, and 0-0.025~keV~cm$^{-3}$ in the
  $T$, $n_{\rm e}$, $S$, and $P$ panels, respectively.  The image size is 
  $11^{\prime} \times 11^{\prime}$. The
  hole in the cluster center for EXO0422 is the region excluded in the
  spectral analysis to avoid the possible contamination from the galaxy
  CIG0422-09 found by Belsole et al. (2005).
  \label{f:exo0422map}}
\end{figure*}

\begin{figure*}
\begin{center}
\includegraphics[angle=0,width=8cm]{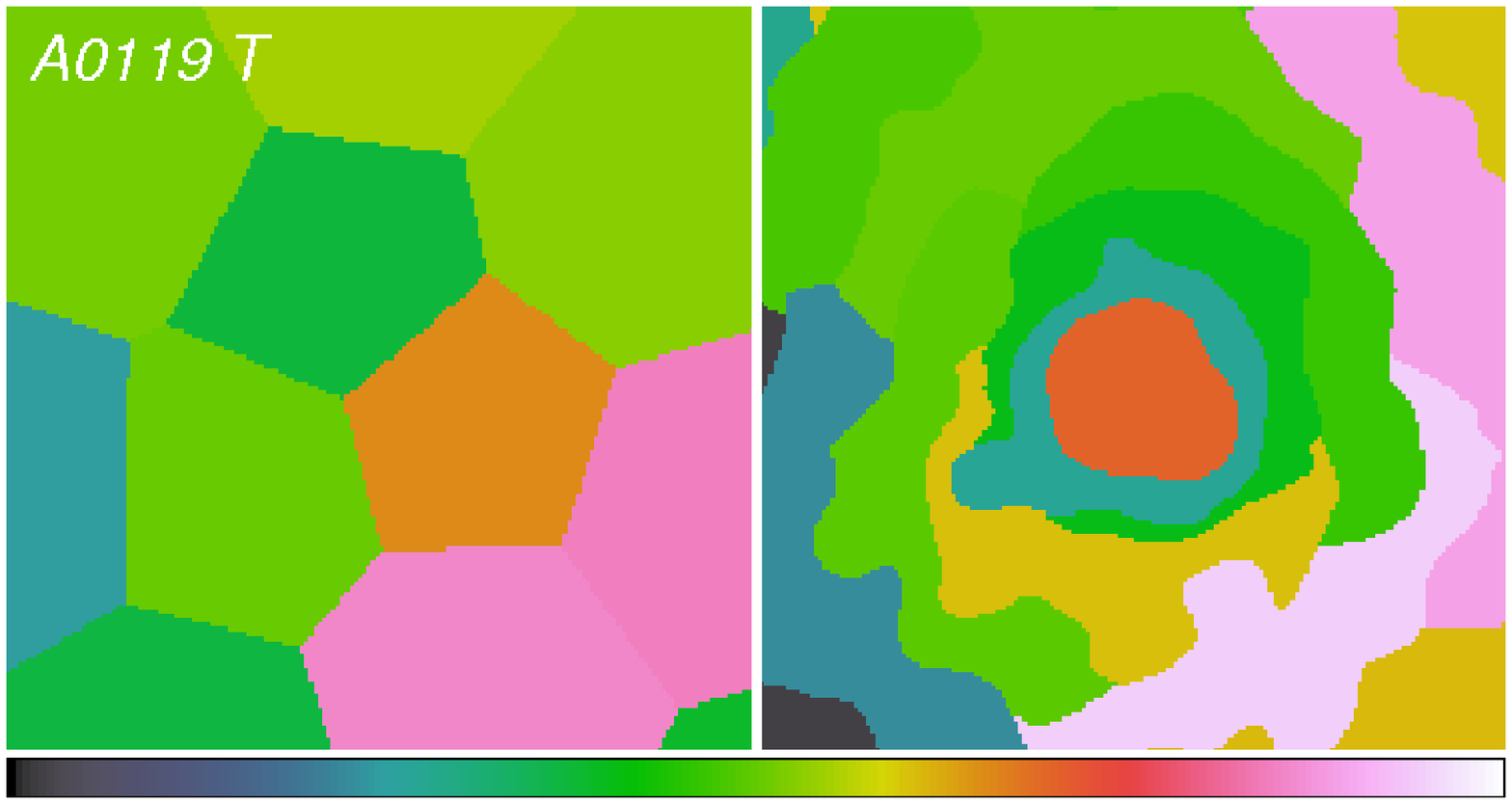}
\includegraphics[angle=0,width=8cm]{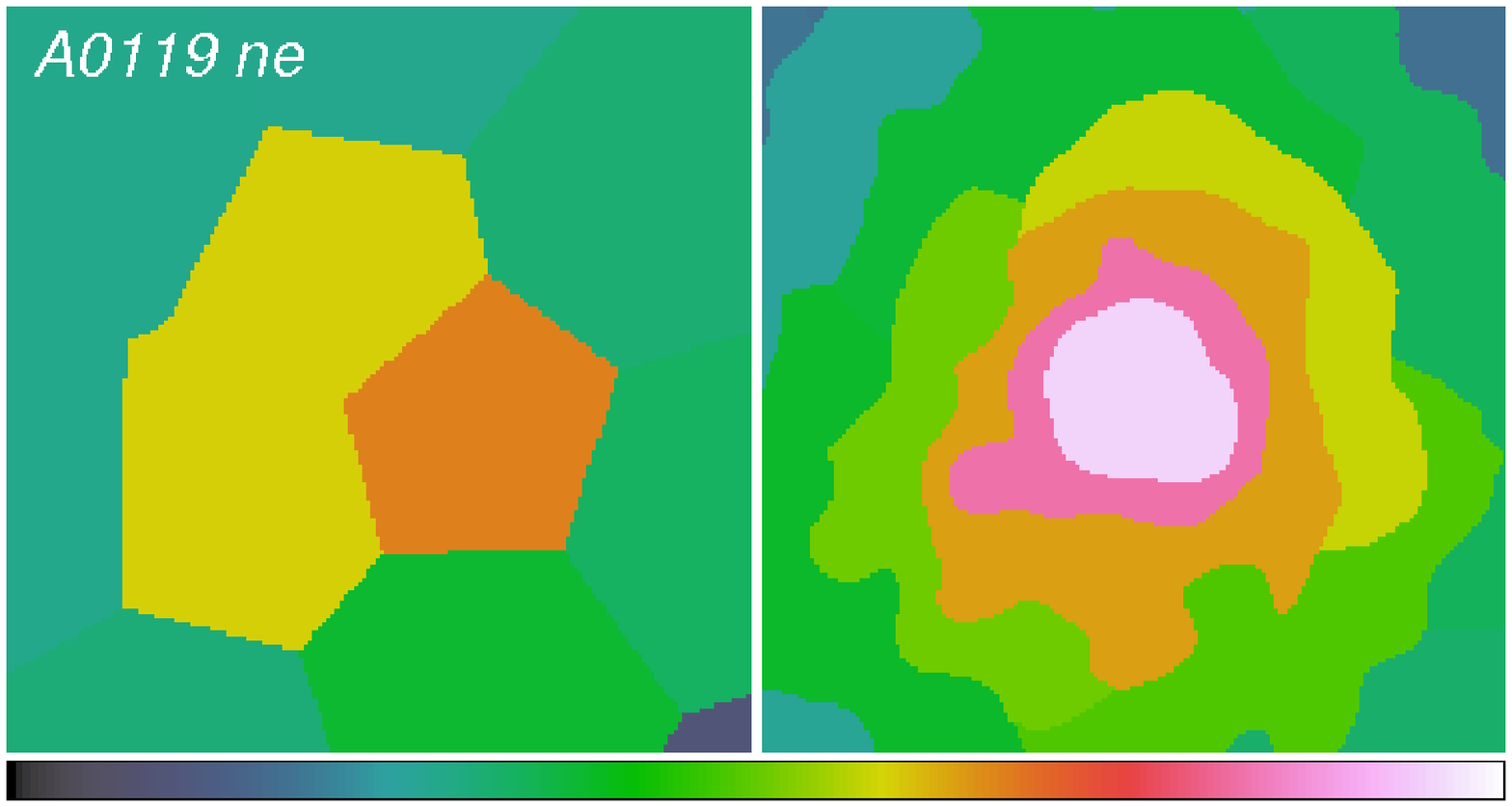}

\includegraphics[angle=0,width=8cm]{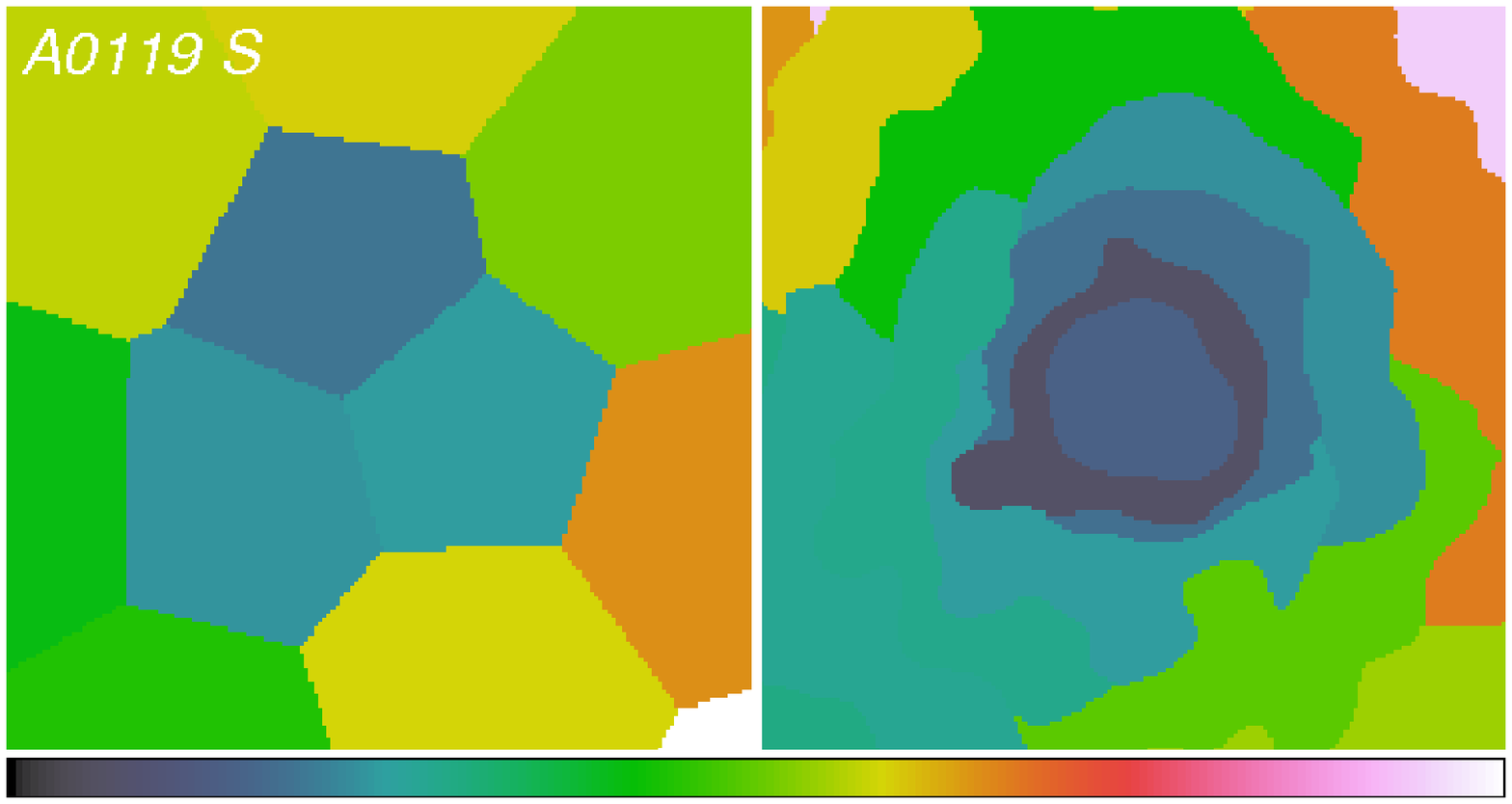}
\includegraphics[angle=0,width=8cm]{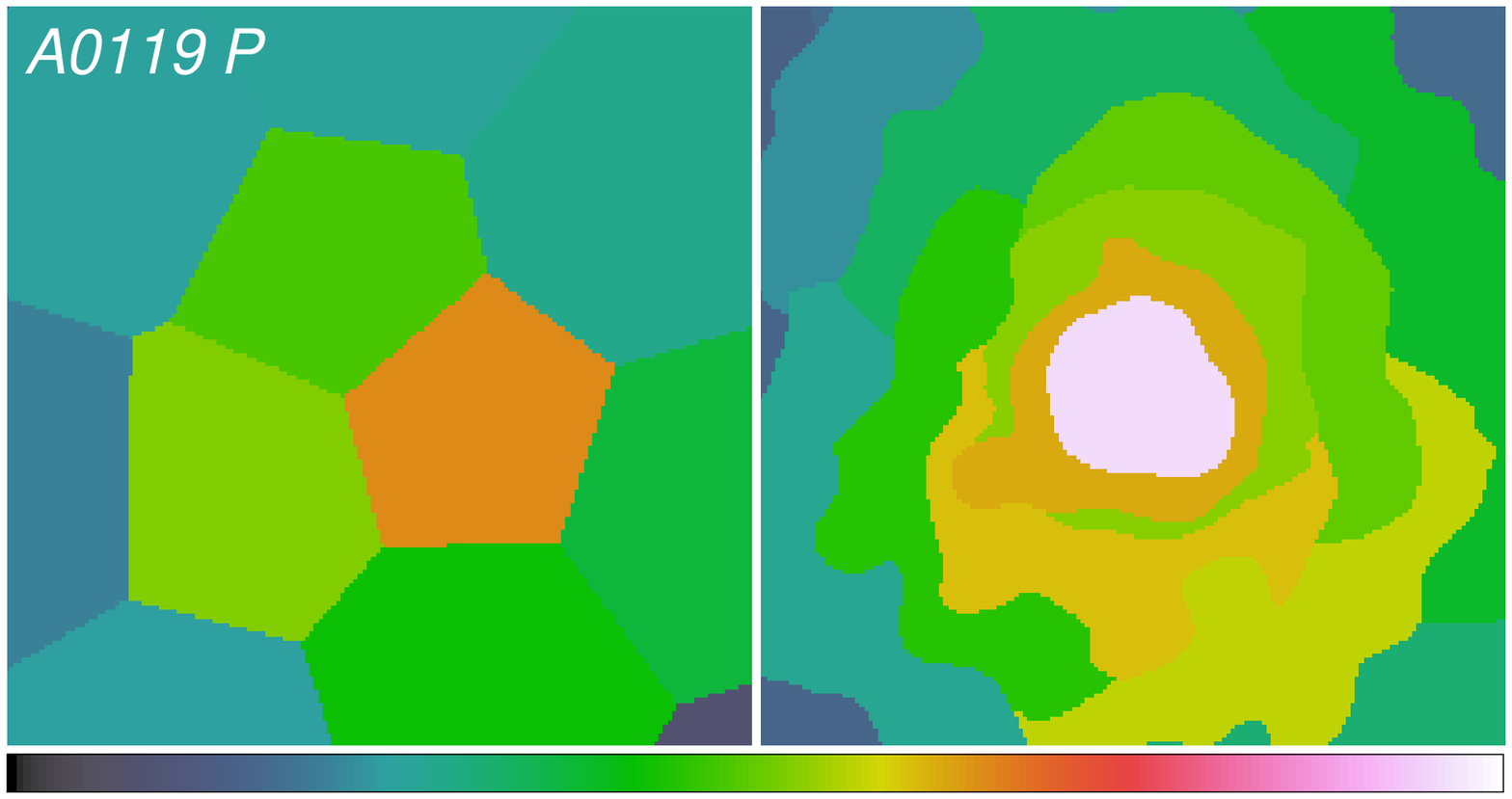}
\end{center}
\caption{Temperature ($T$), electron number density ($n_{\rm e}$), entropy
  ($S$), and pressure ($P$) maps for A0119 with the \emph{Mask-V} on the 
  left and the \emph{Mask-S} on the right in each panel. The 
  color bars are in the range of 3.5-7.2~keV,
  0-0.003~cm$^{-3}$, 200-900~keV~cm$^2$, and 0-0.018~keV~cm$^{-3}$ in the
  $T$, $n_{\rm e}$, $S$, and $P$ panels, respectively.  The image size is 
  $11^{\prime} \times 11^{\prime}$. 
  \label{f:a0119map}}
\end{figure*}

The entropy ($S$) and pressure ($P$) maps can be derived from the temperature
and electron number density maps by $S=kT n_{\rm e}^{-2/3}$ and $P=kT n_{\rm
  e}$. The X-ray spectrally measured temperature, electron number density,
entropy and pressure maps for IIIZw54, A3391, EXO0422 and A0119, respectively,
are shown in Figures~\ref{f:iiizw54map}-\ref{f:a0119map}.

\section{Substructure Diagnostics with ICM $T$, $n_{\rm e}$, $S$ and $P$ Maps}

The fluctuations in the 2D maps and their scatter can be used as
substructure diagnostics. A disturbance in a cluster may appear as a
high amplitude of and/or a discontinuity in the radial profile of the
scatter of the fluctuations. Unrelaxed clusters may show larger
fluctuations and significant correlations between,
e.g., temperature and electron number density fluctuations. The
substructure diagnostics of galaxy clusters are therefore directly
linked to the scatter of the scaling relations due to the bias in
X-ray hydrostatic masses and X-ray observables caused by
substructures.

The \emph{Mask-S} method, whose bins are close to radial annuli, has the
advantage that the interpretation of a comparison to a mean temperature
profile is more straightforward because the range of radii sampled in each bin
is smaller. Therefore, we concentrate more on the results from this method,
particularly when the data quality is low.

\subsection{Scatter of the ICM $T$, $n_{\rm e}$, $S$ and $P$ Fluctuations}
\label{s:scatter} 

The scatter of the fluctuations in the 2D maps from the mean profile
can be used as diagnostics of the ICM substructure. Here we briefly
describe how the scatter and error are calculated. To avoid systematic
errors due to uncertain background subtraction, we only consider bins
of radii $\le 0.6r_{\rm tr}$ ($r_{\rm tr}$ see Table~\ref{t:basic}).

To better show the asymmetries of the clusters, we use a polar
coordinate system with the cluster center as its coordinate center.
The scaled distribution of a 2D map is defined as $D(d, \theta)$, in
which ($d, \theta$) are the angle and distance in the polar coordinate
system. For example $D(d_i, \theta_i)=T(R_i, \theta_i)/T_{0.2-0.5
  r_{500}}$ is the value of the 2D spatial bin ($d_i, \theta_{i}$) in
a 2D temperature map. Note that $T_{0.2-0.5 r_{500}}$ is the cluster
temperature, a volume average of the radial temperature profile
limited to the radial range of $0.2-0.5 r_{500}$ (see Appendix B.2 in
Zhang et al. 2008) and $d_i=R_i/r_{500}$ is the distance between the
cluster center and the geometric center of that 2D bin scaled by the
cluster $r_{500}$. To investigate the asymmetry, we derive an
azimuthal-averaged profile $D(d)$ of the scaled distribution
$D(d,\theta)$. A non-parametric locally weighted
regression\footnote{The procedure calls ``lowess'' in the R package,
  which uses locally weighted polynomial regression (e.g., Becker et
  al. 1988). We used the default smoothing span $f=2/3$.  ``Local'' is
  defined by the distance to the ``floor($f* n$)''{\it th} nearest
  neighbor, and tricubic weighting is used for $x$ which fall within
  the neighborhood.  Note that ``floor'' in the R package takes a
  single numeric argument $x$ and returns a numeric vector containing
  the largest integers not greater than the corresponding elements of
  $x$. More details are at http://CRAN.R-project.org.} (Sanderson et
al.  2005, and references therein) of the averaged profile is used to
derive the mean profile $\langle D(d)\rangle$. The absolute
fluctuation distribution is defined as $F=|D(d, \theta)/\langle
D(d)\rangle-A|$. Here, the renormalization $A$ is not equal to one only
when there is a possible bias in the mean scaled profile. In this
work, we used $A=1$. In the scatter calculation, the weighting of the
absolute fluctuation $F_i$ is the area $w_i$ of the corresponding
$i$th spatial bin in the 2D map.  The cumulative scatter and error are
calculated from the area weighted absolute fluctuations as
$\sqrt{\Sigma F_i^2 w_i}/\sqrt{\Sigma w_i}$, with all bins within $d$.
The differential scatter and error are calculated from the weighted
absolute fluctuations with all bins in the range of $d_1$ and $d_2$,
in which $d=0.5(d_1+d_2)$.

\subsection{Scatter Profiles from the Azimuthal Average}
\label{s:step2}

\begin{figure*}
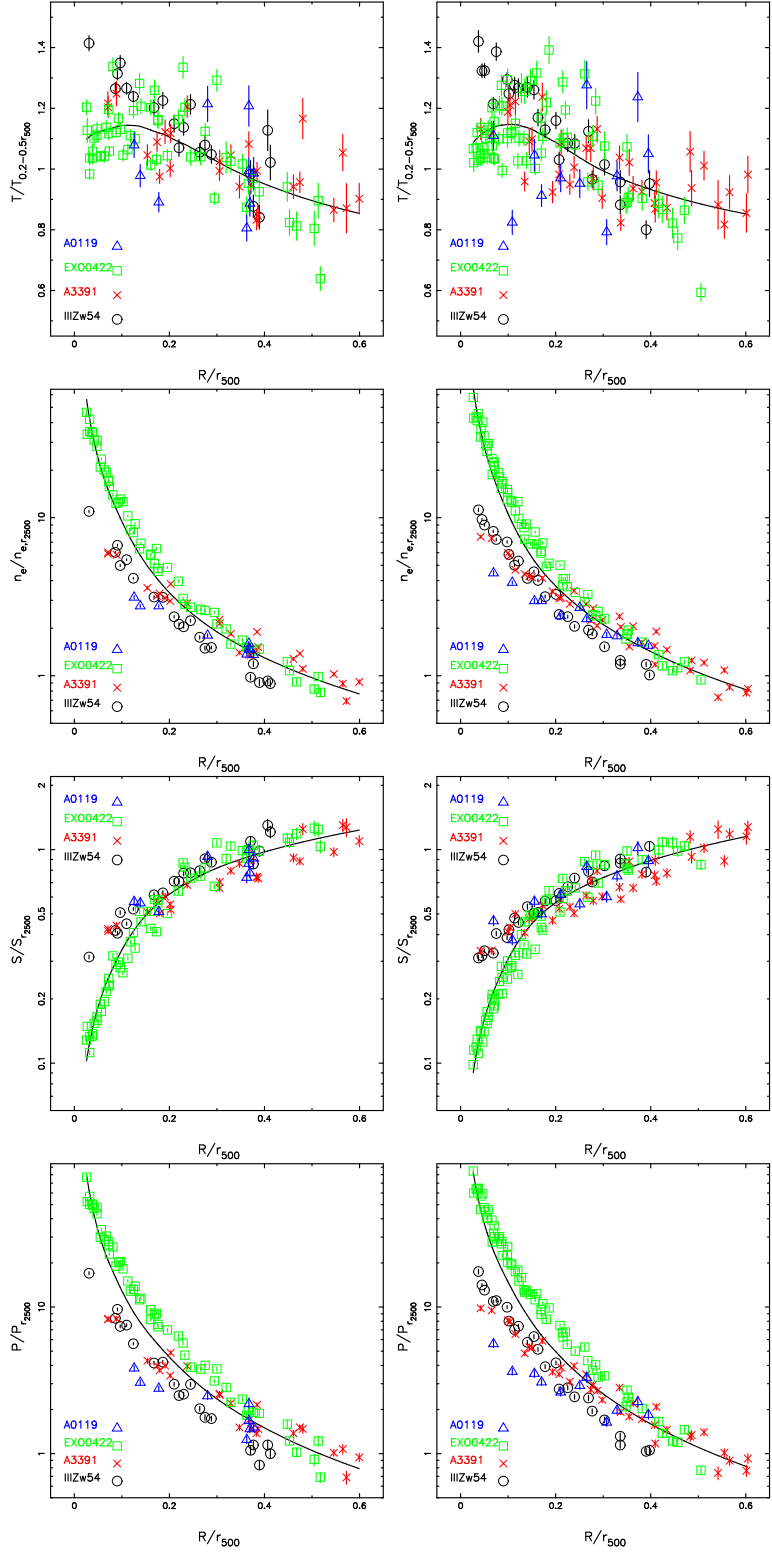

\begin{center}
  \includegraphics[angle=270,width=5.0cm]{plots/f09a.ps}
  \includegraphics[angle=270,width=5.0cm]{plots/f09b.ps}

  \includegraphics[angle=270,width=5.0cm]{plots/f09c.ps}
  \includegraphics[angle=270,width=5.0cm]{plots/f09d.ps}

  \includegraphics[angle=270,width=5.0cm]{plots/f09e.ps}
  \includegraphics[angle=270,width=5.0cm]{plots/f09f.ps}

  \includegraphics[angle=270,width=5.0cm]{plots/f09g.ps}
  \includegraphics[angle=270,width=5.0cm]{plots/f09h.ps}
\end{center}
\vspace{-0.5cm}
\caption{Normalized temperature, electron number density, entropy,
  and pressure distributions (from top to bottom) of the spectrally
  measured 2D maps using \emph{Mask-V} (left) and \emph{Mask-S}
  (right), respectively. The curves are the local regression fits of
  the distributions of the four clusters as a whole.
  \label{f:tdis}}
\end{figure*}

\begin{figure*}
\begin{center}
  \includegraphics[angle=270,width=8.0cm]{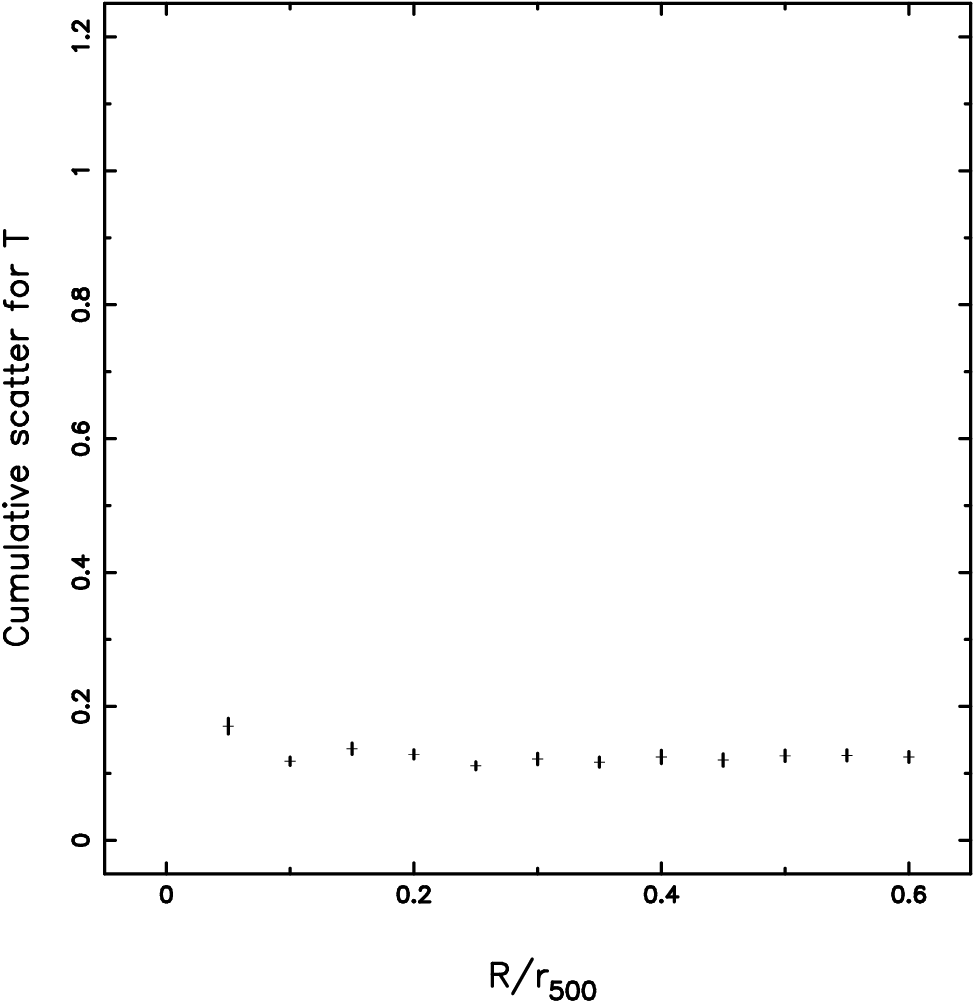}
  \includegraphics[angle=270,width=8.0cm]{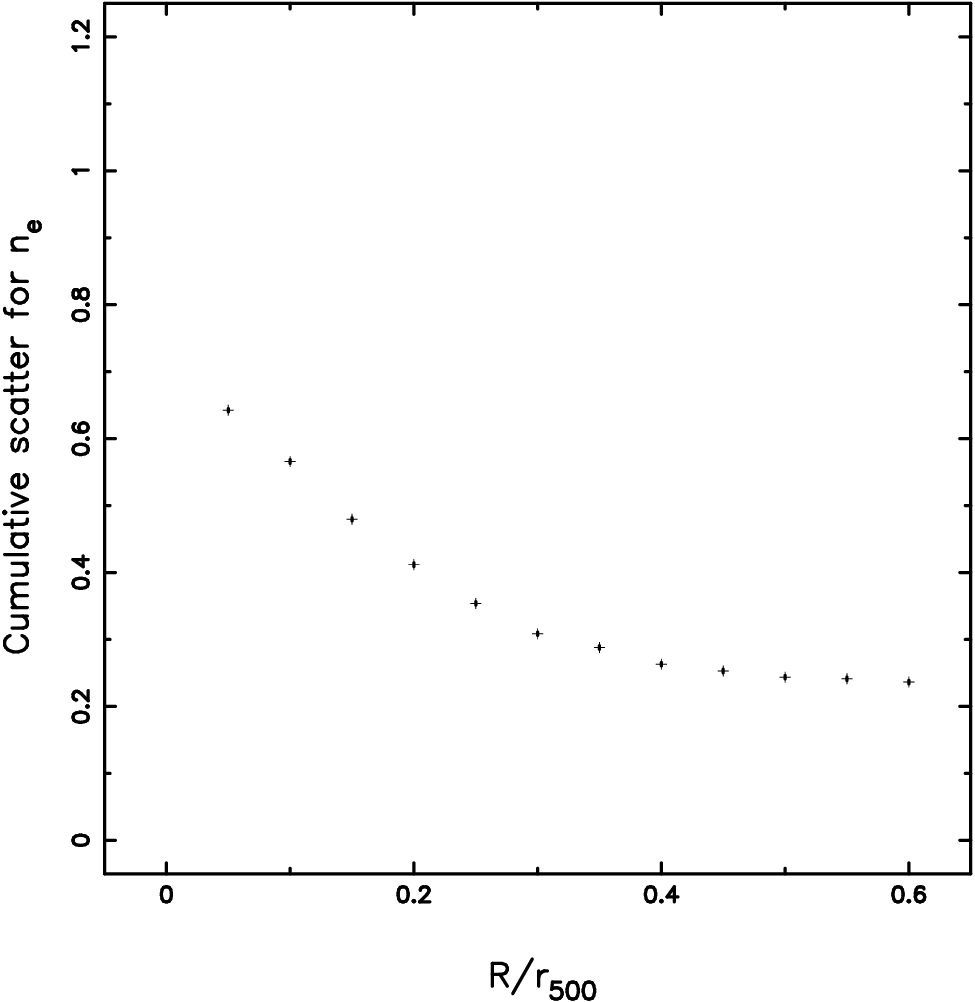}

  \includegraphics[angle=270,width=8.0cm]{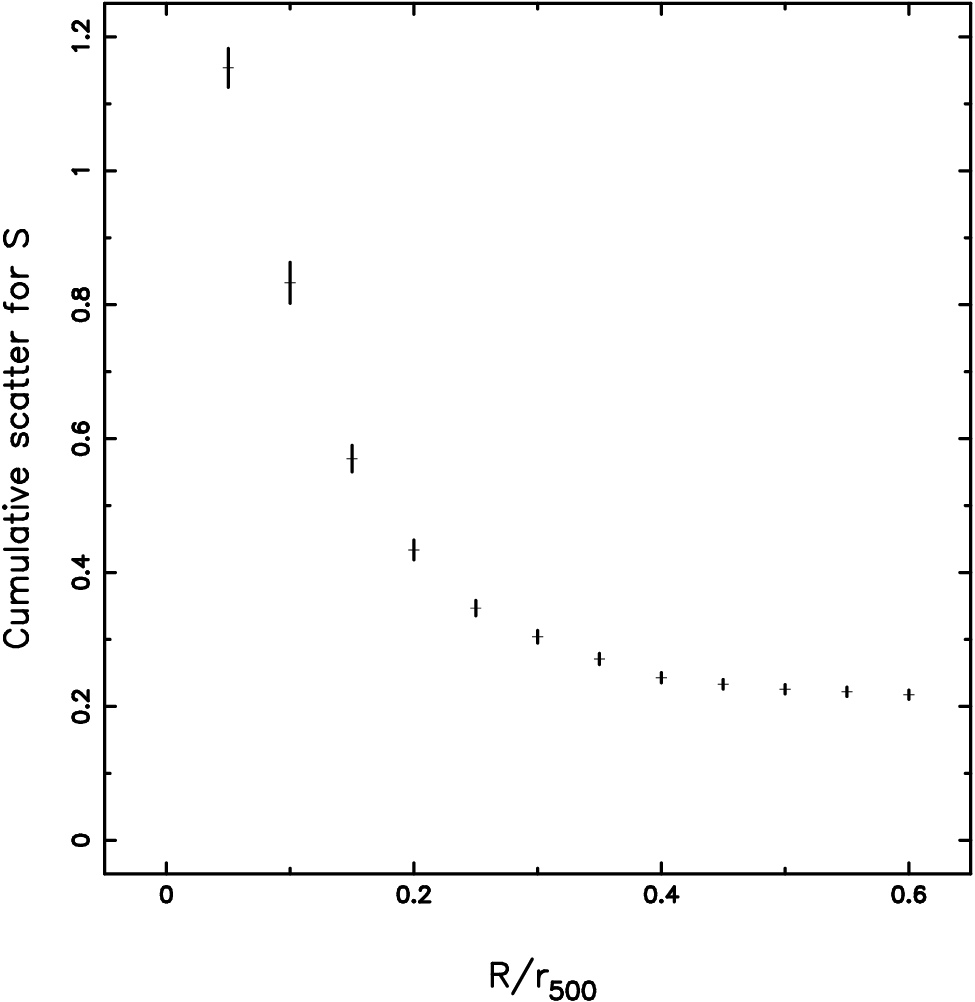}
  \includegraphics[angle=270,width=8.0cm]{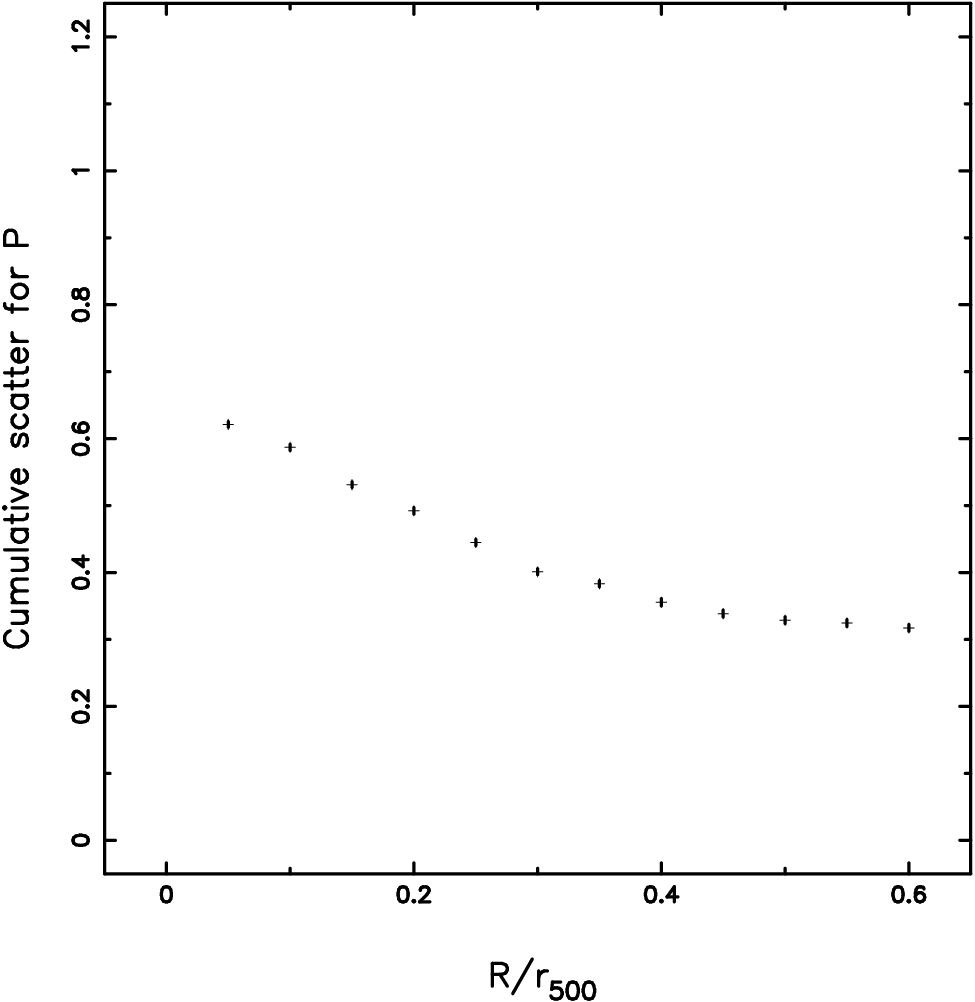}
\end{center}
\vspace{-0.5cm}
\caption{Cumulative scatter of the temperature, electron number
  density, entropy, and pressure fluctuations in the 2D maps for 
  the four clusters as a whole using
  \emph{Mask-S}. Note that for each plot, the
  mean profile $\langle D(d)\rangle$ is determined for the four clusters
  as a whole as described in Sect.~\ref{s:step2}.
  \label{f:tnormsca}}
\end{figure*}

\begin{figure*}
\begin{center}
  \includegraphics[angle=270,width=8.0cm]{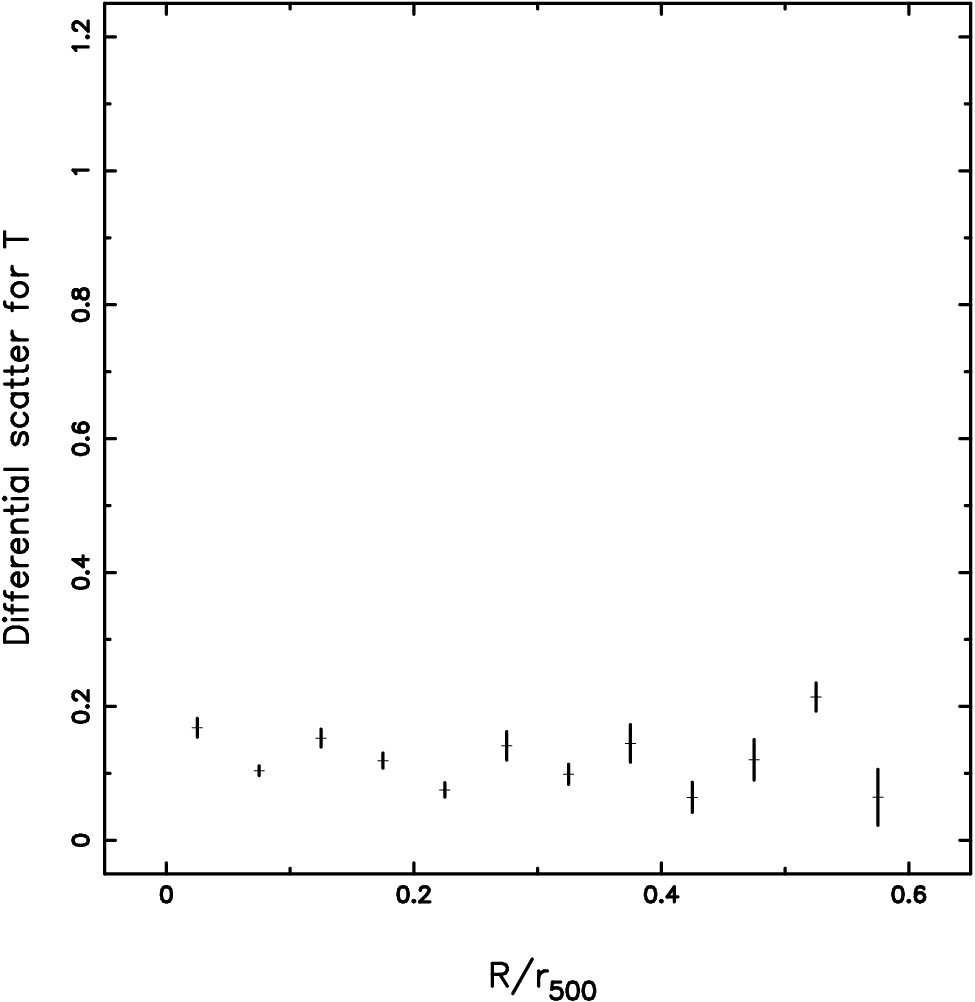}
  \includegraphics[angle=270,width=8.0cm]{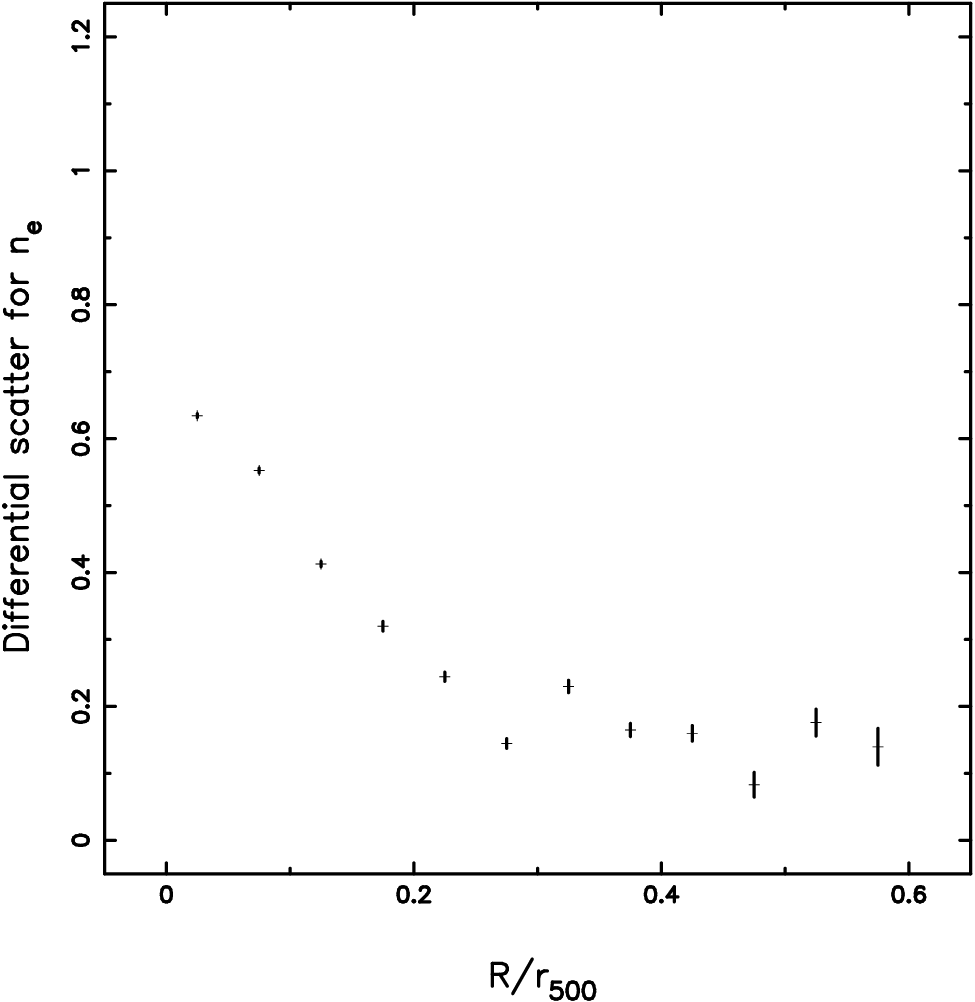}

  \includegraphics[angle=270,width=8.0cm]{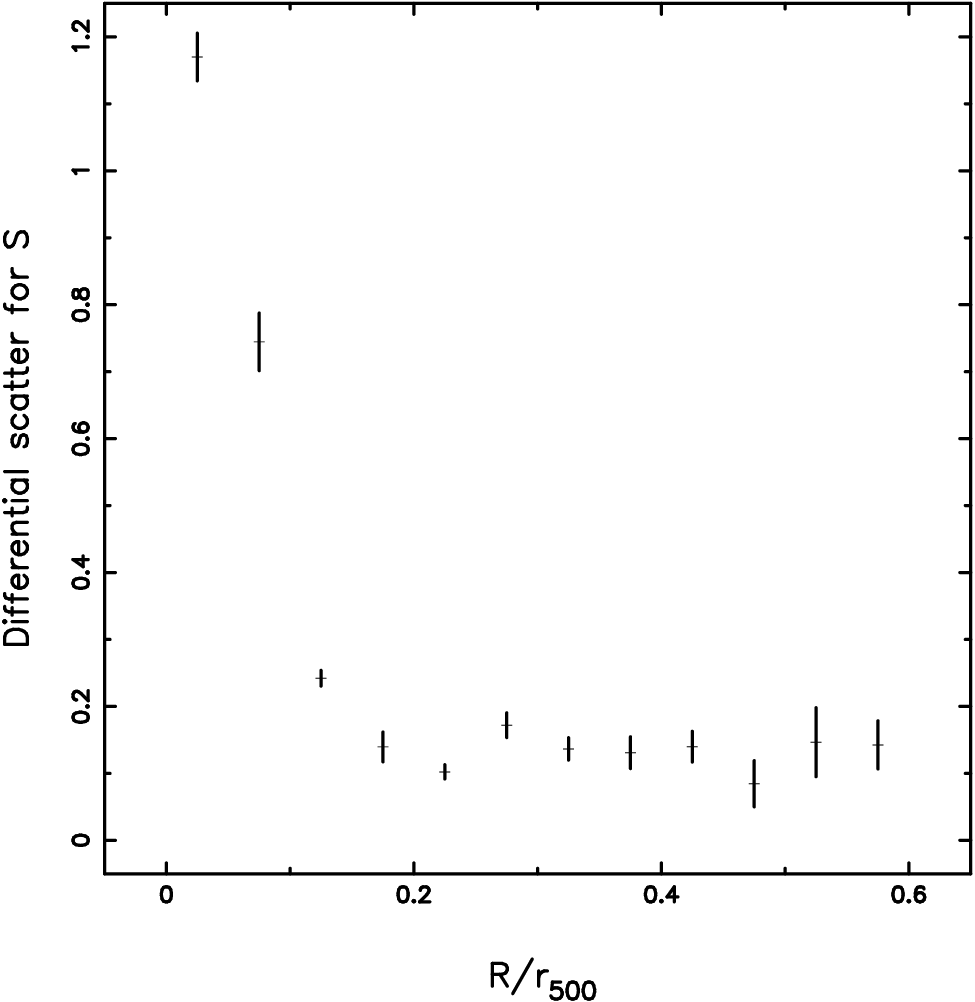}
  \includegraphics[angle=270,width=8.0cm]{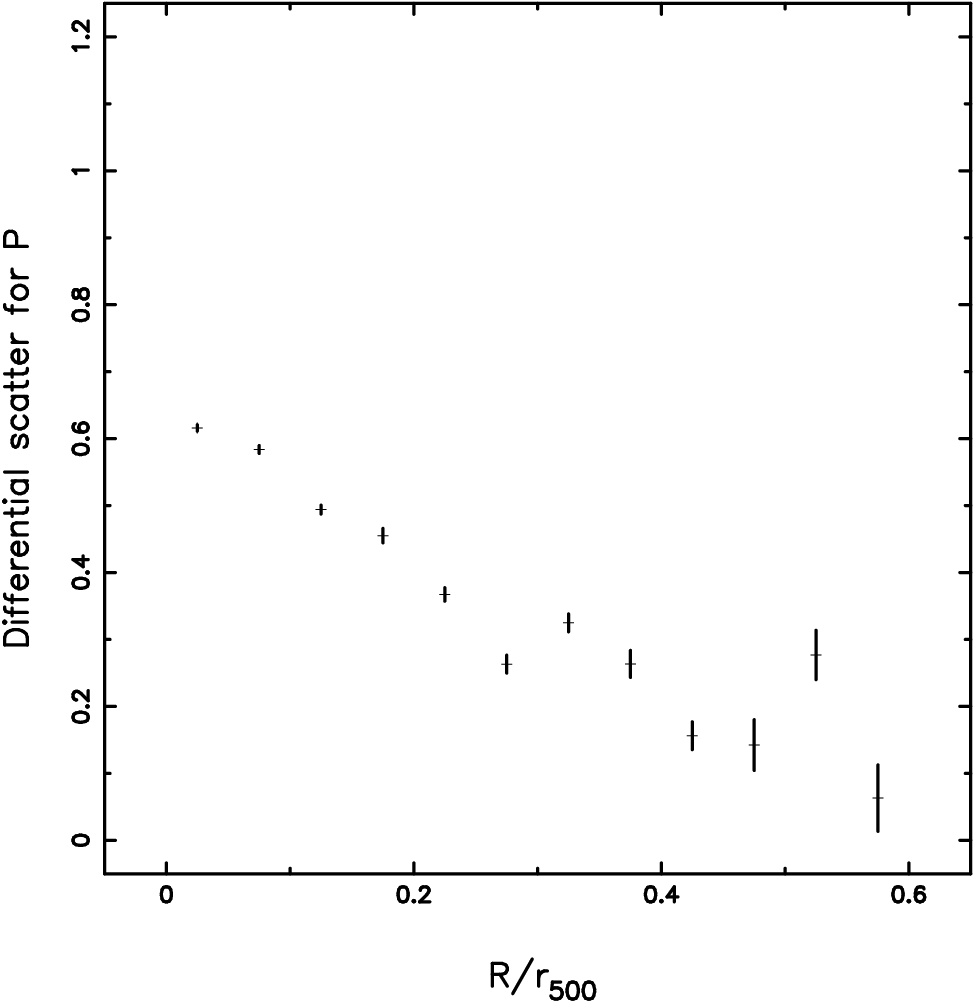}
\end{center}
\vspace{-0.5cm}
\caption{Differential scatter of the temperature, electron number density,
entropy, and pressure 
fluctuations in the 2D maps for the four clusters 
as a whole using \emph{Mask-S}. Note that for each plot, 
the mean profile $\langle D(d)\rangle$ 
is determined for the four clusters as a whole as described in 
Sect.~\ref{s:step2}.
\label{f:tnormsca_dr}}
\end{figure*}

When one single azimuthal-averaged profile $\langle D(d)\rangle$ for the four
clusters as a whole is used, the scatter and error can indicate the degree of
self-similarity of the investigated azimuthal-averaged quantities. We thus
carry out a non-parametric fit on the scaled distribution of temperature,
electron number density, entropy, and pressure, respectively, for the four
clusters as a whole (see Figure~\ref{f:tdis}). The cumulative scatter and error
for the four clusters as a whole are shown in Figure~\ref{f:tnormsca}, and the
differential scatter and error are shown in Figure~\ref{f:tnormsca_dr}.  The
highest amplitude occurs in the cluster core ($\le 0.3r_{500}$) which is
caused by the known difference between cool core and non-cool core clusters.

\begin{figure*}
\begin{center}
  \includegraphics[angle=270,width=8.0cm]{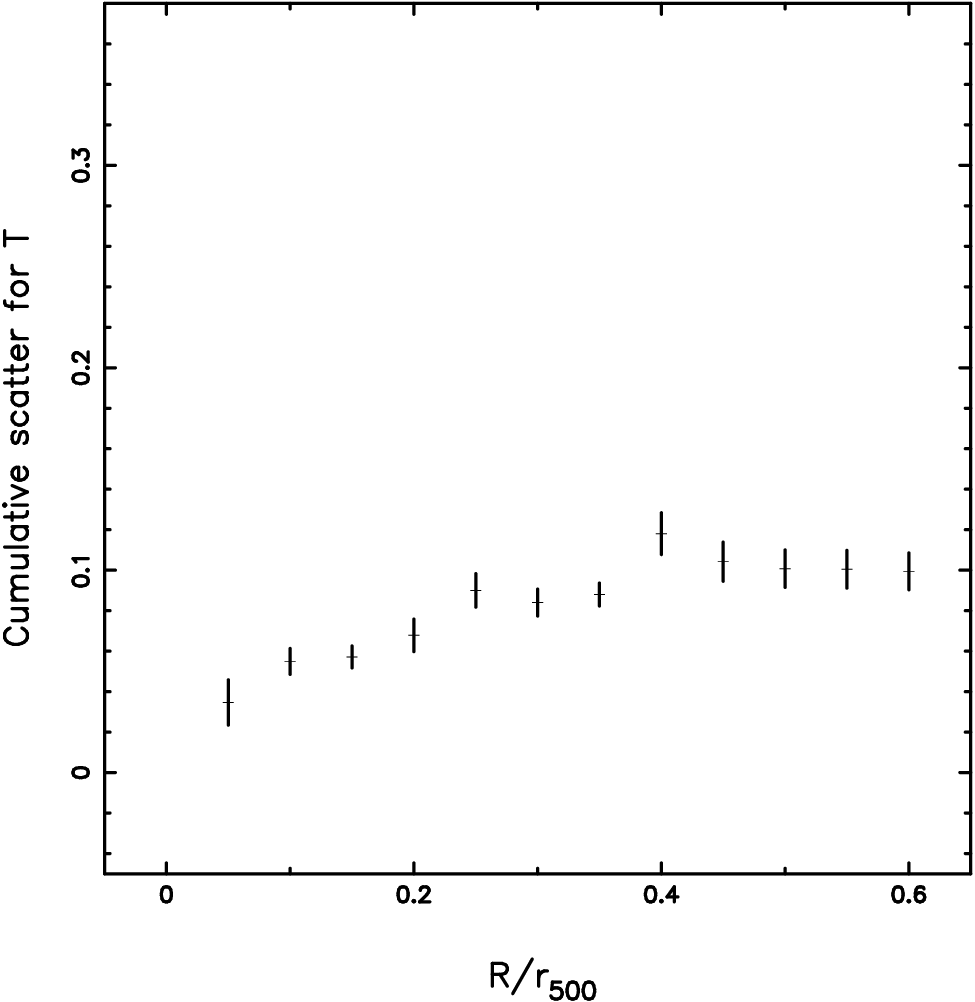}
  \includegraphics[angle=270,width=8.0cm]{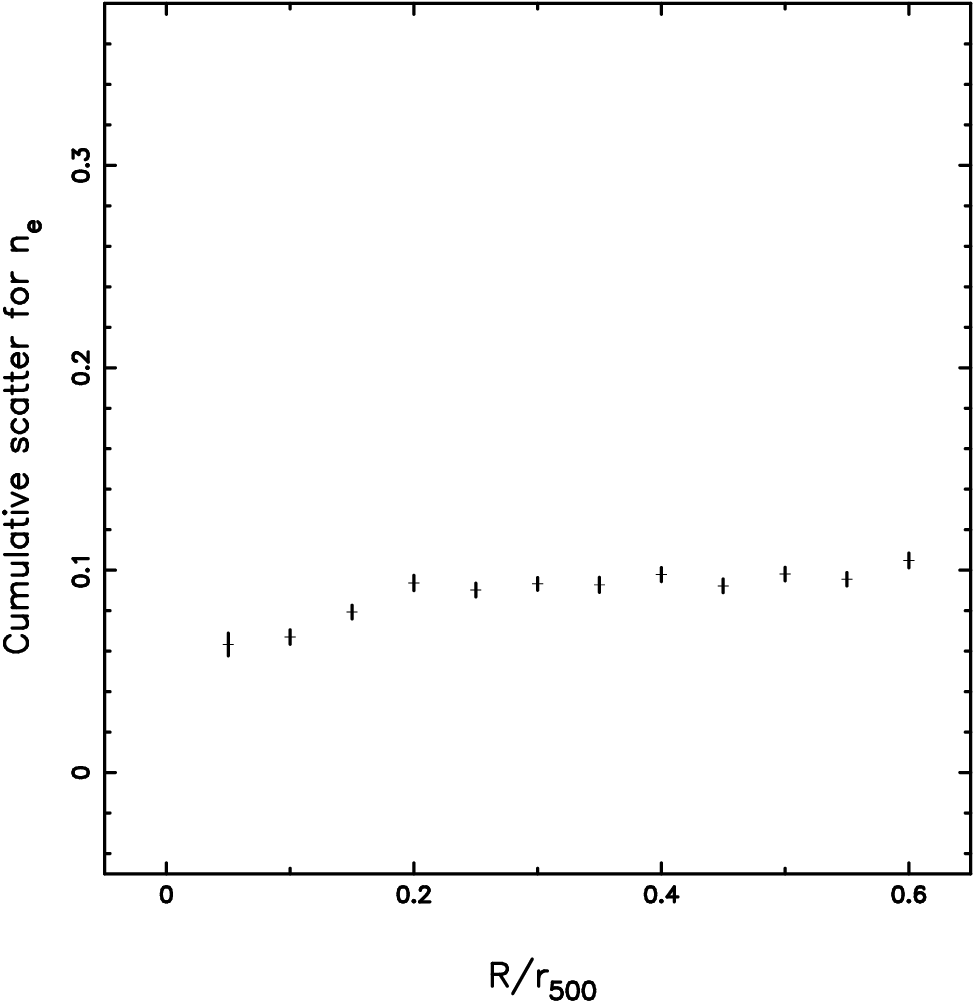}

  \includegraphics[angle=270,width=8.0cm]{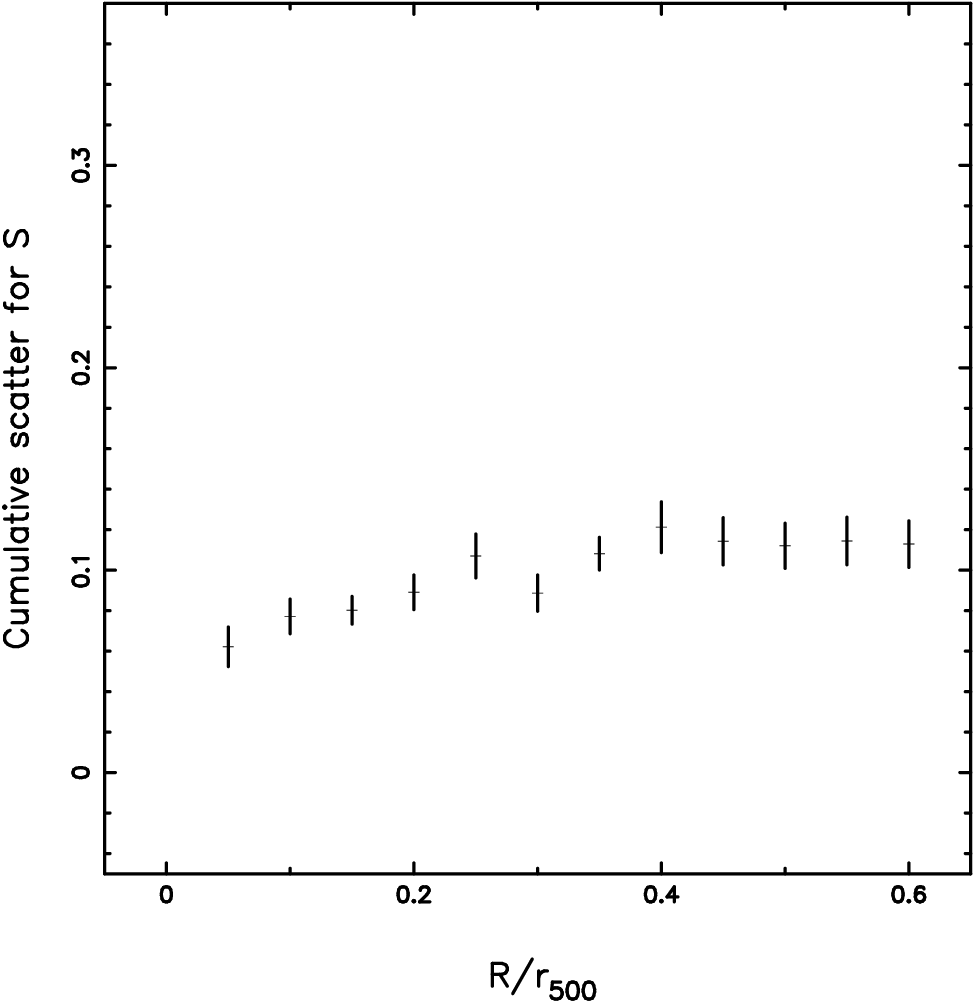}
  \includegraphics[angle=270,width=8.0cm]{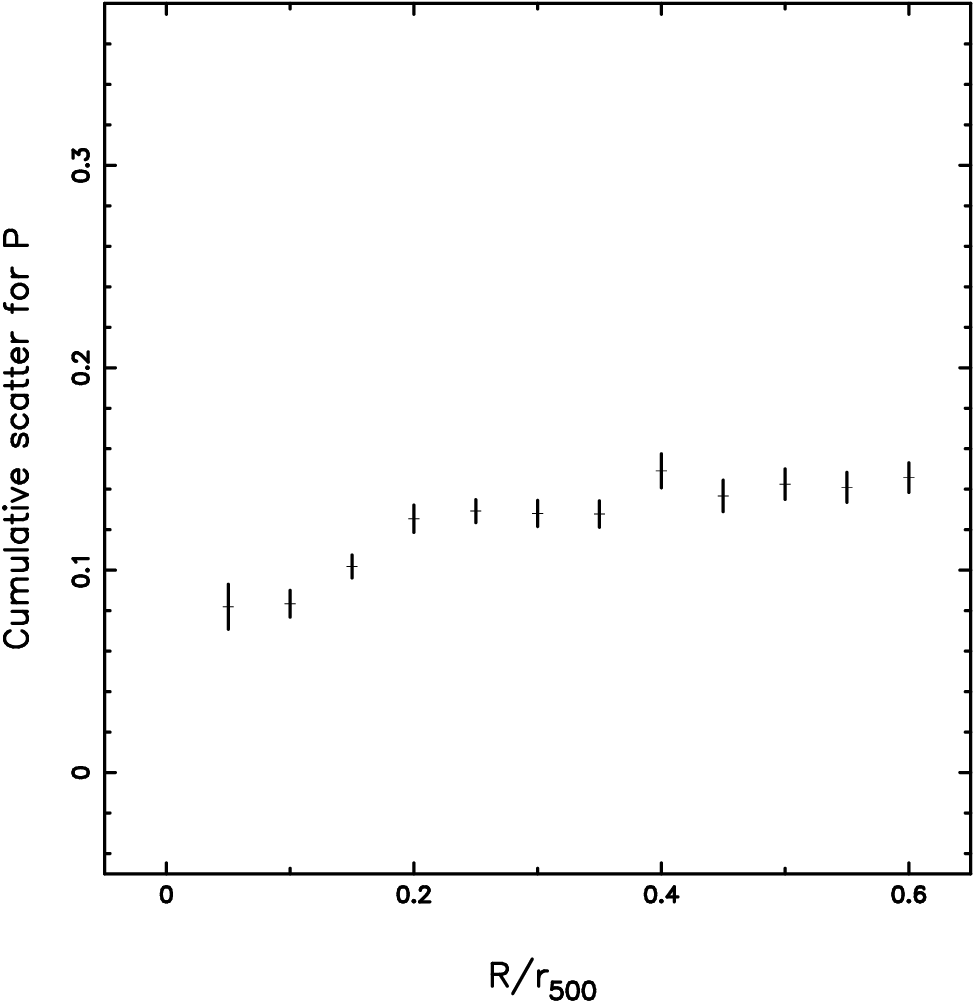}
\end{center}
\vspace{-0.5cm}
\caption{Cumulative scatter of the temperature, electron number
  density, entropy, and pressure fluctuations in the 2D maps
  for the four clusters as a whole using 
  \emph{Mask-S}. Note that for each plot, the
  mean profile $\langle D(d)\rangle$ is individually determined for
  each cluster as described in Sect.~\ref{s:step2}.
\label{f:tnormsca_ind}}
\end{figure*}

\begin{figure*}
\begin{center}
  \includegraphics[angle=270,width=8.0cm]{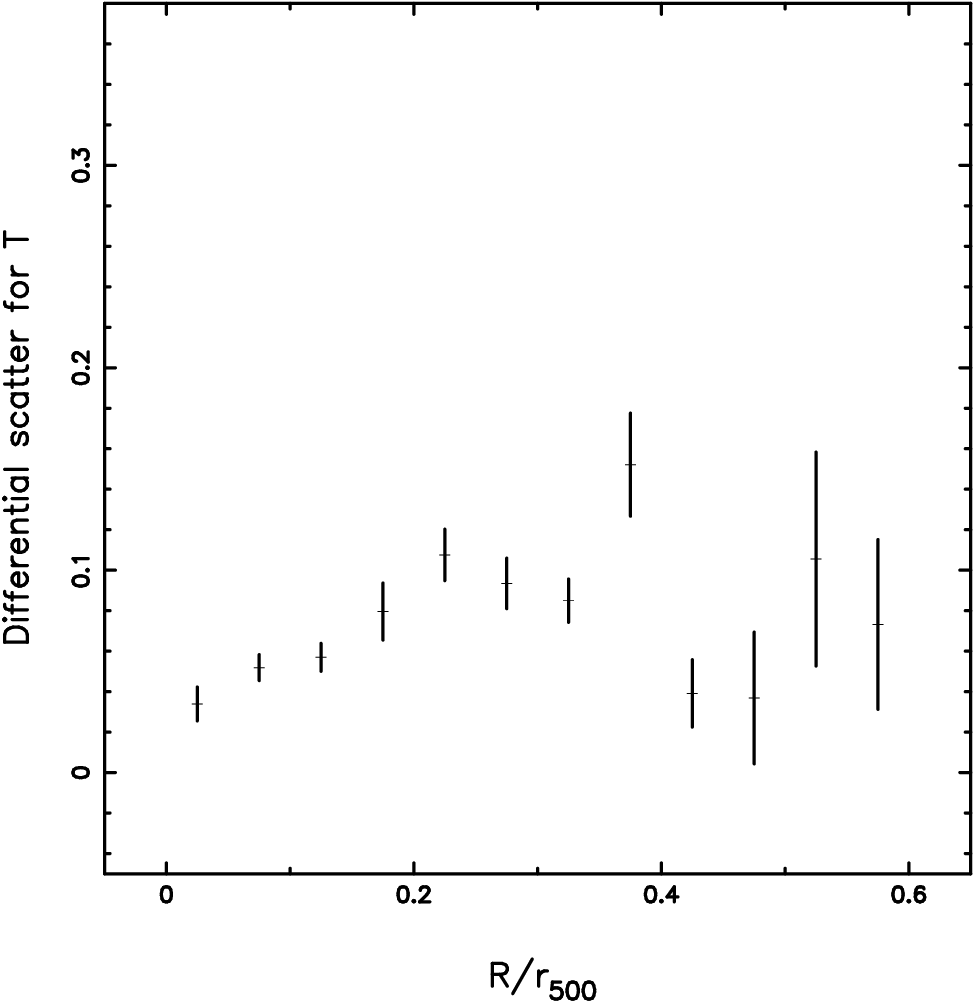}
  \includegraphics[angle=270,width=8.0cm]{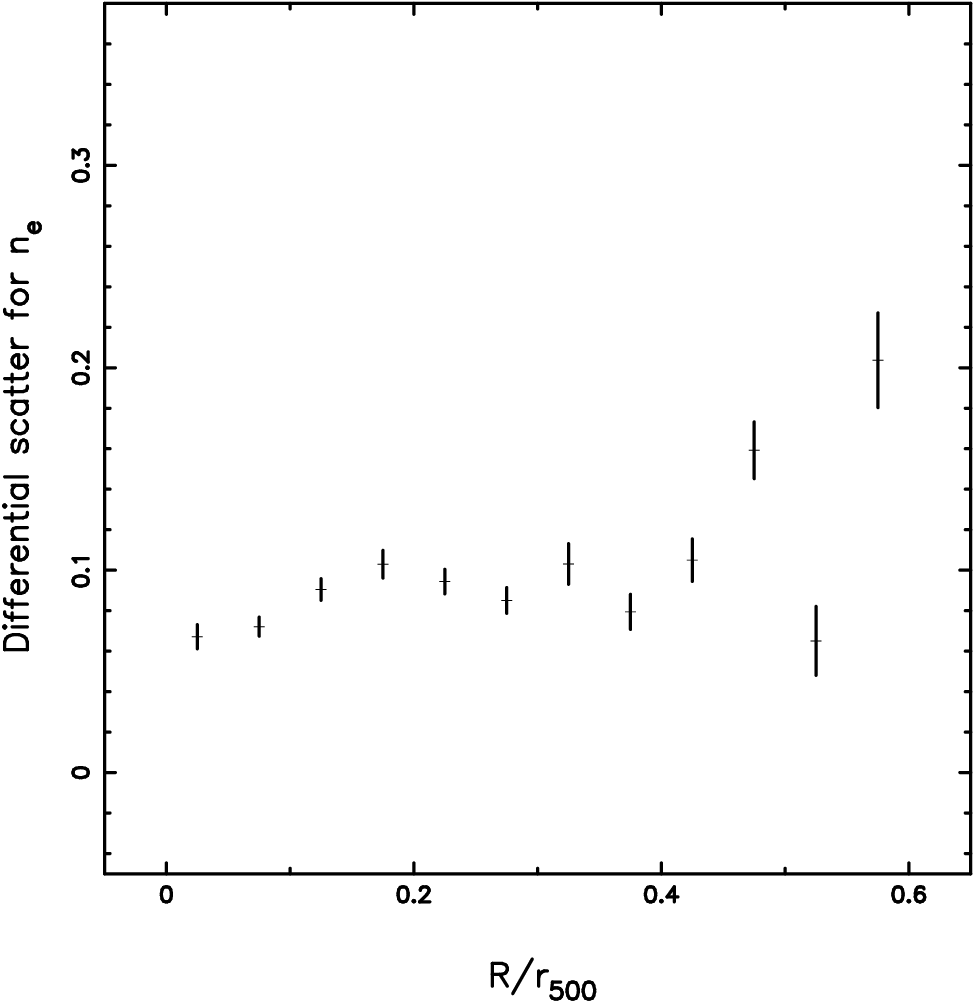}

  \includegraphics[angle=270,width=8.0cm]{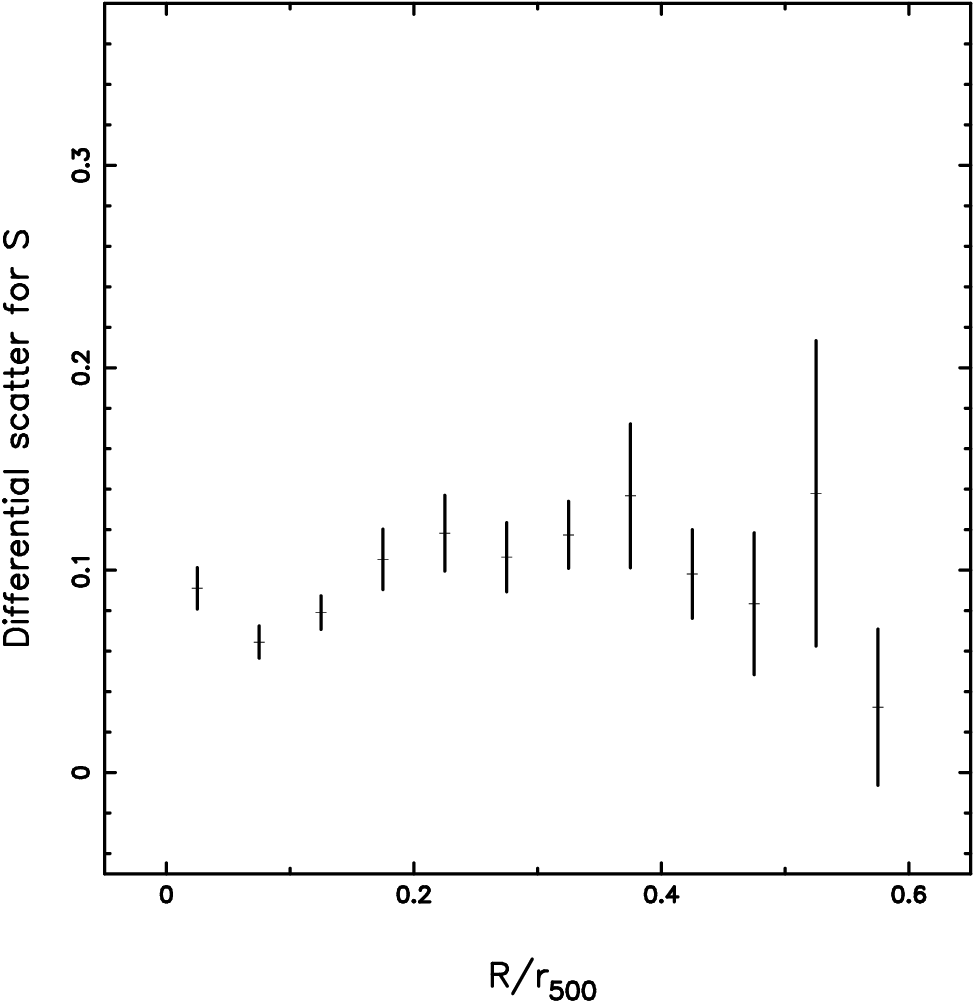}
  \includegraphics[angle=270,width=8.0cm]{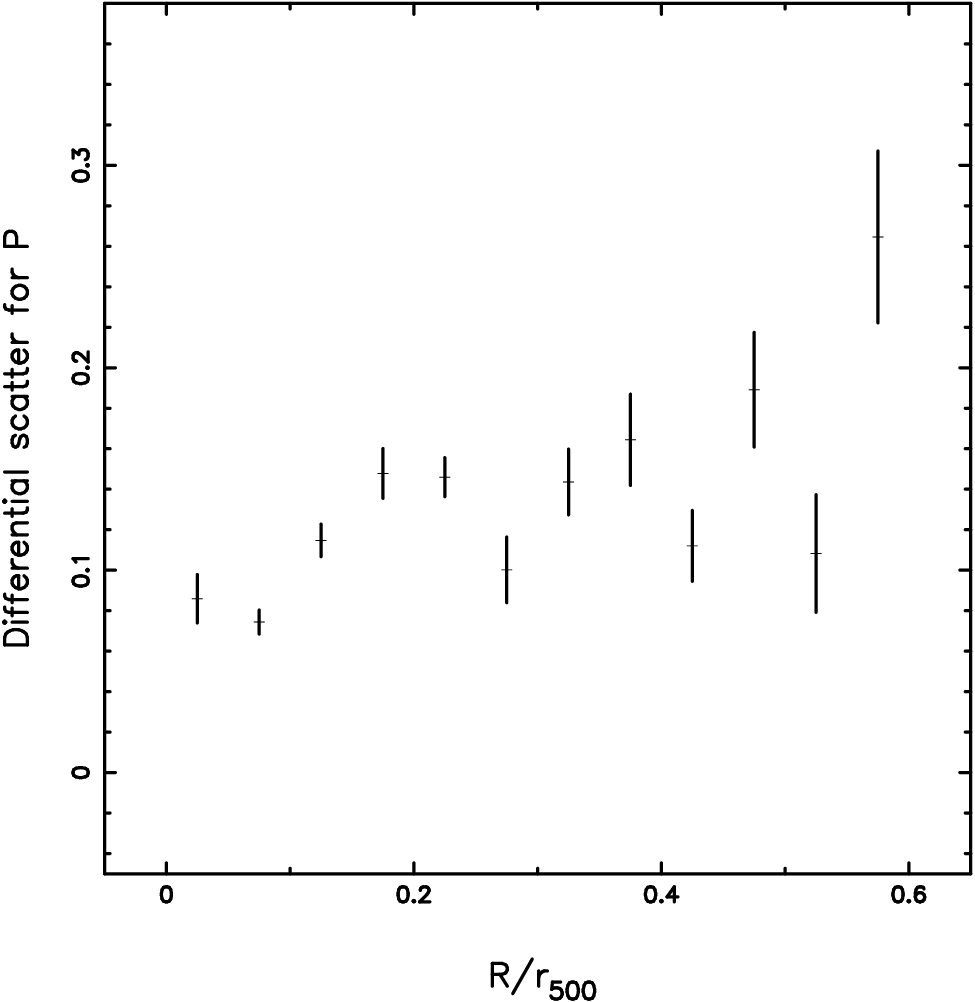}
\end{center}
\vspace{-0.5cm}
\caption{Differential scatter of the temperature, electron number
  density, entropy, and pressure fluctuations in the 2D maps 
  for the four clusters as a whole using \emph{Mask-S}. 
  Note that for each plot,
  the mean profile $\langle D(d)\rangle$ is individually determined
  for each cluster as described in Sect.~\ref{s:step2}.
  \label{f:tnormsca_dr_ind}}
\end{figure*}

To avoid the above scatter due to the difference between cool core and
non-cool core clusters in substructure diagnostics, we carry out a
non-parametric fit to the scaled distribution of temperature, electron
number density, entropy, and pressure, respectively, for each cluster
to derive its own azimuthally averaged scaled distribution $\langle
D(d)\rangle$. The cumulative scatter and error for the four cluster as
a whole using the individual cluster mean scaled profiles are shown in
Figure~\ref{f:tnormsca_ind}, and the differential scatter and error
are shown in Figure~\ref{f:tnormsca_dr_ind}. The cumulative scatter is
quite flat. Its amplitude indicates $\sim 10$\% fluctuations in the
temperature, electron number density, and entropy maps, and $\sim
15$\% fluctuations in the pressure map. The cumulative scatter and
error for each cluster using the individual cluster mean scaled
profiles are shown in Figure~\ref{f:tsca}, and the differential
scatter and error are shown in Figure~\ref{f:tsca_dr}. To derive the
mean profiles for each cluster, broader bins (a bin size of $0.2
r_{500}$) are used due to decreased statistics for individual clusters
compared with the statistics for the four clusters as a whole, where we use
a bin size of $0.1 r_{500}$. The scatter here can be used as
substructure diagnostics for individual clusters as shown in
Sect.~\ref{s:diag}.

\subsection{Substructure Diagnostics in Individual Clusters}
\label{s:diag}

A disturbance in the clusters appears as a high amplitude of and/or a
discontinuity in the radial profile of the scatter of the
fluctuations. The radial studies of the scatter of the fluctuations in
the 2D map for individual clusters thus provide detailed diagnostics
to identify the ICM substructures. 

\begin{figure*}
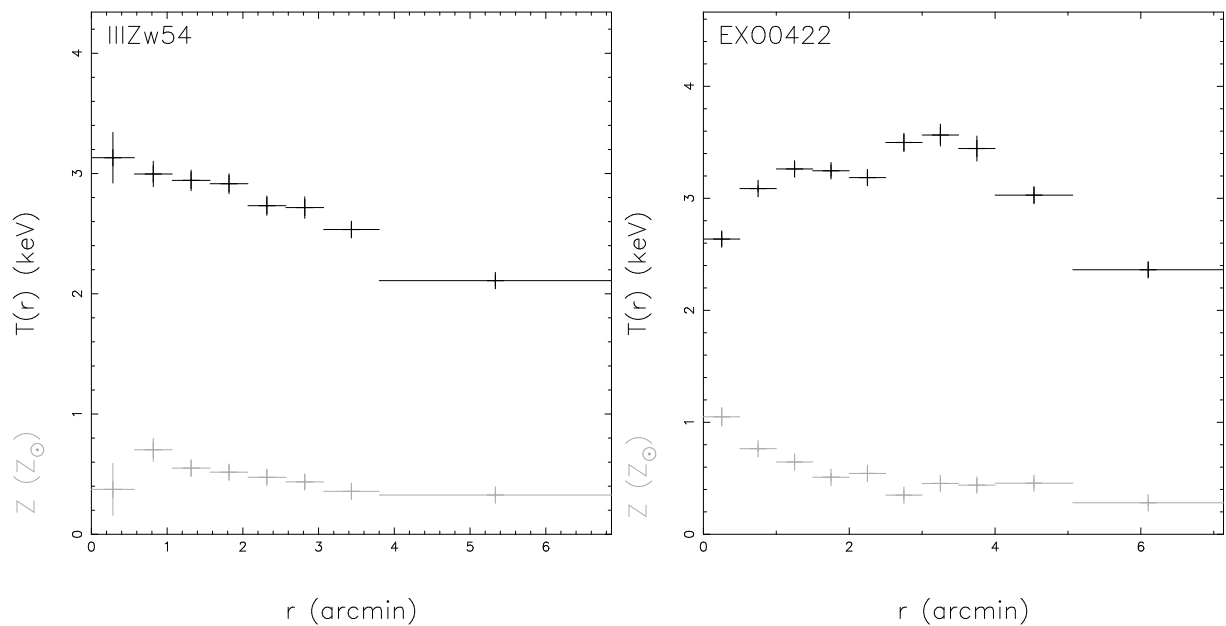

\begin{center}
\includegraphics[angle=270,width=8.0cm]{plots/f02a.ps}
\includegraphics[angle=270,width=8.0cm]{plots/f02b.ps}
\end{center}
\caption{Radial (deprojected) temperature (upper) and metallicity
  (lower) profiles. Note that the very central region ($r\le
  15^{\prime\prime}$) of EXO0422 was excluded in the spectral analysis
  to avoid the possible contamination from the galaxy CIG0422-09 found
  by Belsole et al. (2005).
  \label{f:tprof}}
\end{figure*}

\begin{figure*}
\begin{center}
  \includegraphics[angle=270,width=8.0cm]{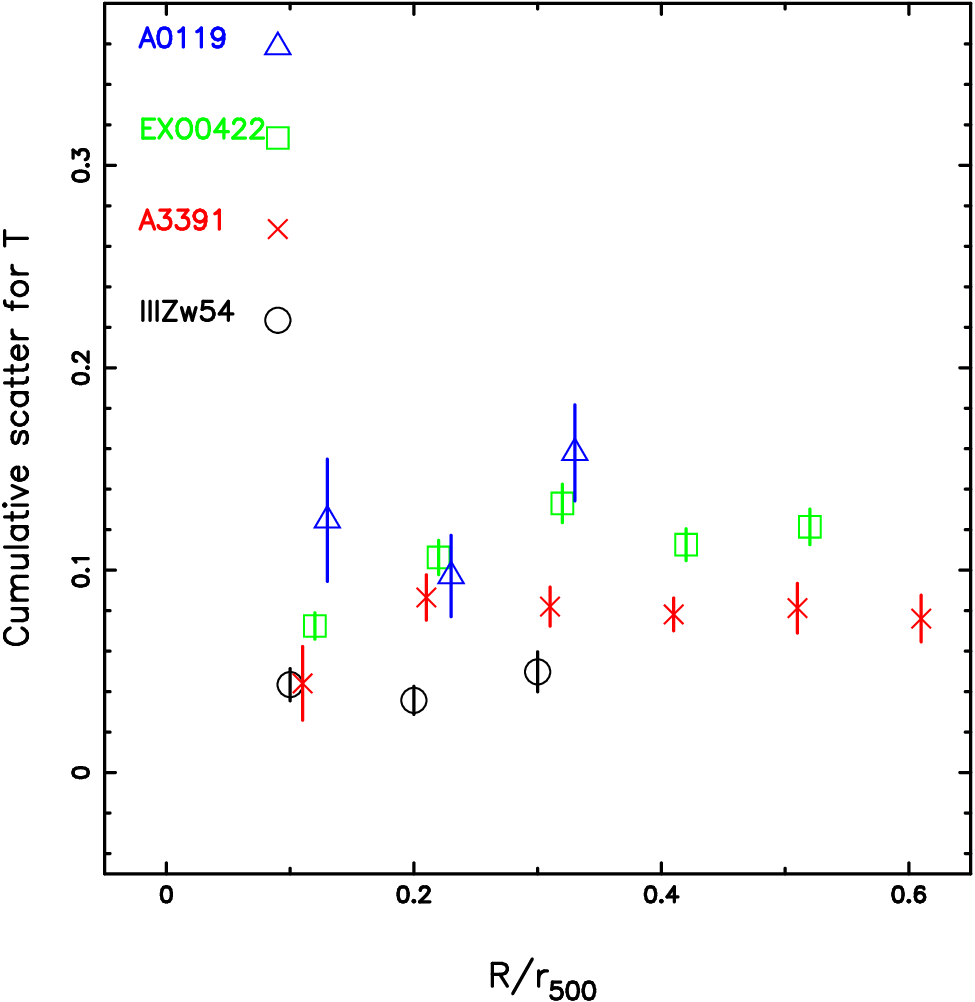}
  \includegraphics[angle=270,width=8.0cm]{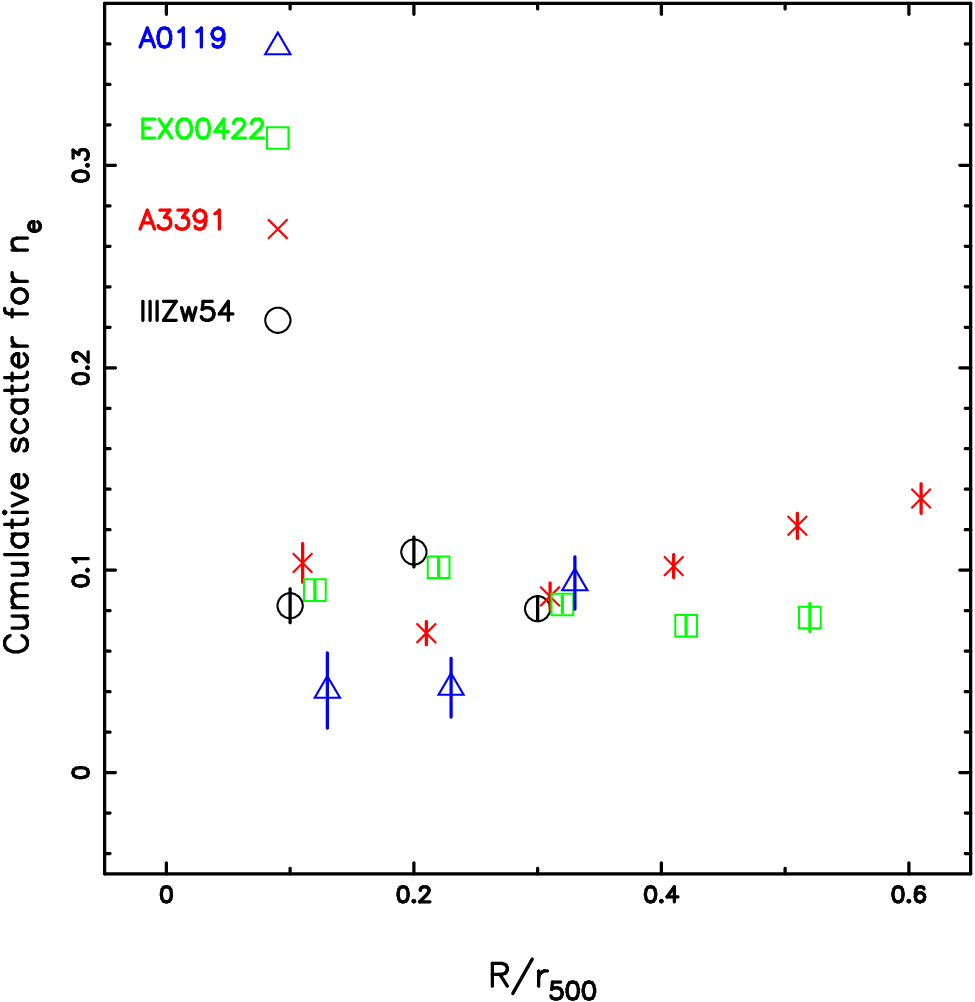}

  \includegraphics[angle=270,width=8.0cm]{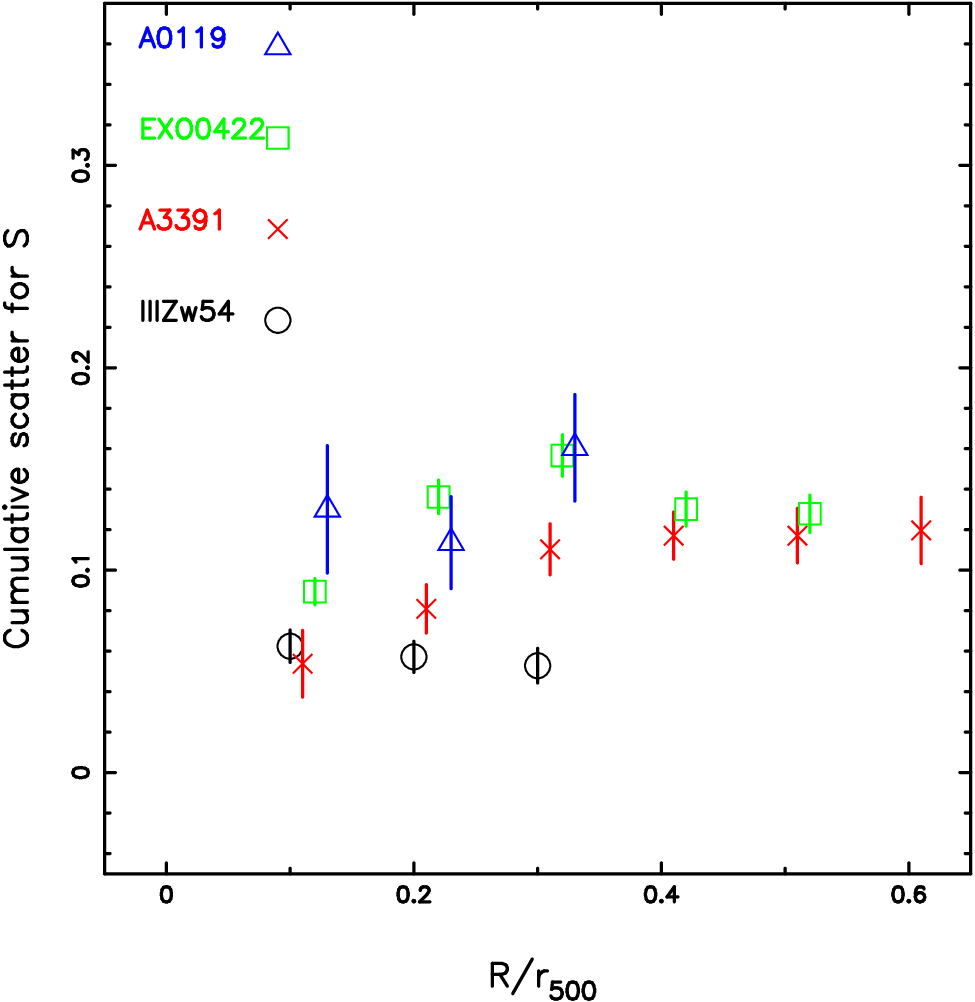}
  \includegraphics[angle=270,width=8.0cm]{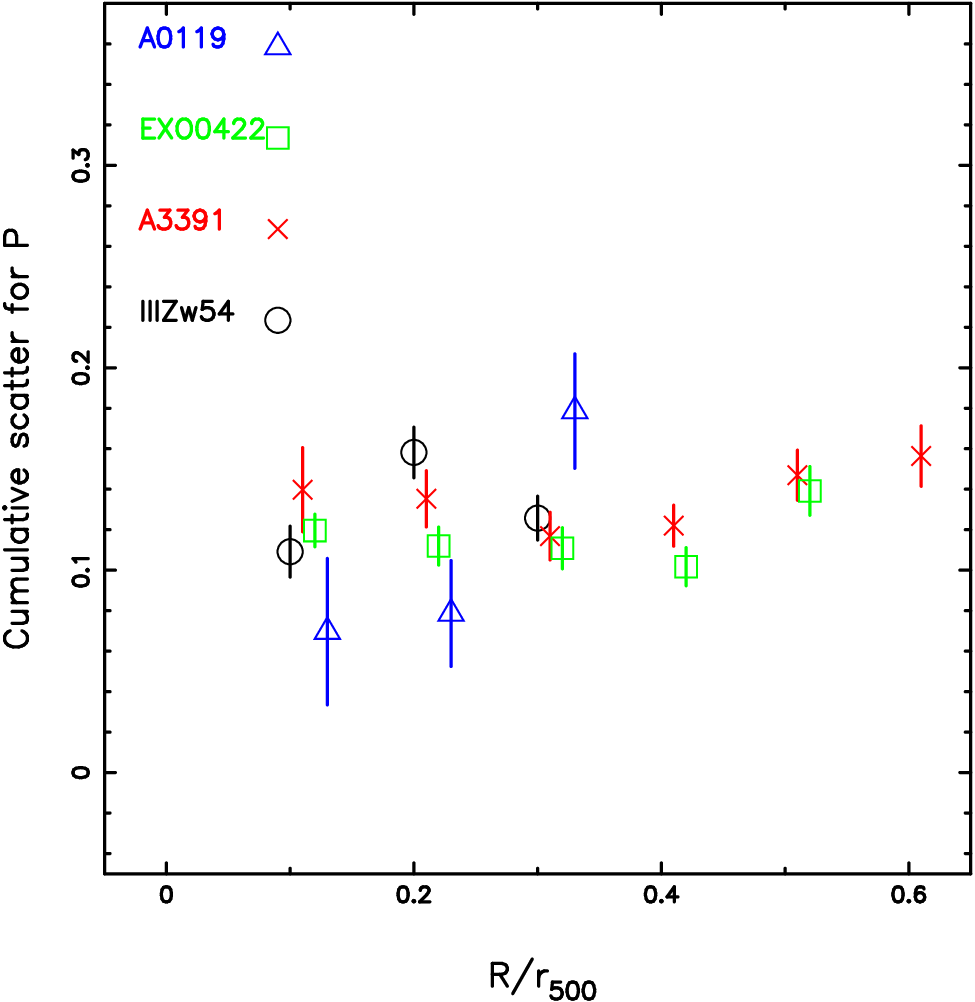}
\end{center}
\vspace{-0.5cm}
\caption{Cumulative scatter of the temperature, electron number
  density, entropy and pressure fluctuations in the 2D maps 
  for each cluster using 
  \emph{Mask-S}. The $X$-axis has been shifted by
  0.005, 0.010, and 0.015 for A3391, EXO0422 and A0119, respectively,
  to avoid the overlap. Note that for each plot, the mean profile
  $\langle D(d)\rangle$ is individually determined for each cluster as
  described in Sect.~\ref{s:step2}.
\label{f:tsca}}
\end{figure*}

{\it IIIZw54}. The 2D maps ($T$, $n_{\rm e}$, $S$ and $P$) show an
azimuthally symmetric appearance (Figure~\ref{f:iiizw54map}). It has a
relatively low amplitude of the scatter, particularly for the
temperature and entropy fluctuations ($\sim 5$\%, see
Figure~\ref{f:tsca}). The cumulative scatter appears very flat.  The
differential scatter (Figure~\ref{f:tsca_dr}), which describes the local
fluctuations, shows an increase for the temperature and entropy in the
radial ranges beyond $0.3r_{500}$ using the \emph{Mask-V} only.  This
indicates that the substructure is roughly round clumps detectable
using the \emph{Mask-V} instead of isophotal annuli detectable using
the \emph{Mask-S}. Both the radial studies of the scatter and the 2D
map appearance shows IIIZw54 is the most relaxed cluster among the four
clusters, with mild entropy clumps beyond $0.3r_{500}$. It is peculiar
that this cluster does not host a cool core (Figure~\ref{f:tprof}).
IIIZw54 therefore is an example of a relaxed non-cool core cluster.

\begin{figure*}
\begin{center}
\includegraphics[angle=0,width=8cm]{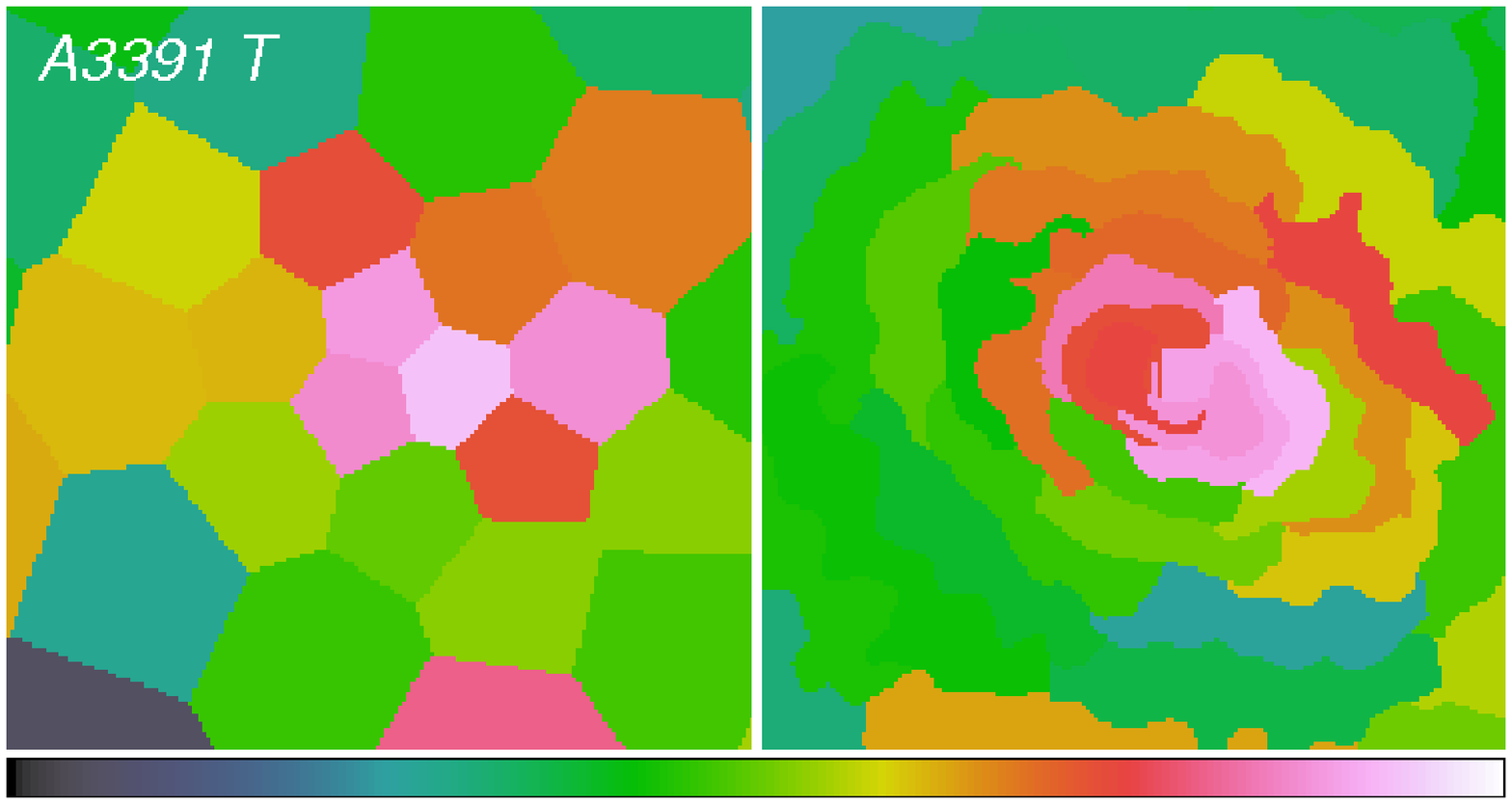}
\includegraphics[angle=0,width=8cm]{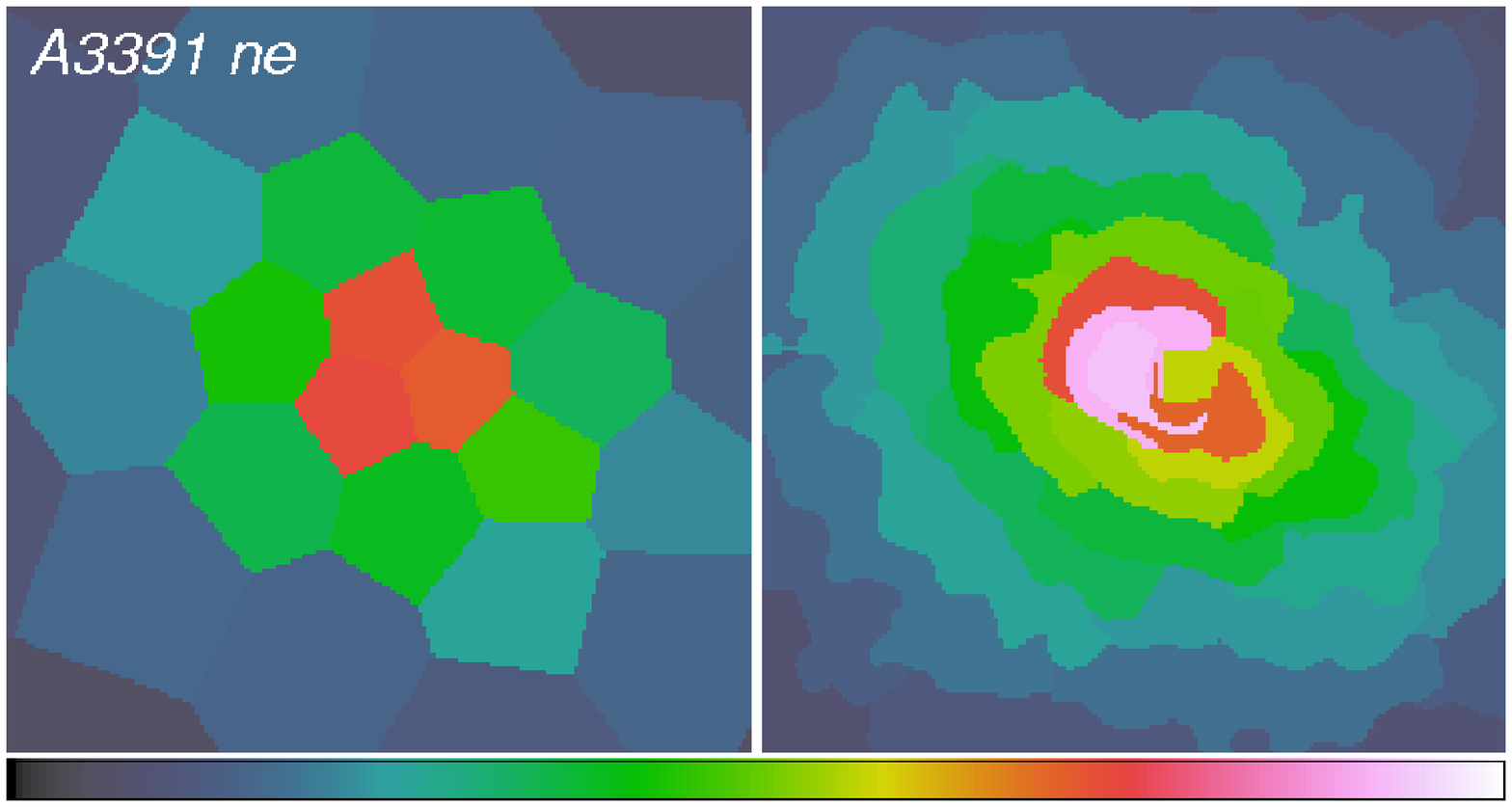}

\includegraphics[angle=0,width=8cm]{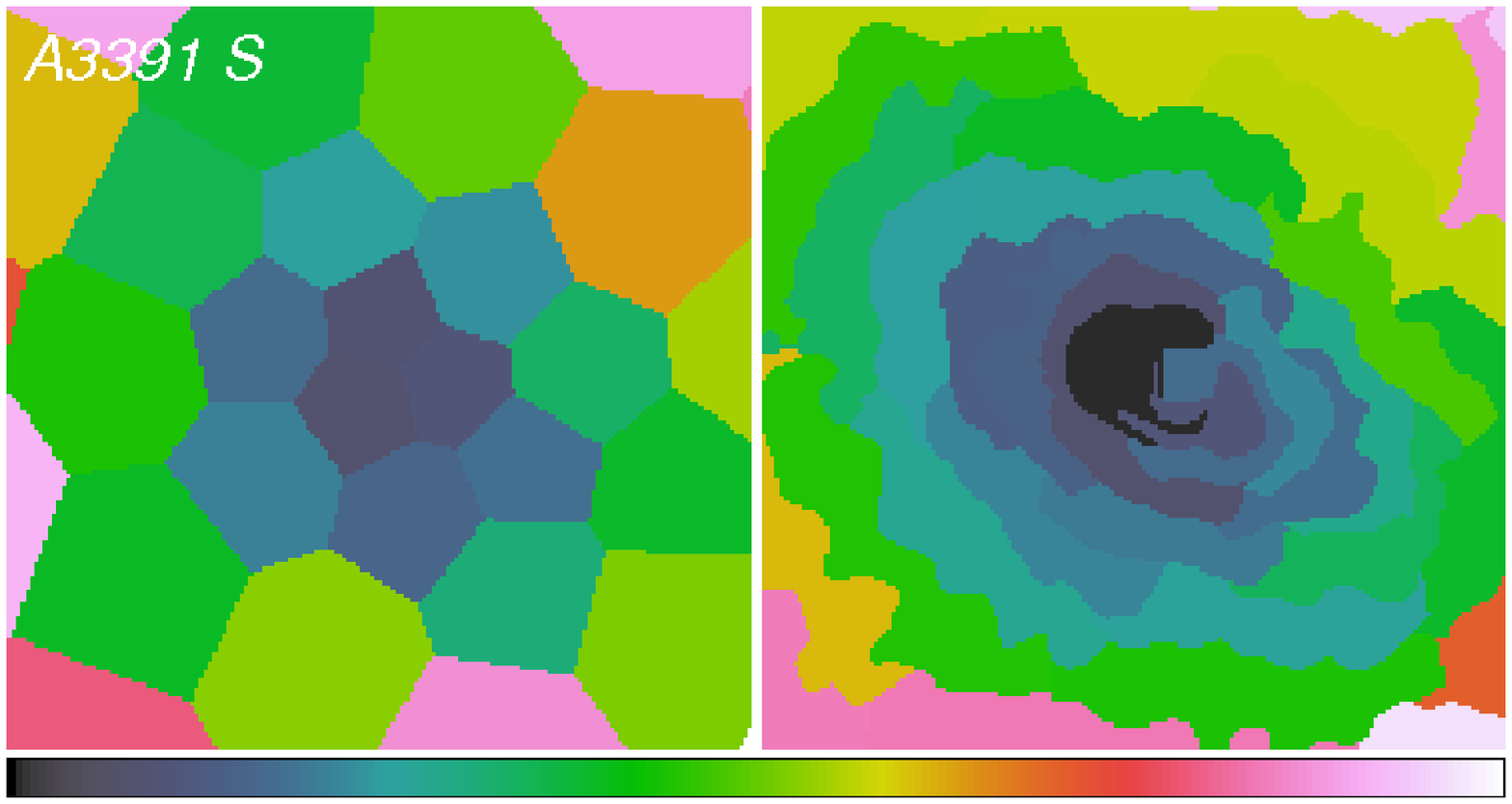}
\includegraphics[angle=0,width=8cm]{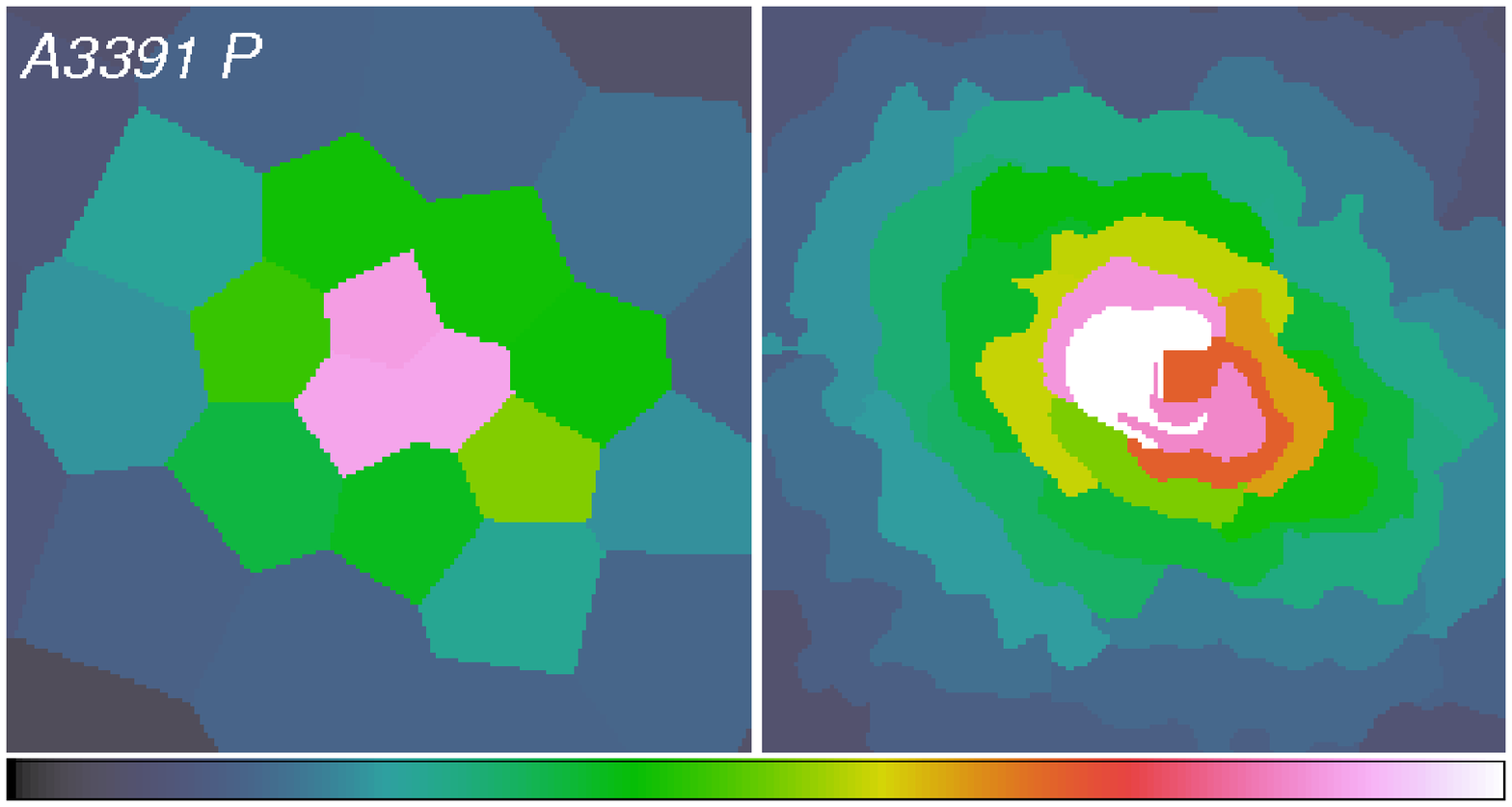}
\end{center}
\caption{Temperature ($T$), electron number density ($n_{\rm e}$), entropy
  ($S$), and pressure ($P$) maps for A3391 with the \emph{Mask-V} on the 
  left and the \emph{Mask-S} on the right in each panel. The 
  color bars are in the range of 3.3-6.5~keV,
  0-0.0046~cm$^{-3}$, 200-850~keV~cm$^2$, and 0-0.025~keV~cm$^{-3}$ in the
  $T$, $n_{\rm e}$, $S$, and $P$ panels, respectively.  The image size is 
  $11^{\prime} \times 11^{\prime}$. 
  \label{f:a3391map}}
\end{figure*}

{\it A3391}. It shows a mild increasing amplitude (from 4\% to 10\%) with 
radius in the cumulative scatter of the fluctuations (Figure~\ref{f:tsca}). The
electron number density and pressure scatter profiles show a discontinuity
around $0.2 r_{500}$ using the \emph{Mask-V}, at which radius the metallicity
also shows significant clumps as well. It has an elliptically
shaped morphology with a bi-sector feature divided by the short axis of the
elliptical as shown in Figure~\ref{f:a3391map}. It is known that A3391 is close
to the interacting cluster A3395. The sector west of the cluster core up to
$\sim 0.25 r_{500}$ shows $\sim 1$~keV higher temperature, together with low
electron number density, higher entropy and low pressure in the maps
(Figure~\ref{f:a3391map}).  These substructure features are consistent with
the observed discontinuity around $0.2 r_{500}$, and all would suggest that
some merging activities are present. A3391 is therefore an unrelaxed non-cool
core cluster. The estimates of both the X-ray observables and the X-ray
cluster mass for such an unrelaxed cluster can be biased due to the observed
substructure.

{\it EXO0422}. This cluster shows the second highest scatter amplitude,
particularly for the entropy. This is somewhat surprising because the X-ray
surface brightness appears azimuthally symmetric and the radial temperature
profile even shows a drop toward the center (Figure~\ref{f:tprof}), typical of
a cool core cluster. However, the temperature maps (Figure~\ref{f:exo0422map})
clearly show a bi-sector feature divided by the southeast-northwest axis
through the cluster core. The northeastern sector shows $\sim $0.5-1~keV
higher temperature and 0.3-0.5~$Z_{\odot}$ higher metallicity than the
southwestern sector. This feature might cause a high cluster temperature
estimate, and thus a high $Y_{\rm X}$ parameter value. However, no such
significant substructure features are shown in the electron number density,
entropy and pressure maps. This indicates the cluster is almost relaxed and
that the total mass estimate for this cluster can hardly be biased. Therefore
EXO0422 is an almost relaxed cool core cluster.  This seems to be a good
example of a cluster with some merging activities which would go unnoticed
without a temperature map. The scatter of the mass-observable relations for
such an almost relaxed cluster can be caused by the bias in its temperature
estimate and thus in its $Y_{\rm X}$ estimate due to the temperature
substructure.

{\it A0119}. It stands out clearly in radial studies of the scatter in
Figures.~\ref{f:tsca} and \ref{f:tsca_dr}.  The scatter of the
temperature fluctuations exhibits a high amplitude, particularly in
the outskirts. We observe an elongation in its X-ray morphology with a
faint emission tail (see also Buote \& Tsai 1996, Hudson et al.
2009). Though this cluster has low-quality data, the maps show clearly
an asymmetric structure with the southwestern sector up to 1-2~keV
hotter than the northeastern sector (Figure~\ref{f:a0119map}), and the
high-temperature zone is located at a central radius of $\sim
2^{\prime}$ ($\sim 0.17r_{500}$) from the cluster center using the
\emph{Mask-S} method. As the $<0.2 r_{500}$ region is excluded in the
cluster temperature determination, the cluster temperature, and thus
the $Y_{\rm X}$ parameter, is less affected by the hot structure in
the cluster core. The pressure map shows a similar feature observed in
the temperature map, while the entropy map shows less significant
features. This suggests that the fluctuations of temperature and
density are likely isentropic, which can be produced by a low Mach
number shock, compression wave, turbulence or triaxiality in the dark
matter distribution. The appearance of A0119 is in favor of being
unrelaxed. Its total mass estimate may thus be significantly affected.
A0119 is an unrelaxed non-cool core cluster.

{\it Summary}. The application of the diagnostics on the four clusters
show the differential scatter of either entropy or temperature is a
sensitive indicator of the substructure. Particularly, the temperature
map is more sensitive to unnoticed substructure which only exists in
the temperature map for an almost relaxed cluster. For an unrelaxed
cluster, the amplitudes of the scatter profiles in the 2D maps are
likely high, with a possible discontinuity in the scatter
profiles. 

\subsection{Density versus Temperature Fluctuations}

A correlation between temperature fluctuations and electron number density
fluctuations may shed light on the origin of the fluctuations, e.g.,  a
constant pressure solution yielding ratios of temperature fluctuations to
electron number density fluctuations, $-1$; and a constant entropy solution
yielding ratios, $2/3$.

To check whether the electron number density fluctuations show a correlation
with the temperature fluctuations, we performed a linear fit ($Y=A+B X $) to
the relation. It shows the highest Pearson correlation coefficient value for
A0119 (Table~\ref{t:dndtfit}), but is still hard to conclude a concrete
correlation. There is no trend of the fluctuations as a function of radius
except for a bump at radii of $0.35-0.45 r_{500}$ for the two unrelaxed
clusters, A3391 and A0119.

\subsection{Scaling Relations versus ICM $T$, $n_{\rm e}$, $S$ and $P$
  Fluctuations}
\label{s:scaling}

\begin{figure*}
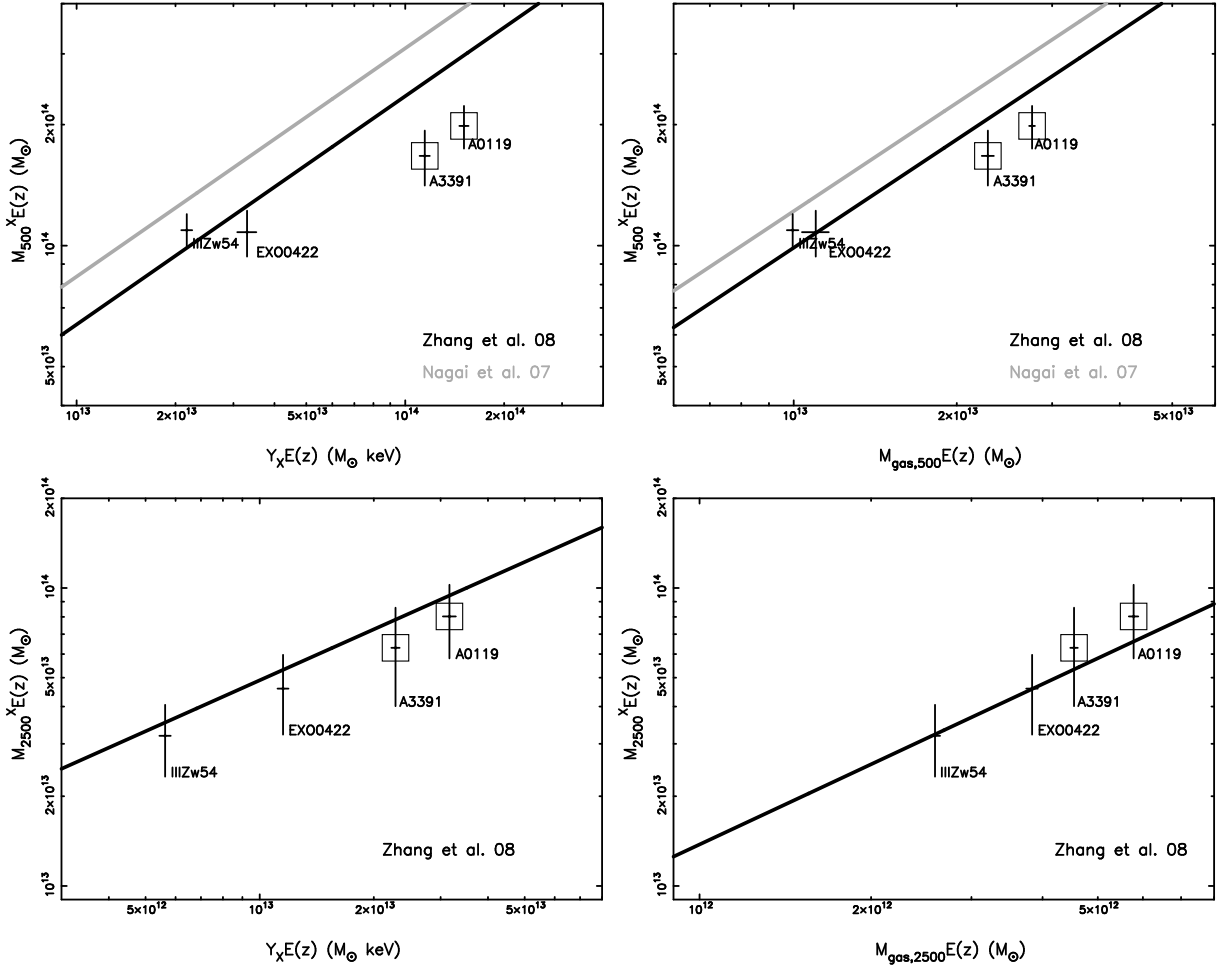

\begin{center}
  \includegraphics[angle=270,width=8.0cm]{plots/f16a.ps}
  \includegraphics[angle=270,width=8.0cm]{plots/f16b.ps}

  \includegraphics[angle=270,width=8.0cm]{plots/f16c.ps}
  \includegraphics[angle=270,width=8.0cm]{plots/f16d.ps}
\end{center}
\caption{Mass-$Y_{\rm X}$ relations (left) and the mass-$M_{\rm gas}$
  relations (right). We compile a sample of 44 LoCuSS clusters (37 LoCuSS
  clusters are from Zhang et al.  2008) and use the best-fit scaling
  relations of a subsample of 22 clusters characterized as relaxed
  (black lines) at $r_{500}$ and $r_{2500}$, respectively, as the
  reference for our studies. The scaling relations at $r_{500}$ from
  simulations in Nagai et al. (2007; gray lines) are shown for comparison.
  Clusters characterized as possibly merging/elliptical in Hudson et al.
  (2009) are denoted by open squares. The cluster masses for the four 
  clusters are determined in Sect.~\ref{s:mass}. 
  \label{f:mymg}}
\end{figure*}

The substructure diagnostics of galaxy clusters are of prime importance to the
understanding of the X-ray mass estimates and the X-ray observables.  In Zhang
et al.  (2008), we found the X-ray gas mass ($M_{\rm gas}$) and the X-ray
analog of the integrated SZ flux ($Y_{\rm X}= M_{\rm gas} \cdot
T_{0.2-0.5r_{500}}$) can be used as low scatter cluster mass indicators
compared with other X-ray observables. Therefore we present the mass-$Y_{\rm X}$
relation and mass-$M_{\rm gas}$ relation here (Figure~\ref{f:mymg}).

In simulations, there is a small intrinsic dispersion between the true mass
and the mass derived from the hydrostatic equilibrium equation for relaxed
clusters (e.g., Nagai et al. 2007). We therefore used the deviation of the
cluster mass from the mass-observable relations for a sample of relaxed
clusters as an indicator of the mass bias for the hydrostatic mass. Note that
the mass-observable relations for relaxed clusters could still be biased by
residual non-thermal support (e.g., Mahdavi et al. 2008, Zhang et al. 2008).

We have compiled a sample of 44 LoCuSS clusters (37 in Zhang et al. 2008), and
used the best-fit scaling relations of the subsample of all 22 relaxed
clusters as the reference. At $r_{2500}$, the four clusters are in good agreement
with the subsample of 22 relaxed LoCuSS clusters.  With the cluster masses
determined in Sect.~\ref{s:mass}, we observe tantalizing hints linking the
scatter of the ICM fluctuations and the hydrostatic mass bias relative to the
expected mass based on the $M$-$Y_{\rm X}$ and $M$-$M_{\rm gas}$ relations,
particularly at $r_{500}$.

A typical example of a relaxed cluster, IIIZw54 (Figure~\ref{f:iiizw54map}),
lies on the mass-observable scaling relations. A3391 is a weakly merging
cluster (Figure~\ref{f:a3391map}), and lies significantly off from the
mass-observable scaling relations. EX0422 is a mild unrelaxed cool core
cluster. Though its pressure map has no significant substructures which means
the cluster mass may be unbiased. The hot substructure in the temperature map
could cause a high cluster temperature estimate and thus a high $Y_{\rm X}$
estimate. As a result, it lies on the mass-$M_{\rm gas}$ scaling relations but
shows a small offset toward the hot side in the $M$-$Y_{\rm X}$ relation,
particularly at $r_{500}$.  A typical example of a dynamically active cluster
is A0119 (Figure~\ref{f:a0119map}). The significant feature of substructures
observed in the pressure map might cause a large bias in the cluster mass
estimate. This cluster indeed shows significant deviations in both the
$M$-$Y_{\rm X}$ and $M$-$M_{\rm gas}$ relations.

To quantify the trend between the mass bias and the scatter of the
fluctuations in the 2D maps, we defined the quasi-mass bias as follows. For
example, the quasi-true mass $M_{\Delta}^{\rm Y_{\rm X}}$ (or $M_{\Delta}^{\rm
  M_{\rm gas}}$) can be derived from $Y_{\rm X}$ (or $M_{\rm gas}$) at
$r_{\Delta}$ via the $M_{\Delta}$-$Y_{\rm X}$ (or $M_{\Delta}$-$M_{\rm gas}$)
scaling relation of a subsample of 22 LoCuSS clusters characterized as
relaxed. The quasi-mass-bias is thus defined as $B_{\rm M_{\Delta}}^{\rm
  Y_{\rm X}}=M_{\Delta}/M_{\Delta}^{\rm Y_{\rm X}} -1$ (or $B_{\rm
  M_{\Delta}}^{\rm M_{\rm gas}}=M_{\Delta}/M_{\Delta}^{\rm M_{\rm gas}} -1$).
We carried out a simple linear fit $Y=A+B X $ to the relation of the
quasi-mass-bias at $r_{\Delta}$ versus the cumulative scatter at the outermost
radius one can measure for individual clusters. The correlation is only
significant (i.e. correlation coefficient $>0.65$) using the scatter of either
entropy or pressure fluctuations and using the quasi-mass-bias at $r_{500}$.
Therefore, we only list the best fit and Pearson correlation coefficient using
the scatter of entropy and pressure fluctuations and the quasi-mass-bias at
$r_{500}$ in Table~\ref{t:bias}. We interpret this result as tentative
evidence for an interesting correlation between mass bias and scatter
amplitude. We will constrain the parameters of these relations in more detail
using a larger cluster sample. These findings shall encourage similar studies
to be carried out using hydrodynamical simulations.

\subsection{Data Quality versus Radial Studies of ICM $T$, $n_{\rm e}$, $S$ and
  $P$ Fluctuations}

As shown above, sufficient photon statistics are required to provide
quantitative diagnostics of the substructure in the ICM and to imply
detailed physics relevant to the systematics of the scaling relations.
EXO0422 has the highest data quality among the four clusters. Therefore
the ICM substructure shown in great details allows us to understand
its small offset in the scaling relations, which would have been missed
without our studies.  A0119 has insufficient photon statistics to
perform such radial studies of the ICM $T$, $n_{\rm e}$, $S$ and $P$
fluctuations. Though we observed significant features indicating the
strong merging activities in the 2D maps of A0119, the radial
profiles of the scatter show large statistical error and could not
reveal possible discontinuities. Therefore $\ge 30$ bins within the
$0.6r_{\rm tr}$ region is required for such radial studies of the
fluctuations of the spectrally measured ICM $T$, $n_{\rm e}$, $S$ and
$P$ maps. In term of net counts, $\ge$~120,000 cluster photons are
required for one nearby cluster.

\section{Conclusions}
\label{s:conclusion}

Substructure diagnostics of galaxy clusters are crucial to the robustness of
the estimates of both the cluster mass and the X-ray observables. Therefore
they have enormous importance to the understanding of the systematics and
scatter of the mass-observable scaling relations. As a result, the knowledge
of the substructure directly affects the precision of the cosmological tests
using the cluster mass function. To probe possible biases in hydrostatic mass
estimates as a function of cluster dynamical state, we developed a precise
background subtraction procedure using both MOS and pn and a spectral analysis
procedure to derive the X-ray maps via spectral measurements in each spatial
bin. With \emph{XMM-Newton} observations of the four morphological different
clusters selected from the HIFLUGCS sample, we report our procedures and
strategies for the ICM substructure studies using the spectrally measured 2D
temperature, electron number density, entropy, and pressure maps with medium
quality \emph{XMM-Newton} data for nearby clusters. Our procedures provide
detailed 2D diagnostics and a new complementary tool, the radial studies of
the fluctuations in the 2D map of ICM temperature, electron number density,
pressure and entropy, to quantify the substructure in galaxy clusters, and
attempt to explain the deviation of the cluster from the mass-observable
scaling relations.

The amplitude of and the discontinuity in the scatter provide
substructure diagnostics due to merging, the physics behind the
scatter of the mass-observable scaling relations. The amplitude
indicates $\sim 10$\% fluctuations in the temperature, electron number
density, and entropy maps, and $\sim 15$\% fluctuations in the pressure
map. The differential scatter can indicate the most disturbed radial
range, e.g.,  $0.35-0.45 r_{500}$ for the unrelaxed clusters, A3391 and
A0119.

The temperature map is particularly unique to identify the substructure of an
almost relaxed cluster which would be unnoticed in the ICM electron number
density and pressure maps.

There is a tantalizing link between the substructure identified using the
scatter of the entropy and pressure fluctuations and the hydrostatic mass bias
relative to the expected mass based on the $M$-$Y_{\rm X}$ and $M$-$M_{\rm
  gas}$ relations particularly at $r_{500}$. A typical relaxed cluster, such
as IIIZw54, lies on the mass-observable scaling relations. A weakly merging
cluster, A3391, lies significantly off from the mass-observable scaling
relations.  An almost relaxed cool core cluster, EXO0422, shows a small offset
in the $M$-$Y_{\rm X}$ relation. A typical dynamical active cluster, A0119,
shows significant mass deviation in the both $M$-$Y_{\rm X}$ and $M$-$M_{\rm
  gas}$ relations. The scatter of the observed scaling relations caused by an
unrelaxed cluster can be due the mass estimate being biased by the pressure
substructure and the temperature estimate biased by the temperature
substructure in this cluster, e.g., A0119. The scatter of the observed scaling
relations can also be caused by an almost relaxed cluster, due to the bias in
its temperature estimate affected by its temperature substructure, e.g.,
EXO0422.

\emph{XMM-Newton} observations with $\ge$~120,000 source photons per cluster
are sufficient to apply our method for detailed diagnostics to identify the
substructures of the clusters. More concrete conclusions require such
substructure studies using a statistically large sample, with $\ge$~120,000
source photons per cluster in their \emph{XMM-Newton} observations; this is
work in progress. It will then be interesting to make a detailed comparison of
a possible scatter - mass-bias correlation with the results of numerical
simulations.

\acknowledgments

The \emph{XMM-Newton} project is an ESA Science Mission with instruments and
contributions directly funded by ESA Member States and the USA (NASA). The
\emph{XMM-Newton} project is supported by the Bundesministerium f\"ur
Wirtschaft und Technologie/Deutsches Zentrum f\"ur Luft- und Raumfahrt
(BMWI/DLR, FKZ 50 OX 0001) and the Max-Planck Society. Y.Y.Z. acknowledges J.  de
Plaa, S.L. Snowden, H. B\"ohringer, and J. S. Sanders for useful discussions.
Y.Y.Z., T.H.R. and D.S.H. acknowledge support by the DFG through Emmy Noether Research
Grant RE\,1462/2 and the Transregional Collaborative Research Centre TRR33
``The Dark Universe'' and by the German BMBF through the Verbundforschung
under grant\,50\,OR\,0601. A.F. acknowledges support from BMBF/DLR under
grant\,50\,OR\,0207 and MPG and support through the funding of the DFG
for the Excellence Cluster Universe EXC153. CLS was supported in part by NASA
\emph{XMM-Newton} grants NNX06AE76G, NNX08AZ34G, and NNX08AW83G, and by
Chandra grant GO7-8129X. This work has been partially supported from NASA
grants NNX07AV73G and NNX07AT29G. For help in the early stage of the project,
we acknowledge H.  B\"ohringer, T. Clarke, T.  Erben, A.  Evrard, Y.  Fujita,
Y.  Ikebe, O.-E.  Nenestyan, E. Pierpaoli, S.  Randall, and P.  Schuecker.



\clearpage

\begin{deluxetable}{lccccrcccccc}
\tabletypesize{\scriptsize}
\rotate
\tablecaption{Cluster Properties and
      \emph{XMM-Newton} Observations.\label{t:basic}}
\tablewidth{0pt}
\tablehead{
\colhead{Name} & \colhead{OBS-ID} & \colhead{} & \colhead{Net exposure (ks)} & \colhead{} &
\colhead{Mode} &  \colhead{X-ray centroid} & \colhead{(J2000)} &
\colhead{$z$}  & \colhead{$N_{\rm H}$} & \colhead{$T_{0.2-0.5r_{500}}$} &
\colhead{$r_{\rm tr}$} 
}
\startdata
        &            & MOS1 & MOS2 & pn             & pn  & R.A.       &
        Decl.   &        &$10^{22}{\rm cm}^{-2}$& keV & arcmin\\
\hline
IIIZw54 & 0505230401 & 23.3 & 22.3 & 30.3           & EFF &$ 03:41:18.729 $&$
+15:24:13.91 $& 0.0311 & 0.1470 & $ 2.17 \pm 0.03 $ & 13.35\\
A3391   & 0505210401 & 23.3 & 24.6 & 18.2           & EFF &$ 06:26:24.222 $&$
-53:41:24.02 $& 0.0531 & 0.0559 & $ 5.02 \pm 0.05 $ & 13.32\\
EXO0422 & 0300210401 & 31.5 & 32.2 & 32.7           & EFF &$ 04:25:51.224 $&$
-08:33:40.34 $& 0.0390 & 0.0808 & $ 2.99 \pm 0.03 $ & 13.29\\
A0119   & 0505211001 & 8.2  & 8.0  & 7.6            & FF  &$ 00:56:17.119 $&$
-01:15:11.98 $& 0.0440 & 0.0328 & $ 5.47 \pm 0.11 $ & 11.00\\
\enddata
\tablecomments{The EFF mode is the extended full frame mode. The MOS data are
  in FF mode. The truncation radius ($r_{\rm t}$) is the radius corresponding
  to an S/N of 3.}
\end{deluxetable}

\begin{deluxetable}{lcccccc}
  \tabletypesize{\scriptsize} \rotate \tablecaption{The Best Linear Fit
    ($Y=A+ B X$) of the Relation of Electron Number Density
    Fluctuations versus Temperature Fluctuations.
\label{t:dndtfit}}
\tablewidth{0pt}
\tablehead{
\colhead{Name} & & \colhead{\emph{Mask-V}} & &  & \colhead{\emph{Mask-S}} & \\ 
}
\startdata
               &  $A$  &  $B$  &  Coefficient  &  $A$  &  $B$  &  Coefficient \\
\hline
IIIZw54 &  $0.029 \pm 0.004$ & $0.03 \pm 0.08$ & 0.154 & $0.026 \pm 0.002$ &
$0.05 \pm 0.03$ & 0.321 \\
A3391   &  $0.038 \pm 0.003$ & $0.06 \pm 0.03$ & 0.335 & $0.043 \pm 0.003$ &
$0.05 \pm 0.03$ & 0.285 \\
EXO0422 &  $0.028 \pm 0.002$ & $0.01 \pm 0.02$ & 0.066 & $0.029 \pm 0.001$ &
$0.03 \pm 0.02$ & 0.323 \\
A0119   &  $0.046 \pm 0.003$ & $0.06 \pm 0.02$ & 0.765 & $0.052 \pm 0.003$ &
$0.08 \pm 0.02$ & 0.830 \\
\hline
All four clusters  &  $0.032 \pm 0.001$ & $0.03 \pm 0.02$ & 0.192 & $0.034 \pm 0.001$ & 
$0.04 \pm 0.01$ & 0.263 \\
\enddata
\end{deluxetable}

\begin{deluxetable}{llcccccc}
  \tabletypesize{\scriptsize} \rotate \tablecaption{The Best Linear Fit ($Y=A+
    B X$) of the Relation of the Quasi-Mass Bias, Column~(1), versus the
    Cumulative Scatter at the Outermost Radius One can Measure for
    Individual Clusters, Column~(2). 
\label{t:bias}}
\tablewidth{0pt}
\tablehead{
\colhead{Bias}                   & Scatter
& & \colhead{\emph{Mask-V}} & 
& & \colhead{\emph{Mask-S}} & \\ 
}
\startdata
                             &                    
&  $A$  &  $B$  &  coeff.  
&  $A$  &  $B$  &  coeff. \\
\hline
$B_{\rm M_{500}}^{\rm Y_{\rm X}}$   & $S$
&  $1.45 \pm 0.38$ &$-14.3 \pm 3.3$ &-0.98 
&  $0.61 \pm 0.22$ &$-6.6  \pm 1.9$ &-0.80 \\
                              & $P$
&  $1.07 \pm 0.29$ &$-8.0 \pm 1.9$ &-0.97 
&  $1.12 \pm 0.31$ &$-9.2 \pm 2.2$ &-0.97 \\
$B_{\rm M_{500}}^{\rm M_{\rm gas}}$ & $S$
&  $1.06 \pm 0.41$ &$-9.8 \pm 3.6$ &-0.95 
&  $0.42 \pm 0.23$ &$-4.0 \pm 2.0$ &-0.69 \\
                              & $P$
&  $0.79 \pm 0.32$ &$-5.5 \pm 2.1$ &-0.94 
&  $0.76 \pm 0.32$ &$-5.9 \pm 2.3$ &-0.90 \\
\enddata
\tablecomments{See details in Sect.~\ref{s:scaling}.}
\end{deluxetable}

\clearpage

\clearpage

\clearpage

\clearpage 

\clearpage

\clearpage 

\clearpage 

\clearpage 

\begin{figure*}
\begin{center}
  \includegraphics[angle=270,width=8.0cm]{plots/f15a.ps}
  \includegraphics[angle=270,width=8.0cm]{plots/f15b.ps}

  \includegraphics[angle=270,width=8.0cm]{plots/f15c.ps}
  \includegraphics[angle=270,width=8.0cm]{plots/f15d.ps}
\end{center}
\vspace{-0.5cm}
\caption{Differential scatter of the temperature, electron number
  density, entropy, and pressure fluctuations in the 2D maps
  for each cluster using 
  \emph{Mask-S}. The X-axis has been shifted by
  0.005, 0.010, and 0.015 for A3391, EXO0422 and A0119, respectively,
  to avoid the overlap. Note that for each plot, the mean profile
  $\langle D(d)\rangle$ is individually determined for each cluster as
  described in Sect.~\ref{s:step2}.
\label{f:tsca_dr}}
\end{figure*}

\end{document}